\documentclass{aastex63}
\usepackage{amsmath}

\received{XXX XX , 2020}
\revised{XXXX XX , 2020}
\accepted{XXXX XX, 2020}
\submitjournal{ApJ}

\shorttitle{Revisiting Attenuation Curves}
\shortauthors{Calzetti et al.}

\begin{document}

\title{Revisiting Attenuation Curves: the Case of NGC 3351
\footnote{Based on observations 
obtained with the NASA/ESA Hubble Space Telescope, at the Space Telescope Science Institute, which is operated by the Association of 
Universities for Research in Astronomy, Inc., under NASA contract NAS 5-26555. }}

\correspondingauthor{Daniela Calzetti}
\email{calzetti@astro.umass.edu}

\author[0000-0002-5189-8004]{Daniela Calzetti}
\affiliation{Department of Astronomy,\\
University of Massachusetts,\\
710 N. Pleasant Street, LGRT 619J, \\
Amherst, MA 01002, USA}

\author[0000-0003-4569-2285]{Andrew J. Battisti}
\affiliation{Research School of Astronomy and Astrophysics,\\
Australian National University\,\
 Cotter Road, Weston Creek, \\
 ACT 2611, Australia}

\author[0000-0003-4702-7561]{Irene Shivaei}
\affiliation{Hubble Fellow}
\affiliation{Steward Observatory,\\
University of Arizona\,\
933 North Cherry Avenue,\\ 
Tucson, AZ 85721, USA}

\author[0000-0003-1427-2456]{Matteo Messa}
\affiliation{Observatoire de Gen\`eve\\
Universit\'e de Gen\`eve\\
Chemin Pegasi, 51\\
1290 Versoix, Switzerland}

\affiliation{Department of Astronomy\\
Stockholm University\\
Roslagstullsbacken 21\\
SE-114 19 Stockholm, Sweden}

\author[0000-0001-6291-6813]{Michele Cignoni}
\affiliation{Dipartimento di Fisica,\\
Universit\`a degli Studi di Pisa,\\
Largo Bruno Pontecorvo 3,\\
 56127 Pisa, Italy}
 
 \affiliation{INFN,\\ Largo B. Pontecorvo 3,\\ 
 56127, Pisa, Italy}

 \affiliation{INAF / Osservatorio di Astrofisica e Scienza dello Spazio,\\ 
 Via Gobetti 93/3, \\
 40129, Bologna, Italy}
 
\author[0000-0002-8192-8091]{Angela Adamo}
\affiliation{Department of Astronomy,\\
Stockholm University,\\
Roslagstullsbacken 21,\\
SE-114 19 Stockholm, Sweden}

\author[0000-0002-5782-9093]{Daniel A. Dale}
\affiliation{Department of Physics and Astronomy,\\
University of Wyoming,\\
1000 E. University,\\
Laramie, WY 82071, USA}

\author[0000-0001-8608-0408]{John S. Gallagher}
\affiliation{Department of Astronomy,\\
University of Wisconsin -- Madison,\\
475 N. Charter Street,\\ 
Madison, WI 53706, USA}

\author[0000-0002-3247-5321]{Kathryn Grasha}
\affiliation{Research School of Astronomy and Astrophysics\\
Australian National University\\
 Cotter Road, Weston Creek, \\
 ACT 2611, Australia}

\author[0000-0002-1891-3794]{Eva K.  Grebel}
\affiliation{Astronomisches Rechen-Institut\\
 Zentrum f\"ur Astronomie der Universit\"at Heidelberg\\
 M\"onchhofstra{\ss}e 12-14\\
 D-69120 Heidelberg, Germany}

\author[0000-0001-5448-1821]{Robert C. Kennicutt}
\affiliation{Steward Observatory\\
University of Arizona\\
933 North Cherry Avenue,\\ 
Tucson, AZ 85721, USA}

\affiliation{George P. and Cynthia W. Mitchell Institute for Fundamental Physics \& Astronomy\\
Texas A\&M University\\
College Station, TX  77843, USA }

\author[0000-0002-1000-6081]{Sean T. Linden}
\affiliation{Department of Astronomy\\
University of Massachusetts\\
710 N. Pleasant Street, LGRT 619J \\
Amherst, MA 01002, USA}

\author[0000-0002-3005-1349]{G\"oran \"Ostlin}
\affiliation{Department of Astronomy\\
Stockholm University\\
Roslagstullsbacken 21\\
SE-114 19 Stockholm, Sweden}

\author[0000-0003-2954-7643]{Elena Sabbi}
\affiliation{Space Telescope  Science Institute\\
3700 San Martin Drive\\
Baltimore, MD 21218, USA}

\author[0000-0002-0806-168X]{Linda J. Smith}
\affiliation{European Space Agency (ESA), ESA Office, Space Telescope  Science Institute\\
3700 San Martin Drive\\
Baltimore, MD 21218, USA}

\author[0000-0002-0986-4759]{Monica Tosi}
\affiliation{INAF /  Osservatorio di Astrofisica e Scienza dello Spazio di Bologna\\
Via Gobetti 93/3\\
40129 Bologna, Italy}

\author[0000-0001-8289-3428]{Aida Wofford}
\affiliation{Universidad Nacional Aut\'onoma de M\'exico
Instituto de Astronom\'ia, AP 106\\
Ensenada 22860, BC, M\'exico}

\begin{abstract}
Multi--wavelength images from the farUV ($\sim$0.15~$\mu$m) to the sub--millimeter of the central region of the galaxy NGC\,3351 are analyzed 
to constrain its stellar populations and dust attenuation. Despite hosting a $\sim$1~kpc circumnuclear starburst ring, NGC\,3351 deviates from the IRX--$\beta$ relation, the relation between the infrared--to--UV luminosity ratio and the UV continuum slope $\beta$ that other starburst galaxies follow.  
To understand the reason for the deviation, we leverage the high angular resolution of archival nearUV--to--nearIR HST images to divide the ring into $\sim$60--180~pc size 
regions and model each individually. We find that the UV slope of the combined intrinsic (dust--free) stellar populations in the central region  is redder  than 
what is expected for a young model population. This is due to the region's complex star formation 
history, which boosts the nearUV emission relative to the farUV.  The resulting net attenuation curve has a UV slope that lies between those of  the starburst attenuation curve \citep{Calzetti+2000} and the Small Magellanic Cloud extinction curve; the total--to--selective attenuation value, R$^{\prime}$(V)=4.93, is larger than both. As found for other star--forming galaxies, the stellar continuum of NGC\,3351 is less attenuated than the ionized  gas, with E(B--V)$_{star}$=0.40 E(B--V)$_{gas}$. 
The combination of the `red' intrinsic stellar population and the new attenuation curve fully accounts for the location of  the central region of NGC\,3351 on the IRX--$\beta$ diagram. Thus, the observed characteristics result from the complex mixture of stellar populations and dust column densities in the circumnuclear region. Despite being a sample of 
one, these findings highlight the difficulty of defining attenuation curves of general applicability outside the regime of centrally--concentrated starbursts. 
\end{abstract} 


\section{Introduction} \label{sec:intro}

Although dust represents $\sim$1\% or less of the interstellar medium mass in a typical galaxy, it has a disproportionate impact on  its light emission and, as a consequence, 
on the galaxy's physical parameters that are derived from measurements of that light. Stellar mass and star formation rate (SFR) estimates, for example, can be underestimated by large factors, up to $\sim$2$\times$ for the mass and $\sim$10$\times$ for the SFR, if the ultraviolet--to--near~infrared spectral energy distribution  (UV--to--NIR SED) of a galaxy is not corrected for the effects of dust attenuation\footnote{In this paper we distinguish between dust extinction and dust attenuation. Dust extinction describes the optical properties and amount of the dust along the line of sight. Extinction is measured for a point source. Dust attenuation combines the effects of extinction with those of the geometrical distribution of dust, gas, and stars, including dust scattering into the line of sight. Dust attenuation is characteristic of extended sources \citep{Calzetti+1994}}. Roughly 25\% of the cosmic SFR of galaxies is detected directly in the UV in the redshift range 0--2.5, while the remaining $\sim$75\% is re--processed by dust into the Far--Infrared  (FIR) \citep{Lutz2014, MadauDickinson2014, Casey+2014, Casey+2018}.

Efforts to characterize the effects of dust  attenuation in galaxies span almost four decades, since the IRAS satellite began to show that galaxies are systematic infrared emitters \citep{Soifer+1984} and models began to reveal that the dust is mixed with the stellar 
populations in more complex geometries than simple foreground screens \citep{Witt+1992}.  The general effect of dust absorption is to dim and (often, but not always) redden the UV--to--NIR emission of stellar populations; the dust--absorbed stellar light is re--emitted in the IR--to--mm wavelength range, beyond $\approx$5~$\mu$m.

UV--to--NIR attenuation curves have been derived for classes of galaxies using the `pair--method': attenuated SEDs are compared with intrinsic (unattenuated or dust--free) SEDs within the same galaxy class to derive the attenuation curve \citep[e.g.,][]{Calzetti+1994, Wild+2011, Reddy+2015, Scoville+2015, Battisti+2016, Battisti+2017a, Teklu+2020, Shivaei+2020a}. In this context, a `class' is a set of galaxies whose intrinsic stellar population SEDs can be reasonably assumed to be the same across the class. The pair--method is widely used to derive extinction curves from SEDs of individual stars, so the method is borrowed for application to the more complex conditions of entire galaxies. Different authors have made different choices in regard to the source of intrinsic SED: either observational, derived from the data with the lowest  amount of measured dust attenuation \citep{Calzetti+1994, Calzetti+2000, Wild+2011, Reddy+2015, Battisti+2016, Teklu+2020, Shivaei+2020a}, or theoretical, i.e., derived from models of the expected intrinsic stellar population \citep{Meurer+1999, Overzier+2011, Scoville+2015}. Both choices are justifiable within the context of applications discussed in each paper.  

A second requirement of the pair--method is that progressively more attenuated and redder SEDs can be parametrized by one variable. In the case of extinction in front of a point source, the variable is the color excess:
\begin{equation}
E(B-V)={A(B)-A(V) \over k(B)-k(V)} = A(B)-A(V),
\end{equation}
which is proportional to the dust column density. A(B) [A(V)] is the total extinction in the B [V] band, and  k(B) [k(V)] is the extinction curve in the same band, usually scaled to k(B)--k(V)$\equiv$1 \citep{Cardelli+1989, Mathis1990, Fitzpatrick1999, Fitzpatrick+2007, Fitzpatrick+2019}. The extinction curve encodes the properties of the dust \citep{Weingartner+2001, Draine2003}. For nearby galaxies with  L$_{IR}\lesssim$a few$\times$10$^{11}$~L$_{\odot}$ and whose emission is dominated by a central starburst\footnote{The definition of starburst  galaxy is not rigorous, and for the purpose of this work we adopt that of \citet{Heckman2000}. It is based on several, non--independent criteria: (1) a $\sim$0.1--1~kpc  size region of active star formation located in the center of the galaxy with (2) a burst intensity, as measured  by the SFR surface density, that is about 10$^3$  times greater than the typical  SFR surface density of galaxy disks, and (3) a  gas consumption timescale of $\approx$10$^8$ yr or smaller. Conversely, star--forming galaxies are less active than starbursts, with  lower SFR  surface densities, star formation distributed across their disks and gas consumption timescales $>$ Gyr.}, \citet{Calzetti+1994, Calzetti+1996}  found that the relevant parametrization is the color excess of the ionized gas, $E(B-V)_{gas}$. This parameter, measured from hydrogen recombination line ratios, has the convenience that the intrinsic line ratio is determined by quantum physics. In the central starburst and star--forming regions of nearby galaxies, $E(B-V)_{gas}$ is correlated with $\beta$, the slope of the observed UV continuum spectrum; this correlation has enabled the derivation of attenuation curves, for starbursts \citep{Calzetti+1994, Calzetti+2000} and star--forming galaxies \citep{Battisti+2016}. The same parameter, $E(B-V)_{gas}$, will be used as reference value in our analysis and we will refer to  the  starburst attenuation curve of \citet{Calzetti+1994, Calzetti+2000} as `SB' in the rest of the paper.

\citet{Meurer+1999} found that the SB attenuation curve reproduces the trend between the observed FIR--to--UV luminosity ratio, a 
measure of the fraction of stellar light from recent star formation absorbed by dust, and $\beta$. This correlation, called the IRX--$\beta$ relation, is potentially a powerful predictor of total SFR when only a limited amount of information, e.g.,  a UV color or slope, is available. Observations confirm that the relation is broadly applicable out to redshift z$\sim$2--3 \citep{Reddy+2006, Reddy+2010, Reddy+2012, Overzier+2011, Forrest+2016, McLure+2018, Shivaei+2020b} or even higher redshift \citep{Bouwens+2016, Fudamoto+2020, Bouwens+2020} for massive galaxies; it can be used to recover the intrinsic average stellar properties (SFRs, masses) of high redshift galaxies, when matched dust emission measurements are not available. 

However, significant  deviations from the canonical IRX--$\beta$ relation of \citet{Meurer+1999} have also been found. Luminous and Ultraluminous Infrared galaxies, with L$_{IR}>$a few$\times$10$^{11}$~L$_{\odot}$,  populate  the region {\em above} the IRX-$\beta$ relation: at a given UV slope, these galaxies are overluminous in the FIR relative to the measured UV emission \citep{Goldader+2002, Reddy+2006, Reddy+2010, Overzier+2011, Casey+2014}. This effect can be readily accounted for with mixing of dust and stars, and other geometrical effects, in the galaxies \citep{Calzetti2001, Popping+2017}. 

\begin{figure}[t]
\plotone{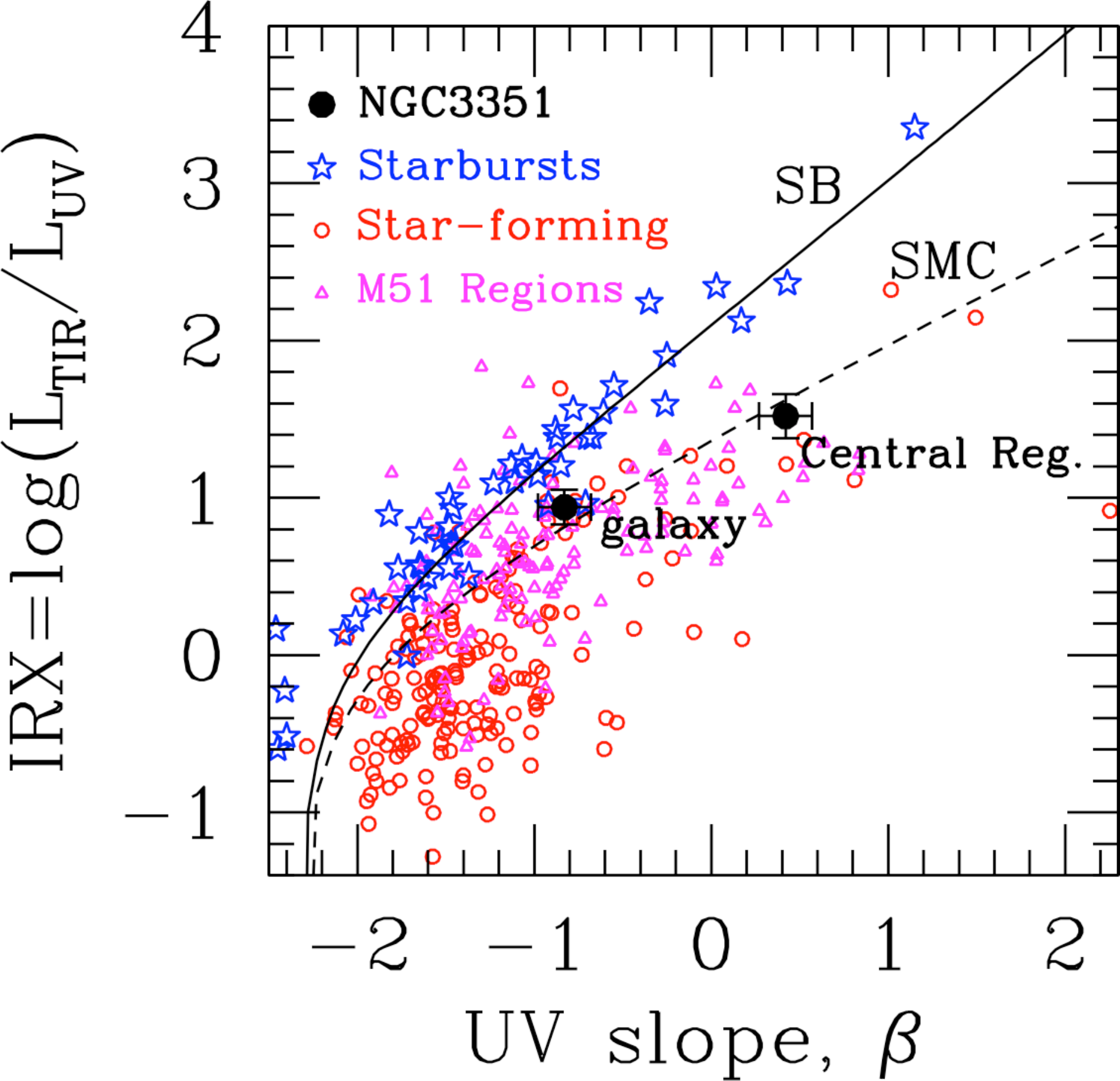}
\caption{The ratio of IR--to--UV luminosity, IRX, as a function of the  UV spectral slope $\beta$ for local galaxies. The IR--to--UV luminosity is defined as the ratio between the dust emission in the infrared wavelength range 3-1,100~$\mu$m (TIR) and the observed luminosity in the farUV (FUV), typically $\approx$0.15~$\mu$m. Here, L$_{UV}$ is the FUV filter from GALEX. The black circles with 1~$\sigma$ error--bars mark the  position of NGC~3351 (galaxy) and its central, 0.5~kpc radius, starburst region (Central Reg.). The locations on this plot of nearby starburst galaxies \citep[blue stars,][]{Calzetti+1994, Meurer+1999}, star--forming galaxies from the Local Volume Legacy survey \cite[red circles,][]{Dale+2009}, and $\sim$0.5~kpc star--forming regions in the galaxy NGC~5194  \citep[magenta triangles,][]{Calzetti+2005} are shown. The galaxies in these samples have L$_{TIR}\lesssim$a~few$\times$10$^{11}$~L$_{\odot}$, with the star--forming galaxies being typically fainter than the starbursts. The relation traced by the starburst attenuation curve  of \citet[][SB]{Calzetti+1994,Calzetti+2000} is shown as a continuous curve, and that of the SMC extinction curve of \citet{Pei1992} as a dashed curve. The attenuation/extinction curves impart a high level of non--linearity to UV spectra, thus the measured slope $\beta$ depends on the chosen UV wavelength range. For this plot, the UV range is that of the GALEX filters. \label{fig:IRX_beta}}
\end{figure}

Conversely, the z=0 star--forming galaxies that populate the Main Sequence of Star Formation \citep{Cook+2014} are usually located {\em below} the IRX--$\beta$ relation, i.e., they have low IR emission for their UV slope \citep[Figure~\ref{fig:IRX_beta};][]{Buat+2002, Buat+2005,  Cortese+2006, Dale+2009}. The same is true for regions within galaxies \citep{Calzetti+2005, Boquien+2012}. At z$\gtrsim$2,  low--mass starburst galaxies begin to deviate from the z=0 canonical IRX--$\beta$ relation as well, and by z$\sim$4--5 they tend to be systematically below it, which may be indicative of a trend towards steeper attenuation curves for increasing redshift, possibly linked to decreasing age, mass, and/or  metallicity \citep{Shivaei+2015, Capak+2015, Bouwens+2016, Salmon+2016, Pope+2017, Reddy+2018, Fudamoto+2020, Shivaei+2020b}. 

The deviations of the Main Sequence galaxies from the z=0 canonical IRX-$\beta$ relation require more than simple dust/star geometrical models to be explained. In addition to the deviation of their mean trend from the locus  marked by the SB curve, star--forming galaxies and regions show a large, seemingly irreducible, scatter in the IRX--$\beta$ plane \citep[e.g.,][]{Kong+2004, Calzetti+2005, Dale+2009, Conroy+2010, Battisti+2016}. Complex mixes of stellar populations with different ages and optical depths \citep{Kong+2004, Calzetti+2005, Boquien+2012, Grasha+2013, Nersesian+2020}, and/or variations, typically a steepening, in the slope of the dust attenuation curve \citep{Noll+2009, Reddy+2010, Reddy+2018, Salim+2019, Shivaei+2020b} have both been invoked. A decrease in the total--to--selective normalization, R(V)=A(V)/E(B--V), of the attenuation curve produces a similar, although milder, effect to steepening \citep{Reddy+2018}. 

The significant deviations from the IRX--$\beta$ relation are mirrored by comparable deviations from the SB attenuation curve, measured directly from the UV--to--NIR SEDs of star forming galaxies both at low and high redshift. Several authors report measuring dependency   
on the galaxy's inclination angle \citep[e.g.,][]{Conroy+2010, Wild+2011, Chevallard+2013, Battisti+2017b}, and on the galaxies stellar populations' age, mass, metallicity, and SFR surface densities \citep{Battisti+2016, Reddy+2018, Salim+2018, Teklu+2020, Shivaei+2020a, Nersesian+2020}. 

Over the past $\sim$2 decades, progress in modeling the emission from stellar populations and dust in galaxies has enabled use of multi--wavelength SED fitting as a tool to effectively gauge and remove the effects of dust attenuation \citep{daCunha+2008, Wild+2011, Utomo+2014, Reddy+2012b, Shivaei+2015, Shivaei+2016, DeBarros+2016, Battisti+2016, Salmon+2016, Leja+2017, Boquien+2019, Hunt+2019}. The basic approach consists of modeling the galaxy's SED from the UV to the FIR/mm, adopting an energy balance technique to reconstruct the intrinsic stellar population SED from the attenuated one in the UV--to--NIR and the dust emission in the Mid--IR to mm wavelength region \citep[e.g.,][]{daCunha+2008, Conroy2013, Boquien+2019}.  Several assumptions are built into these fitting algorithms, with the most consequential ones being the star formation history (SFH), the dust attenuation recipe, and the number of attenuation components included in the model. The attenuation components correspond to regions where the amount of dust is assumed to be different from the galaxy's average, based on certain characteristics of the regions (e.g., the age of the local stellar population); an example is the two--component model, one for star forming regions and one for the diffuse medium of galaxies, by \citet{CharlotFall2000}. 

Those SED fitting algorithms have shown that there is  a high degree of degeneracy between the adopted SFHs and the attenuation curve(s), especially for extended regions or entire galaxies where multiple generations of stars are present. 
The observed UV colors of a young stellar population with an attenuation curve that has a steep UV slope can often be exchanged for those of an older stellar population with a shallower UV attenuation curve \citep{Calzetti2001, Battisti+2016, Popping+2017, Narayanan+2018}. Thus, analyses of multi--wavelength SEDs of galaxies require trade--offs between allowing the broadest range in SFHs and attenuation curves and converging to a number of manageable minima in the fits; this is, for instance, pushing  investigations of physically--motivated assumptions for the SFHs \citep{Leja+2019}. The trade--offs also depend on the wavelength coverage and density of the data (e.g., sparse photometry versus continuous spectroscopy). For this reason, models and simulations suggest that the broad range of IRX--$\beta$ values found in star forming galaxies at low and high redshift can result from multiple, non--mutually exclusive, effects: complex geometries and a range of attenuation curves, as well as complex SFHs \citep{Seon+2016, Popping+2017, Narayanan+2018, Trcka+2020, Salim+2020}.

With the advent of the James Webb Space Telescope, Euclid, the Nancy Grace Roman Space Telescope, and the Extremely Large Telescopes, large, deep, and wide--field samples of high redshift galaxies will be secured, all the way to the epoch of Reionization of the Universe and beyond. Observations will capture the restframe UV and optical SEDs of these galaxies, but complementary restframe IR ($>$10~$\mu$m) data may not be generally available to capture the dust emission. Even with the sensitivity of ALMA, observing efficiency limitations due to small fields--of--view, difficulty in covering multiple rest--frame IR wavelengths, and low dust contents will hamper large surveys of typical (L$^*$) star--forming galaxies at z$\gtrsim$1. Establishing the observational characteristics and dependencies of dust attenuation on broad parameters, including the SFH,  thus becomes key for extracting accurate physical quantities from observations of high--redshift galaxies. The present work aims at helping address this issue.  

The goal of this paper is to separate the effects of SFH from those of dust attenuation by dissecting the central  $\sim$1~kpc region of a nearby starburst galaxy, NGC~3351. Despite its classification as a starburst, 
the galaxy, and its circumnuclear ring of star formation, are underproducing the dust infrared emission relative to expectations for starburst galaxies (Figure~\ref{fig:IRX_beta}). We use both low--resolution and high--resolution imaging data  from the UV to the 
sub--mm to model the SEDs of individual, tens--to--hundreds~pc regions within the central starburst of this galaxy. The regions's sizes, comparable to those of HII regions and complexes, are small enough that simplifying assumptions about their individual SFHs can be made. By summing each region's contribution, we can therefore reconstruct the SFH of the 1~kpc starburst region, and use this as a prior to derive the attenuation curve. Albeit with several limitations, the small--region fitting approach enables us to account for most of the observational characteristics of the starburst, including its low infrared luminosity. We provide suggestions for interpreting the IRX--$\beta$ locus in the case of complex stellar populations like those found in the center of NGC~3351.

The outline of the paper is as follows.  Section~\ref{sec:galaxy} describes the general characteristics of the galaxy NGC\,3351  and of its central starburst region. Section~\ref{sec:nomenclature}  gives an overview of the nomenclature attributed to separate components of the central starburst region. The data  used in this work are presented in  Section~\ref{sec:data}, and the region selection and photometry are in Section~\ref{sec:photometry}. The models used for comparison with the data and the SED fitting approach are described in Section~\ref{sec:models}. Sections~\ref{sec:central}, \ref{sec:regions}, and \ref{sec:residual} present the results of  the best fits of  individual components, while Section~\ref{sec:attenuation} derives the attenuation curve  for the starburst center of NGC\,3351. Finally, the results are discussed in Section~\ref{sec:results} and a summary is given in Section~\ref{sec:conclusions}. 

\section{The Galaxy NGC~3351}\label{sec:galaxy}

NGC~3351 is a barred spiral located at a Cepheid--based distance of 9.33~Mpc \citep{Freedman+2001}, with inclination of 40$^o$ and suffering from a small amount of 
foreground Galactic extinction, E(B$-$V)=0.024 (from NED\footnote{NED=NASA Extragalactic Database, $http://ned.ipac.caltech.edu/$}). A $\sim$750~pc--diameter circum--nuclear ring of star formation is fed by the bar \citep{Regan+2006}, and is the brightest feature in this galaxy, in stellar, ionized gas, dust, and molecular gas light \citep{Knapen2005, Regan+2006, Leroy+2009}. The ring is responsible for $\sim$85\% of the UV light  in the central $\sim$1~kpc region, thus for the vast majority of the recent star formation in the area. Our analysis concentrates on this 1~kpc Central Region where the starburst ring is located, and on regions along the ring itself  (Figure~\ref{fig:Pictures}). The oxygen abundance in the region is 12$+$log(O/H)= 8.67--8.76, close to solar metallicity \citep{Moustakas+2010}, and with little variation from location to location along the starburst ring \citep{Diaz+2007}. 

Although the Central Region was observed in the UV by IUE \citep{Kinney+1993}, it was not analyzed by \citet{Calzetti+1994} as part of their starburst sample. The galaxy's classification as `hotspot' \citep{Sersic+1965} and the potential presence of a non--thermal source in the nucleus \citep{Kinney+1993} justified the original exclusion. However, more recent measurements in the mid-IR with Spitzer IRS \citep{Goulding+2009} and in the X--ray with Chandra \citep{Grier+2011} place a tight upper limit to the presence of an AGN in the nucleus of NGC 3351, indicating the emission is dominated by star formation. Additionally, analysis of near--IR images from the VLT suggests that the nucleus is the site of recent star formation \citep{Lin+2018}. The location of the nucleus is marked by  a magenta star in Figure~\ref{fig:Pictures}.

The locations of NGC~3351 and of its Central Region on the IRX--$\beta$ plane are shown in Figure~\ref{fig:IRX_beta}, together with the location of the starburst galaxies from \citet{Calzetti+1994} and \citet{Meurer+1999}, nearby star--forming galaxies from \citet{Dale+2009},  and star--forming regions in the local galaxy NGC~5194=M~51 \citep{Calzetti+2005}. The total infrared luminosity (TIR) is  the integrated luminosity from 3~$\mu$m to 1100~$\mu$m, and captures the  entire dust emission; we adopt this terminology to indicate the dust emission, in lieu of the equally common `FIR'.  The Central Region is a starburst according to the definition of \citet[][see footnote~2 in the Introduction]{Heckman2000}: the circumnuclear star--forming ring is centrally  concentrated within the inner $\sim$750~pc, has a gas  consumption  timescale of about 400--600~Myr and a SFR surface density of $\sim$0.8~M$_{\odot}$~yr$^{-1}$~kpc$^{-2}$, which is about 800 times larger  than the mean SFR surface  density in the disk of NGC~3351. The Central  Region qualifies as a starburst also based on its SFR and stellar mass (SFR$\sim$0.3--0.5~M$_{\odot}$~yr$^{-1}$, M$_{\star}\simeq$1.2$\times$10$^9$~M$_{\odot}$): its SFR is  about 4--7 times higher ($\sim$2--4~$\sigma$)  than the SFR of an equal--mass galaxy  along the Main Sequence of Star Formation in the local Universe \citep{Cook+2014}\footnote{The offset is smaller, about a factor 2--3, if the Main Sequence relation of \citet{Peng2010} or \citet{Hunt2020} is used, still $\sim$2--3~$\sigma$ above both relations, since these have smaller scatter than the relation of \citet{Cook+2014}.}. Yet its IRX value is lower by almost an order of magnitude than expected for starburst galaxies, and close to the locus marked by the extinction curve of the Small Magellanic Cloud \citep[SMC;][]{Pei1992}. For this reason, the Central Region in NGC~3351 represents an important case study for the interplay between dust attenuation and stellar population ages. 

\section{Nomenclature\label{sec:nomenclature}}

The derivation and analysis of dust attenuated and intrinsic UV--to-NIR SEDs will require photometric and model comparisons among different areas within the Central Region. For clarity, we list here the nomenclature we adopt throughout this paper in reference to different components within this Region, and provide a summary list in Table~\ref{tab:nomenclature}.

\begin{figure}[ht!]
\plotone{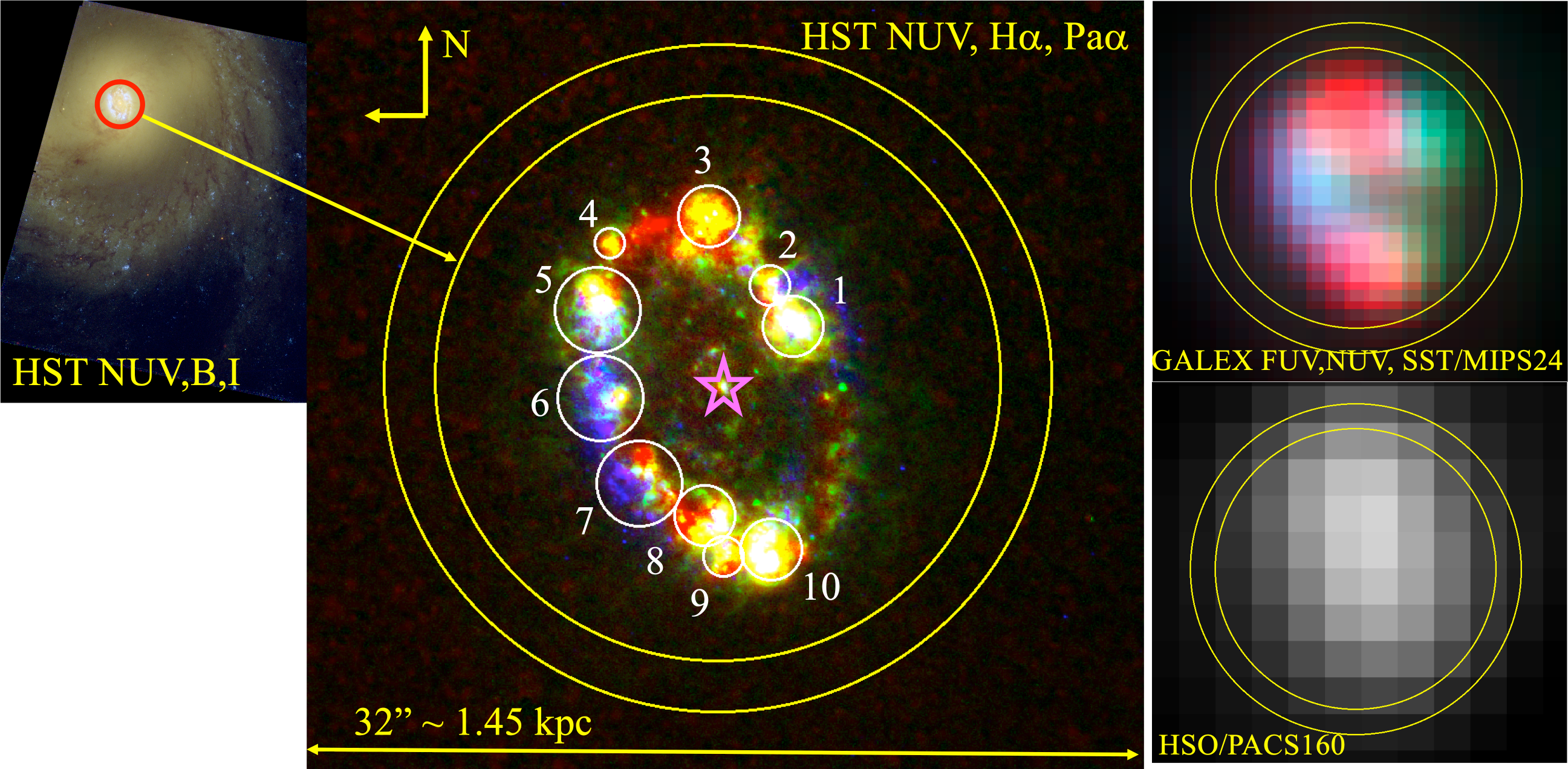}
\caption{Multi--wavelength images of NGC~3351. {\bf LEFT:} Three--color combination (NUV=blue, B=green, I=red) of a $\sim$7.4~kpc region of the galaxy, obtained with the HST/WFC3, which includes the central region, marked with a red circle \citep{Calzetti+2015}. {\bf CENTER:} The central $\sim$1.45~kpc area of NGC~3351 highlighting the circumnuclear starburst  ring  in a three--color composite with HST NUV (0.27~$\mu$m, blue), H$\alpha$ (0.6563~$\mu$m, green), and Pa$\alpha$ (1.8756~$\mu$m, red) images. The two yellow circles mark the the photometric aperture (inner) and background annulus (outer) used for the Central Region photometry, with radius of 11$^{\prime\prime}$ (=498~pc) and thickness of 2$^{\prime\prime}$ (=90~pc), respectively. The ring regions, i.e. individual areas of emission, are numbered from 1 to 10 and are marked with white circles; these are referred to as R1, ...., R10 in the rest of the paper. Section~\ref{sec:photometry} describes how the apertures for the ring regions are defined. The location of the galaxy's nucleus, which is a site of recent star formation, is marked with a magenta star. North is up, East  is left. {\bf RIGHT:} The same region as the one in the center panel is shown in a three color composite of GALEX FUV (blue), NUV (green), and Spitzer/MIPS24 (red), all three with comparable 4$^{\prime\prime}$--6$^{\prime\prime}$ resolution (top) and in a grey--scale Herschel/PACS160 image (bottom). In both panels the two yellow circles are the same as those in the center panel. \label{fig:Pictures}}
\end{figure}

The Central Region is the 0.5~kpc--radius area surrounding the nucleus of NGC\,3351, and includes the starburst ring (Figure~\ref{fig:Pictures}). Its multi--wavelength photometry is the {\em galaxy--background subtracted} measurement of this area from the GALEX FUV to the Herschel 500~$\mu$m (Section~\ref{sec:data}). The observed photometry and derived model SEDs and other quantities from this photometry will have subscript {\em Central}.

The 10 regions named R1, .... R10 are located along the circumnuclear ring of star formation within the Central Region (numbered 1 through 10 in Figure~\ref{fig:Pictures}), and will  be referred to as `ring  regions'. Their photometry is measured from the HST  0.27--to--1.9~$\mu$m images only, with {\em local background subtraction}. The sum of their observed photometry and of model SEDs, both attenuated and intrinsic, and both spectra and photometry, is indicated as $\Sigma$(R1--R10). We give this subscript to physical quantities derived from this sum. 

The ring regions occupy a small fraction, about 13\%, of the area of the Central Region, and represent anywhere between 63\% and 7\% of its light in different bands. The residual light, i.e., the difference in the photometry between the Central Region and the sum of the 10 ring regions, is indicated with the subscript {\em Res.}, and called Res. Region  or  Residual Region throughout the paper. The difference is directly measured from observations at  the HST wavelengths. The extrapolation to the GALEX FUV and NUV wavelengths is performed by subtracting the model photometry  of $\Sigma$(R1--R10)  from the observed photometry of the Central Region,  which carries significant uncertainties. 

The sum of the best--fit model spectra and photometry, both attenuated and intrinsic, of the 10 ring  regions and the Res. Region is called {\em Sum}. This is not equivalent to the models of {\em Central}, where the best fit  is obtained from the integrated photometry of the central 0.5~kpc--radius area; conversely, {\em Sum} is the  sum of the {\em best fit models of the individual (R1+...+R10+Res.) regions}.

\begin{deluxetable*}{ll}
\tablecaption{Nomenclature\label{tab:nomenclature}}
\tablewidth{0pt}
\tablehead{
\colhead{Name or } & \colhead{Meaning} \\
\colhead{Subscript} & \colhead{   } 
}
\startdata
Central     & Background--subtracted photometry of the central $\sim$0.5~kpc radius region in NGC\,3351, \\   
                 & and the model SEDs derived from this photometry.\\
R1,  R2,..., R10  & Observed photometry and model SEDs of individual, local--background subtracted, regions \\
                  & along the circumnuclear ring of star formation.\\
$\Sigma$(R1--R10) & Sum of observed photometry and of model SEDs (spectra and photometry, both attenuated \\
                  &and intrinsic) of the 10 ring regions.\\
Res. or  Residual & Photometry and models SEDs of the difference light [Central - $\Sigma$(R1--R10)].\\
                             & Extrapolation of  Res. photometry to GALEX FUV and NUV is performed from combination \\
                              & of observations and models.\\
Sum        & The sum, obtained as $\Sigma$(R1--R10)$+$Res.,  of the {\em model} spectra and {\em synthetic} photometry.\\
\enddata
\tablecomments{The adopted nomenclature for regions and subregions discussed in this paper. See Section~\ref{sec:nomenclature} for more  details.}
\end{deluxetable*}

\section{Imaging Data}\label{sec:data}

\subsection{HST Images}

The high angular resolution images from the Hubble Space Telescope are used to perform the analysis on $\sim$50--150~pc scale (60--180~pc de--projected) regions along the starburst ring of NGC3351, as well as provide integrated photometry for the Central Region in the 0.27--1.9~$\mu$m wavelength range (Figure~\ref{fig:Pictures}). The images were obtained as part of several programs: the broad--band NUV, U, B, V, and I are from the HST  Treasury program LEGUS, Legacy ExtraGalactic UV Survey \citep[GO--13364; PI: Calzetti,][]{Calzetti+2015}; the medium--band V and the narrow--band H$\alpha$ are from GO--13773 \citep[PI: Chandar,][]{Hannon+2019}; and the near--IR images, broad--band H and narrow--band P$\alpha$ and 1.9~$\mu$m from SNAP--9360 \citep[PI: Kennicutt,][]{Calzetti+2007}. The field--of--views of the images are large enough to include the entire Central Region. All images were retrieved from the Hubble Legacy Archive\footnote{https://hla.stsci.edu}, where they are available fully processed and calibrated in units of count/s; the count rates are converted to flux density using the calibration keywords available from the image headers. The list  of instruments, filters, and native resolution for each image are listed in Table~\ref{tab:central}.

\begin{deluxetable*}{lllCc}
\tablecaption{Central Region Photometry\label{tab:central}}
\tablewidth{0pt}
\tablehead{
\colhead{Facility$^1$} & \colhead{Instrument$^2$} & \colhead{Band$^3$} & \colhead{Native  Resol.$^4$} &
\colhead{Log(Luminosity)}\\
\colhead{Name} & \colhead{Name} & \nocolhead{Name} & \colhead{($^{\prime\prime}$, pc)} &
\colhead{(erg~s$^{-1}$)} 
}
\decimalcolnumbers
\startdata
GALEX &              & FUV (0.1524~$\mu$m) &  4.2, 190 & 41.63$\pm$0.06 \\
GALEX &              & NUV (0.2297~$\mu$m) &  6.2, 280 & 41.90$\pm$0.06 \\
HST      & WFC3   &  F275W (0.2710~$\mu$m, NUV) & 0.09, 4 & 41.99$\pm$0.05 \\
HST      & WFC3   &  F336W (0.3355~$\mu$m, U) & 0.09, 4 & 42.05$\pm$0.05 \\
HST      & WFC3   &  F438W (0.4327~$\mu$m, B) & 0.09, 4 & 42.27$\pm$0.05 \\
HST      & WFC3   &  F555W (0.5308~$\mu$m, V) & 0.09, 4 & 42.37$\pm$0.04 \\
HST      & WFC3   &  F547M (0.5447~$\mu$m, V) & 0.09, 4  & 42.36$\pm$0.05 \\
HST      & WFC3   &  F657N (0.6567~$\mu$m, H$\alpha+$[NII]) & 0.09, 4 & 42.54$\pm$0.04 \\
HST      & WFC3   &  F814W (0.8030~$\mu$m, I) & 0.09, 4  & 42.54$\pm$0.04 \\
HST      & NICMOS/3  & F160W (1.607~$\mu$m, H) & $\sim$0.31, 14 & 42.54$\pm$0.06 \\
HST      & NICMOS/3  & F187N (1.876~$\mu$m, Pa$\alpha$) & $\sim$0.31, 14 & 42.56$\pm$0.06 \\
HST      & NICMOS/3  & F190N (1.899~$\mu$m) & $\sim$0.31, 14 & 42.47$\pm$0.07 \\
SST      & IRAC     & 3.6 (3.56~$\mu$m)               & 1.9, 86  & 41.87$\pm$0.04 \\
SST      & IRAC     & 8.0 (7.96~$\mu$m)               & 2.8, 127 & 42.07$\pm$0.05$^a$ \\
SST      & MIPS     & 24 (23.8~$\mu$m)               & 6.5, 294 & 42.24$\pm$0.05$^a$ \\
HSO     & PACS     & 70 (71.8~$\mu$m)               & 5.7, 258  & 42.85$\pm$0.04 \\
HSO     & PACS     & 100 (103.~$\mu$m)               & 7.1, 321 & 42.75$\pm$0.05 \\
HSO     & PACS     & 160 (167.~$\mu$m)               & 11.2, 507 & 42.42$\pm$0.05 \\
\enddata
$^1$ GALEX=Galaxy  Evolution Explorer \citep{Martin+2005}; HST=Hubble Space Telescope; SST= Spitzer Space Telescope; HSO=Herschel Space Observatory.\\ 
$^2$  WFC3=Wide Field Camera 3. NICMOS=Near-Infrared Camera and MultiObject Spectrometer --  observations were performed with the Camera 3 on NICMOS, which  operated slightly out--of--focus.  IRAC=InfraRed Array Camera. MIPS=Multiband Imaging Photometer for Spitzer. PACS=Photodetecting Array Camera and Spectrometer.\\
$^3$ Filter names and, in parenthesis, the pivot wavelength. The HST filters are equated to the equivalent Johnson's filters, 
where applicable; for narrow--band HST filters, the main lines targeted are listed.\\
$^4$ The native resolution is given as the Full Width at Half Maximum (FWHM)  of the Point Spread Function (PSF) in arcseconds and  in parsec for the subtended physical scale.\\
$^a$ The luminosities at 8~$\mu$m and 24~$\mu$m refer to the dust emission only in these bands, after subtraction of the stellar contribution.\\
\tablecomments{The luminosities for the Central Region are given as $\lambda$L($\lambda$) and are scaled to the resolution of the HST/WFC3 images. These 
are a factor  1.57 larger than the luminosities measured at the resolution of the HSO/PACS160. GALEX  and HST luminosities are corrected for 
foreground MW extinction; SST and HSO luminosities are color--corrected. The adopted distance is 9.33~Mpc.}
\end{deluxetable*}

\subsection{Other Images}

Imaging data in the FUV and NUV from GALEX, and mid/far--IR maps from the Spitzer Space  Telescope (SST) and the Herschel Space Observatory (HSO) extend the wavelength coverage of the Central Region, albeit at  much 
lower angular resolution than HST (Figure~\ref{fig:Pictures} and Table~\ref{tab:central}). The GALEX images are from the GALEX  Ultraviolet Atlas of Nearby Galaxies \citep{GildePaz+2007}, the SST images from SINGS, the Spitzer Infrared Nearby  Galaxies Survey  \citep{Kennicutt+2003}, and the HSO images from KINGFISH, Key Insights on Nearby Galaxies: a Far--Infrared Survey with Herschel \citep{Kennicutt+2011}. 
All images are already processed and flux calibrated or have calibration keywords available from archives.

For the bulk of the analysis in this  paper, we limit the HSO imaging to the PACS instrument, to keep the angular resolution of the data well below the size of the Central Region (Table~\ref{tab:central}). However, 
in order to model the shape of the IR SED of the Central Region, we add imaging data from HSO/SPIRE at 250, 350, and  500~$\mu$m when fitting the dust emission SED at $\lambda\ge$8~$\mu$m in Section~\ref{sec:central}. 
The SPIRE data have lower angular resolution than those of the main dataset discussed above, but they enable a better characterization of the shape of the IR SED at long wavelengths. Once the shape is parameterized, we use the higher angular resolution data from SST/IRAC~8~$\mu$m to HSO/PACS~160~$\mu$m to derive the total IR luminosity of the Central Region. 

\begin{deluxetable*}{lllC}
\tablecaption{Central Region Photometry at SPIRE500 Resolution\label{tab:SPIRE500}}
\tablewidth{0pt}
\tablehead{
\colhead{Facility$^1$} & \colhead{Instrument$^2$} & \colhead{Band$^3$} & \colhead{Log(Luminosity)}\\
\colhead{Name} & \colhead{Name} & \nocolhead{Name} & \colhead{(erg~s$^{-1}$)} 
}
\decimalcolnumbers
\startdata
SST      & IRAC     & 3.6 (3.56~$\mu$m)     & 42.13$\pm$0.04 \\
SST      & IRAC     & 8.0 (7.96~$\mu$m)    & 42.11$\pm$0.04$^a$ \\
SST      & MIPS     & 24 (23.8~$\mu$m)     & 42.25$\pm$0.04$^a$ \\
HSO     & PACS     & 70 (71.8~$\mu$m)      & 42.90$\pm$0.04 \\
HSO     & PACS     & 100 (103.~$\mu$m)   & 42.81$\pm$0.05 \\
HSO     & PACS     & 160 (167.~$\mu$m)   & 42.48$\pm$0.05 \\
HSO     & SPIRE     & 250 (250.~$\mu$m)   & 41.86$\pm$0.07 \\
HSO     & SPIRE     & 350 (360.~$\mu$m)   & 41.29$\pm$0.08 \\
HSO     & SPIRE     & 500 (520.~$\mu$m)   & 40.59$\pm$0.10 \\
\enddata
$^1$ SST= Spitzer Space Telescope; HSO=Herschel Space Observatory.\\ 
$^2$  IRAC=InfraRed Array Camera. MIPS=Multiband Imaging Photometer for Spitzer. PACS=Photodetector Array Camera and Spectrometer. SPIRE=Spectral  and Photometric Imaging Receiver.\\ 
$^3$ Filter names and, in parenthesis, the pivot (central for SPIRE) wavelength.\\
$^a$ The luminosities at 8~$\mu$m and 24~$\mu$m refer to the dust emission only in these bands, after subtraction of the stellar contribution.\\
\tablecomments{The luminosities for the Central Region, expressed as $\lambda$L($\lambda$), measured in $\sim$36$^{\prime\prime}$ radius 
apertures, after all images have been convolved to the HSO/SPIRE500 resolution, using the kernels of \citet{Aniano+2011}. No aperture corrections are applied to 
the listed luminosity values. All luminosities are color--corrected. The adopted distance is 9.33~Mpc.}
\end{deluxetable*}

\subsection{Processing}\label{subsec:processing}

All images have been obtained with sufficient depth that the regions in Figure~\ref{fig:Pictures} are measured with S/N$\gg$10 at all wavelengths.
We align, register, and resample all images to  the field--of--view of the HST/WFC3 images, to facilitate photometry. Measurements are performed at  two angular resolutions: the HST/WFC3 one for the ring regions in the wavelength range 0.27--1.9~$\mu$m and the HSO/PACS160 one for the integrated photometry of the Central Region in the broader wavelength range 0.15--160~$\mu$m  (Table~\ref{tab:central}). Degradation of the higher angular resolution images to the HSO/PACS160 resolution is performed by convolving the images with the kernels of \citet{Aniano+2011}. However, all measurements quoted in this paper for the Central Region refer to photometry rescaled to the HST/WFC3 resolution, i.e., the highest  resolution in our dataset. For this purpose, we multiply the fluxes measured at the HSO/PACS160 resolution by a factor 1.57 \footnote{The factor is calculated by comparing the photometry in the WFC3 images at full and degraded resolution, for the Central Region aperture. For the ring regions, the HST/NICMOS3 photometry is rescaled to the WFC3--equivalent. To calculate the rescaling factors, we degrade the WFC3 images using a gaussian convolution kernel  to  match the NICMOS3 PSF; factors are calculated for each ring region size, and range from 1.06 for the smallest region (0.6$^{\prime\prime}$ radius) to 1.00 for the largest region (1.7$^{\prime\prime}$ radius).}; the scaling factor is accurate to within 5\%--7\%, which also accounts for small mis--alignments between the images. This choice is made to provide measurements that are consistent with each other and as close as possible to the `total flux' in each region. The apertures used for the photometry of the 10 ring regions are sufficiently  large that the small resolution variations among the WFC3 bands have negligible impact on the measured fluxes.  In what follows, we often refer to the fluxes/luminosities as `aperture--corrected'; while not strictly true (aperture corrections are for photometry to infinite radius), the PSF of the HST/WFC3 images is sufficiently narrow that for the measurements we perform in this work negligible aperture corrections  would be required.

Pure emission--line images are derived by subtracting the stellar continuum from the narrow--band HST images WFC3/F657N (H$\alpha+$[NII]) and NIC3/F187N (Pa$\alpha$). The stellar continuum for the optical image is constructed from the interpolation between the F547M and the F814W, both tracers of stellar emission with only weak emission lines; the stellar continuum for the infrared line is  obtained by direct re--scaling (factor 0.94, from the ratio of the two filters' transmission efficiencies) of the adjacent narrow--band F190N. Both continuum--subtracted images are then multiplied by the respective filter widths (0.0121~$\mu$m for F657N and 0.0188~$\mu$m for F187N) and corrected for the filter transmission curve at the galaxy's redshift (z=0.002595, from NED), in order to derive line fluxes. The optical line is further corrected for the [NII] contribution, from the ratio [NII]($\lambda$0.6584~$\mu$m)/H$\alpha$=0.37 \citep{Moustakas+2010}. We adopt a constant value of [NII]/H$\alpha$ for the Central Region, owing to its relatively constant metallicity \citep{Diaz+2007}. The final result is two emission--line images at H$\alpha$($\lambda$0.6563~$\mu$m) and Pa$\alpha$($\lambda$1.8756~$\mu$m), respectively. All UV, optical, and near--IR images are corrected for the MW foreground extinction, E(B--V)=0.024.

Color--corrections, at the level of 10\% for the HSO/PAC70 image and less than 5\% for all other SST and HSO images, are applied to all images from these facilities. For the purpose of obtaining dust--emission--only fluxes, the stellar contribution is removed from both the 8~$\mu$m and 24~$\mu$m images, using the formulae of \citet{Helou+2004} \citep[see also,][]{Calzetti+2007, Calapa+2014}: f$_{\nu, D}$(8) = f$_{\nu}$(8) - 0.25 f$_{\nu}$(3.6), and f$_{\nu, D}$(24)=f$_{\nu}$(24) - 0.035 f$_{\nu}$(3.6), where the flux densities are in units of Jy, and the subscript `D' indicates the dust--only emission component. The contamination of the 3.6~$\mu$m image by the 3.3~$\mu$m Polycyclic Aromatic Hydrocarbons (PAH) emission feature is small \citep[$\sim$5\%--15\%,][]{Meidt+2012}, and the IRAC~3.6 can be used as a stellar continuum tracer. In the center of NGC~3351, stellar continuum contribution to the fluxes at 8~$\mu$m and 24~$\mu$m is 5\% and $<$2\%, respectively. 

As mentioned in the  previous section, characterization of the shape of the IR SED requires use of the SPIRE images from the HSO, which cover three sub-mm bands with PSFs$\simeq$18$^{\prime\prime}$, 25$^{\prime\prime}$, and 36$^{\prime\prime}$ at 250, 350, and 500~$\mu$m, respectively. In order to  include these lower resolution data in the IR SED fit, we create a second set of photometric data by degrading all images at  8~$\mu$m$\le  \lambda \le$500~$\mu$m to the SPIRE/500 resolution, using the kernels of \citet{Aniano+2011}. Table ~\ref{tab:SPIRE500} lists the photometry from the degraded images measured in a 36$^{\prime\prime}$ radius aperture centered on  the nucleus of NGC~3351. The listed luminosities are those directly measured in the aperture; to convert these values to total luminosity, an aperture correction of 0.09 dex should be added to each. We stress that these measurements are only used to fit the shape of the IR SED, and the parameters that define it, in Section~\ref{sec:central_dust_emission}. We use the higher angular resolution IR data, from SST and HSO/PACS in {Table~\ref{tab:central}} to derive the IR luminosity of the Central Region.

\begin{figure}
\plottwo{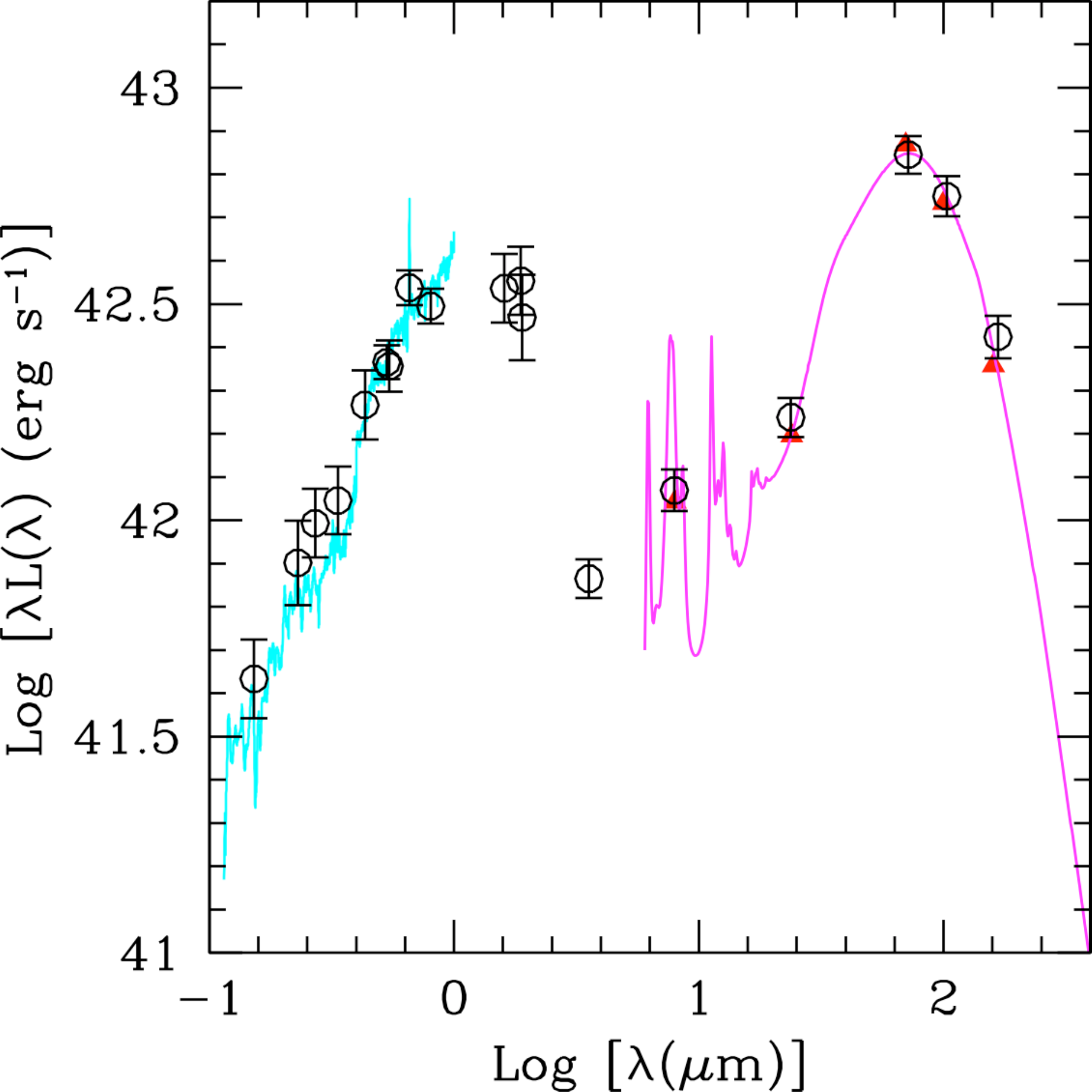}{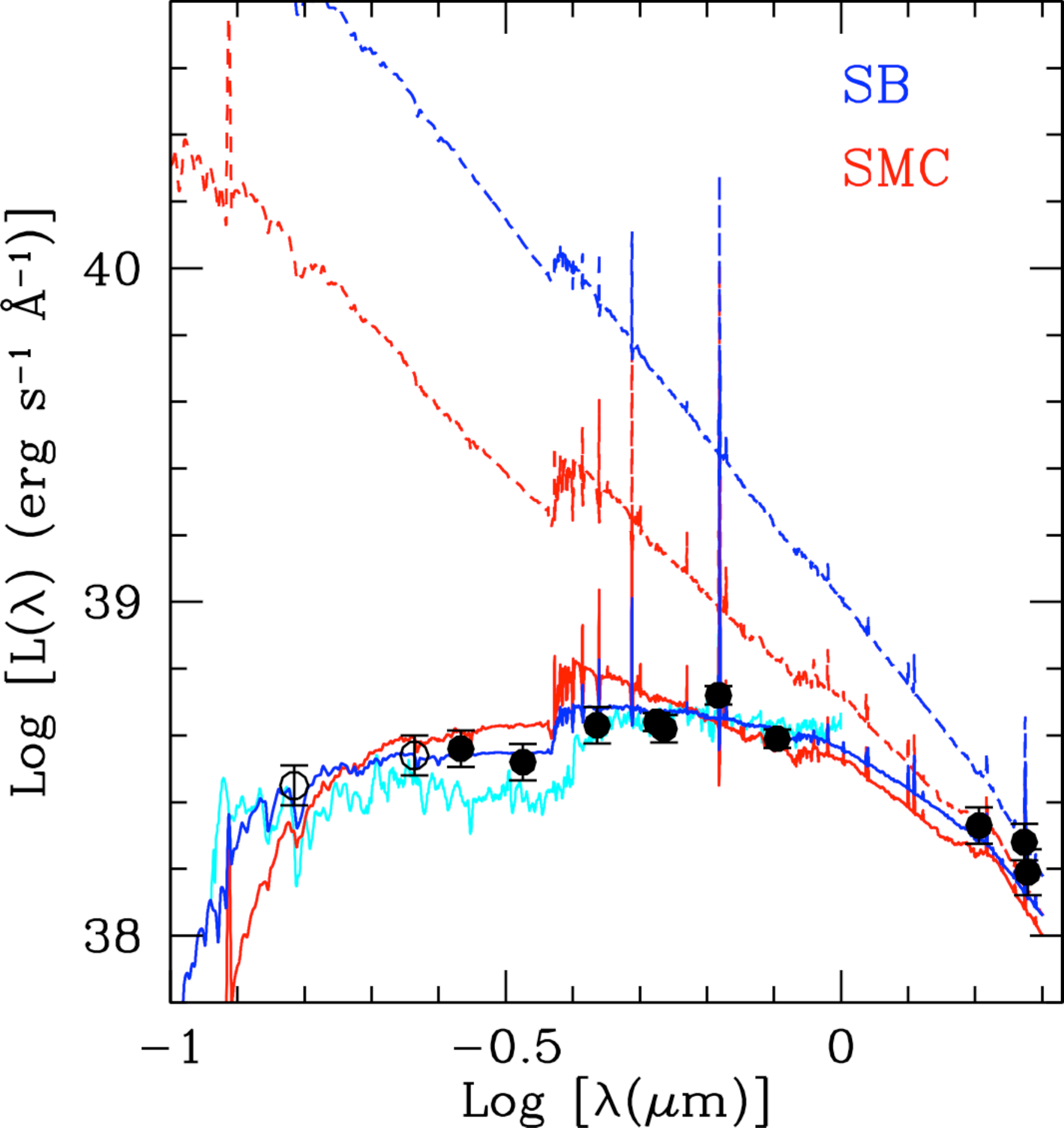}
\caption{{\bf LEFT:} The FUV--to--IR SED  of the Central Region in NGC~3351. The black circles with 1~$\sigma$ error bars are the measurements listed in Table~\ref{tab:central}. The cyan UV--optical spectrum is the unscaled combination of IUE and ground--based data from \citet{StorchiBergmann+1995}. The magenta line and red triangles show the best fit  \citet{DraineLi2007} model and photometry to the IR data, from Section~\ref{sec:central}. {\bf RIGHT:} A zoom--in of the left panel into the FUV--to--NIR SED  (0.15--1.9~$\mu$m), now expressed as luminosity density, shown for the aperture--corrected GALEX measurements (empty black circles) and the HST measurements (filled black circles), together with the spectrum of \citet{StorchiBergmann+1995} (cyan line). The dust attenuated and intrinsic best--fit models are shown separately for stellar populations with an SB attenuation curve (blue continuous and dashed lines, respectively) and an SMC  extinction curve (red continuous and dashed lines, respectively; Sections~\ref{subsec:popmodels} and \ref{subsec:attmodels}); the attenuation/extinction curves are the same ones used for the model lines in Figure~\ref{fig:IRX_beta}. The SB curve yields a significantly better fit ($\chi^2_{red}$=1.2) than the SMC curve ($\chi^2_{red}$=9.1) for the FUV--to--NIR SED. See Section~\ref{sec:central} for more details.  \label{fig:central_SEDs}}
\end{figure}

\section{Regions and Aperture Photometry}\label{sec:photometry}

The size of the  photometric aperture for the Central Region is selected to encompass the circumnuclear  starburst ring, not only in the HST images but also in the SST and HSO/PACS images (Figure~\ref{fig:Pictures}, right). Thus, while the ring is an inclined ellipse about 750~pc in diameter, the Central Region is a circle about 1~kpc in diameter (Figure~\ref{fig:Pictures}). This choice aims at minimizing aperture corrections in the lowest resolution images, which are performed as described in the previous section, while at the same time minimizing the contribution to our measurements from the rest of the galaxy. The aperture--corrected photometric values, measured in a 11$^{\prime\prime}$ radius aperture with an annulus 2$^{\prime\prime}$ wide for background subtraction, are listed in the last column of Table~\ref{tab:central}, together with their 1~$\sigma$ uncertainty. The apertures used in this  paper are circular on the sky, but, because of the 40$^o$ inclination angle of the galaxy, they  are ellipses in spatial coordinates, with the major axis a factor 1.3 larger than the minor axis. This has minor impact on our analysis, and we will continue to refer to our apertures as  `circular' in the rest of this work. The background subtraction removes the diffuse stellar population contributed by the galaxy:  the removed component has a mainly red SED, contributing between 24\%  of the total flux in the FUV and NUV and 45\%  at 3.6~$\mu$m. The resulting FUV--to--IR SED of the Central Region is shown in the left panel of Figure~\ref{fig:central_SEDs}, together with an unscaled UV--optical spectrum from \citet{StorchiBergmann+1995}. Despite differences in the measurement setups, including that the spectrum of \citet{StorchiBergmann+1995} uses a $\sim$10$^{\prime\prime}\times$20$^{\prime\prime}$ extraction aperture, our photometry and the UV--optical spectrum are remarkably close in absolute flux and shape across all common bands, with differences of less than 40\% (Figure~\ref{fig:central_SEDs}, right).

\begin{deluxetable*}{lcr}
\tablecaption{Central Region's Line and Infrared Luminosities\label{tab:linecentral}}
\tablewidth{250pt}
\tablehead{
\colhead{Parameter} & \colhead{Wavelength} & \colhead{Log(Luminosity)}\\
\colhead{ }  & \colhead{$\mu$m} & \colhead{(erg~s$^{-1}$)} 
}
\decimalcolnumbers
\startdata
L(H$\alpha$) & 0.6563 &   40.20$\pm$0.04 \\
L(Pa$\alpha$) & 1.8756 &   39.74$\pm$0.09 \\
E(B--V)$_{gas}$ (mag)$^1$ & & 0.52$\pm$0.10\\
L(H$\alpha$)$_{corr}$ & 0.6563 & 40.73$\pm$0.10\\
L(TIR)$_{mod}$ & 3--1100 & 43.15$\pm$0.08\\
L(TIR)$_{DH02}$ & 3--1100 & 43.07$\pm$0.06\\
\enddata
$^1$ The color excess, E(B--V)$_{gas}$, is derived with the assumption that the intrinsic ratio H$\alpha$/Pa$\alpha$=7.82, and the selective attenuation k(H$\alpha$)--k(Pa$\alpha$)=2.08 \citep{Fitzpatrick1999}; it is already corrected for the foreground Galactic contribution.\\
\tablecomments{The emission line luminosities for the Central Region are measured on stellar continuum--subtracted images (and [NII]--corrected for H$\alpha$), in the same region used to derive the photometry in Table~\ref{tab:central}. The extinction--corrected H$\alpha$ luminosity, L(H$\alpha$)$_{corr}$ adopts the value k(H$\alpha$)=2.54 for the extinction curve. The total IR luminosities, L(TIR), 
are derived with two methods: the \citet{DraineLi2007} models (L(TIR)$_{mod}$) and the formula from \citet{DaleHelou2002} (L(TIR)$_{DH02}$). The two values differ by about 20\%, well within the 1~$\sigma$  uncertainty.}
\end{deluxetable*}

Line emission luminosities at the wavelengths of H$\alpha$ and Pa$\alpha$ for the Central Region are listed in Table~\ref{tab:linecentral}, together with the color excess derived by assuming Case B recombination for a solar metallicity gas, which has an intrinsic luminosity ratio L(H$\alpha$)/L(Pa$\alpha$)=7.82 \citep{Osterbrock+2006}, and selective attenuation between the two lines: k(H$\alpha$)--k(Pa$\alpha$)=2.08 \citep{Fitzpatrick1999}. The  luminosity we derive for H$\alpha$ is about 40\% lower than the value published by \citet{Moustakas+2010} for a similar region's size, but $\sim$20\% higher than the same--region luminosity we calculate from the H$\alpha$ image of \citet{Dale+2009}. Both values for \citet{Moustakas+2010} and \citet{Dale+2009} are from ground--based data, spectroscopy and imaging, respectively. Our value of H$\alpha$ is in--between those two, and we speculate that the discrepancies with those authors (and between the two authors' values as well)  can be ascribed to potential flux calibration uncertainties in ground--based data. For the color  excess, E(B--V)$_{gas}$, our value of 0.52 is also in--between the values from the spectroscopic data  of \citet[][0.64]{Moustakas+2010} and \citet[][0.50]{StorchiBergmann+1995}, and the three values agree within their combined 1~$\sigma$ uncertainties. We also note that, in these  two other works, the E(B--V) is calculated from the ratio of H$\alpha$ to H$\beta$(0.4861~$\mu$m).

As the goal of this study is  to divide the Central Region into simple stellar populations (as close as possible to either instantaneous or constant star formation) that are easy to model and, at  the same time, represent the bulk of the star formation in that region, we select localized areas of star  formation along the circumnuclear ring in the HST images as follows. Peaks of Pa$\alpha$ emission (S/N$>$30) are identified, and circular apertures\footnote{Experiments run with apertures of different shapes, e.g., polygons, yield similar results. We elect  to keep  our apertures circular for ease of local background subtraction.} are grown around them until they either reach down to flux levels with S/N$\sim$5/pixel or encounter a neighboring aperture. We keep the apertures non--overlapping or minimally overlapping to ensure that photometry yields independent measurements. We identify a total of 11 such P$\alpha$ peaks. Subsequently, the HST/F275W  (NUV) image is inspected and the apertures are further grown to reach levels of S/N$\sim$5/pixel in this band. This step is only required for R5, R6, and R7, during which some recentering is necessary, to enclose connected regions of UV emission. We also collapse two adjacent peaks of P$\alpha$ into a single aperture (R8) to enclose the underlying  contiguous NUV emission. The areas enclosed by the apertures are then inspected for presence of off--center local peaks of F814W(I)--band emission not accompanied by either line or UV emission, in order to limit inclusion  of potentially old stellar populations; this leads in two  cases (R3  and R4) to a slight contraction of the  size of the apertures. Finally, regions with measured multi--wavelength fluxes that are barely above the surrounding background are removed. For instance, the area between R3 and R4 (Figure~~\ref{fig:Pictures}, center) contributes less than 4\% of the emission along the ring at all wavelengths; the other excluded areas are fainter. At the end of this process, we identify 10 separate regions, marked and numbered in the center panel of Figure~\ref{fig:Pictures} and indicated as R1,...,R10 throughout this paper (Table~\ref{tab:nomenclature}).   

Cumulatively, the 10 regions along the ring include 75\% (63\%) of the NUV emission from the ring (Central Region, Table~\ref{tab:regions}), although their contribution decreases steadily  with increasing wavelength, becoming less than 10\% in the NIR. They account for almost 1/2 of the gas emission in the Central Region; the remaining 1/2 of  the flux is likely to be mostly associated to  these regions as well, albeit  spread over  a larger area than enclosed by our apertures. This is  suggested by the morphology of the emitting gas, which closely tracks the ring but with a broader distribution (a thicker ring). 

Photometry for  R1,...,R10 is measured in the HST images only, since the  other images do not have sufficient resolution for the aperture  sizes employed in this part of the analysis, which range from 0.6$^{\prime\prime}$ to 1.7$^{\prime\prime}$  in radius. Photometric measurements are listed in Table~\ref{tab:regions}. The HST data  ensure 10 photometric measurements of the stellar and ionized gas emission for each region, but they cannot provide dust emission measurements. For the latter, we have  to rely exclusively on the lower resolution images. The aperture photometry is  background subtracted using annuli with width between 0.2$^{\prime\prime}$ and 0.5$^{\prime\prime}$. For the calculation of the background level in the annulus around each aperture we only use those pixels that are not included in any of the adjacent apertures. With this in mind, the width of the annulus is chosen to ensure that the number of background pixels is no less than $\sim$50\% than the pixels in the aperture, to ensure reasonable statistics. 

\citet{Diaz+2007} obtained ground--based long slit spectroscopy in the range 0.365--0.965~$\mu$m of the circumnuclear ring with three separate slit locations and orientations, identifying seven regions of concentrated emission. Their seven regions (R1 to R7) are close to the location of seven of ours (in order: R3, R1, R10, R8, R7, R6, and R5). We compare both our H$\alpha$ emission and color excess values with the values derived by those authors, in order to assess the robustness of our measurements. For the line emission flux, our values range between 72\% and 103\% of the \citet{Diaz+2007} values for six of the regions, with a median value of 82\%; for the seventh region (our R8 = Diaz's R4) we only recover 58\% of those authors' published value, but this region is also the one with the largest offset relative to the location of R4 in \citet{Diaz+2007}. Differences in aperture location, extraction apertures, and background choices between our photometric apertures and the spectroscopic ones of \citet{Diaz+2007} can account for most of the observed flux discrepancies. The color excess, E(B--V)$_{gas}$, values are calculated from several hydrogen recombination lines at optical wavelengths in \citet{Diaz+2007} and from H$\alpha$/Pa$\alpha$ in our case; the values agree within 0.10~mag ($\sim$1~$\sigma$), with a median discrepancy of 0.05~mag, for 6  of the regions. In the one exception (our R3 = Diaz's R1), we derive E(B-V)$_{gas}$=0.82 and \citet{Diaz+2007} derive E(B-V)$_{gas}$=0.46 (after correction for foreground Milky Way extinction); the large difference can be explained by R3 being a dust--buried region in the ring, which would recover increasingly larger values of E(B--V)$_{gas}$ for redder hydrogen recombination lines;  this hypothesis is reinforced by the co--location of R3 with the strongest peak at 24~$\mu$m in the galaxy.

Although the ring regions enclose the majority of the NUV light, they represent a minor contribution to the SED of the Central Region longward of the B--band. For this reason, we also investigate and model the SED of the residual emission (Res. Region), i.e., of the luminosity difference between the Central Region, L$_{Central}$($\lambda$), and the sum of the 10 ring regions, $\Sigma$(R1--R10) = $\Sigma_{n=1}^{10}$ L$_n$($\lambda$):
\begin{equation}
L_{Res.}(\lambda)=L_{Central}(\lambda) - \Sigma_{n=1}^{10} L_n(\lambda).
\end{equation}
The Res. SED is very red, suggesting an old underlying population (column (14) of  Table~\ref{tab:regions}). The Res. Region includes the star--forming nuclear cluster  (magenta star in Figure~\ref{fig:Pictures}); we keep the nuclear cluster's emission included in the Res. SED, since it contributes less than 1\% of the Res. light at all wavelengths. We will later briefly discuss the SED properties of the nuclear cluster as a stand--alone source, but not separate it for the main goal of this study.

\section{Synthetic Photometry and Fitting Approach}\label{sec:models}

This section summarizes the stellar population, dust attenuation, and dust emission models applied in this study to the Central  Region and to its sub--regions, as defined above. In order to handle  
the angular resolution mismatch between the HST and the longer wavelength images, we do not attempt to model the entire FUV--to--IR SED in a self--consistent manner  
\citep[as adopted by, e.g., MAGPHYS and CIGALE,][]{daCunha+2008, Boquien+2019}, but keep the FUV--to--NIR 
portion of the SED separate from the longer wavelength one; we thus independently model the dust--attenuated stellar population SEDs and the dust emission SED. In this case, the energy 
balance comparison between dust absorption in the FUV--to--NIR and dust emission in the IR becomes a post--facto check.  This approach also 
mimics the situation often encountered in studies of distant galaxies, due to the lower sensitivity and angular resolution of FIR  instruments compared to the optical and NIR ones. 
 
\begin{longrotatetable}
\begin{deluxetable*}{lrrrrrrrrrrrcr}
\tablecaption{Photometry of Individual Regions\label{tab:regions}}
\tablewidth{750pt}
\tabletypesize{\scriptsize}
\tablehead{
\colhead{Property} &\colhead{Units} &\colhead{R1}& 
\colhead{R2} & \colhead{R3} & 
\colhead{R4} & \colhead{R5} & 
\colhead{R6} & \colhead{R7} & 
\colhead{R8} & \colhead{R9} & \colhead{R10} & \colhead{${\Sigma(R1-R10)\over Central}$} & \colhead{Res.} \\ 
\colhead{(1)} & \colhead{(2)} & \colhead{(3)} & \colhead{(4)} & 
\colhead{(5)} & \colhead{(6)} & \colhead{(7)} &
\colhead{(8)} & \colhead{(9)} & \colhead{(10)} & \colhead{(11)} & \colhead{(12)}
& \colhead{(13)} & \colhead{(14)}
} 
\startdata
$\alpha$ &  HH:MM:SS & 10:43:57.58 & 10:43:57.64 & 10:43:57.81  & 10:43:58.07 & 10:43:58.10 & 10:43:58.09 & 10:43:57.99 & 10:43:57.82 & 10:43:57.76 & 10:43:57.64 &        & 10:43:57.78\\
$\delta$  &  DD:MM:SS & +11:42:15.3 & +11:42:16.9 &+11:42:19.6 & +11:42:18.5 & +11:42:15.9 & +11:42:12.5  & +11:42:09.2 & +11:42:08.0 & +11:42:06.4 & +11:42:06.7 &        & +11:42:13.3\\
Radius$^1$    & $\prime\prime$, pc & 1.19, 54.  & 0.79, 36. & 1.19, 54.           & 0.59, 27.     & 1.66, 75.      & 1.66, 75.       &  1.66, 75.     &  1.19, 54.     &  0.79, 36.     &  1.19, 54.     &       &  11.0, 498. \\
F275W & erg~s$^{-1}$~\AA$^{-1}$ & 37.68$\pm$0.04 & 36.94$\pm$0.05 & 36.43$\pm$0.05 & 35.17$\pm$0.10 & 37.61$\pm$0.04  & 37.72$\pm$0.04 & 37.37$\pm$0.04 & 37.06$\pm$0.05 & 36.30$\pm$0.05 & 37.57$\pm$0.04  & 0.63 &38.13$\pm$0.17   \\
F336W & erg~s$^{-1}$~\AA$^{-1}$ & 37.51$\pm$0.04 & 36.86$\pm$0.05 & 36.54$\pm$0.05 & 35.47$\pm$0.07 & 37.49$\pm$0.04 & 37.60$\pm$0.04 & 37.27$\pm$0.04 & 36.97$\pm$0.05 & 36.11$\pm$0.05  & 37.48$\pm$0.04  & 0.53 &38.19$\pm$0.12   \\
F438W & erg~s$^{-1}$~\AA$^{-1}$ & 37.29$\pm$0.04 & 36.75$\pm$0.05 & 36.63$\pm$0.05 & 35.60$\pm$0.06 & 37.41$\pm$0.04 & 37.53$\pm$0.04 & 37.23$\pm$0.04 & 36.92$\pm$0.05 & 35.96$\pm$0.06 & 37.36$\pm$0.04   & 0.33 & 38.46$\pm$0.08  \\
F555W & erg~s$^{-1}$~\AA$^{-1}$ & 37.03$\pm$0.04 & 36.53$\pm$0.05 & 36.62$\pm$0.05 & 35.58$\pm$0.06 & 37.24$\pm$0.04 & 37.38$\pm$0.04 & 37.03$\pm$0.04 & 36.78$\pm$0.05 & 35.84$\pm$0.06 & 37.20$\pm$0.04   & 0.22 & 38.53$\pm$0.06  \\
F547M & erg~s$^{-1}$~\AA$^{-1}$ & 36.98$\pm$0.05 & 36.51$\pm$0.05 & 36.57$\pm$0.05 & 35.56$\pm$0.06 & 37.21$\pm$0.04 & 37.35$\pm$0.04 & 37.02$\pm$0.04 & 36.75$\pm$0.05 & 35.83$\pm$0.06 & 37.16$\pm$0.04   & 0.21  &  38.52$\pm$0.06 \\
F657N & erg~s$^{-1}$~\AA$^{-1}$ & 37.28$\pm$0.04 & 36.66$\pm$0.05 & 37.19$\pm$0.04 & 36.21$\pm$0.05 & 37.37$\pm$0.04 & 37.34$\pm$0.04 & 37.10$\pm$0.04 & 37.02$\pm$0.05 & 36.31$\pm$0.05 & 37.44$\pm$0.04   & 0.27  & 38.58$\pm$0.06  \\
F814W & erg~s$^{-1}$~\AA$^{-1}$ & 36.39$\pm$0.05 & 36.03$\pm$0.05 & 36.45$\pm$0.05 & 35.26$\pm$0.09 & 36.98$\pm$0.05 & 36.92$\pm$0.05 & 36.83$\pm$0.05 & 36.34$\pm$0.05 & 35.68$\pm$0.06 & 36.87$\pm$0.05  &  0.11  & 38.54$\pm$0.06  \\
F160W & erg~s$^{-1}$~\AA$^{-1}$ & 35.38$\pm$0.06 & 34.94$\pm$0.07 & 36.18$\pm$0.05  & 34.53$\pm$0.10 & 36.67$\pm$0.05 & 36.40$\pm$0.05 & 36.31$\pm$0.05 & 35.82$\pm$0.06 & 35.02$\pm$0.07 & 36.34$\pm$0.05  &  0.07  & 38.30$\pm$0.07  \\
F187N & erg~s$^{-1}$~\AA$^{-1}$  & 36.15$\pm$0.05 & 35.67$\pm$0.06 & 36.69$\pm$0.05  & 35.42$\pm$0.07 & 36.56$\pm$0.05 & 36.40$\pm$0.05 & 36.31$\pm$0.05 & 36.16$\pm$0.06 & 35.58$\pm$0.06 & 36.56$\pm$0.05  & 0.11  & 38.23$\pm$0.07  \\
F190N & erg~s$^{-1}$~\AA$^{-1}$  & 35.30$\pm$0.06 & 34.82$\pm$0.07 & 36.09$\pm$0.06 & 34.31$\pm$0.10 & 36.40$\pm$0.05  & 36.27$\pm$0.05 & 36.15$\pm$0.05 & 35.80$\pm$0.06 & 35.03$\pm$0.07 & 36.25$\pm$0.05  & 0.07  & 38.16$\pm$0.07  \\
L(H$\alpha$) & erg~s$^{-1}$                 & 39.08$\pm$0.04 & 38.47$\pm$0.05  & 39.02$\pm$0.04 & 38.08$\pm$0.05 & 39.01$\pm$0.04  & 38.74$\pm$0.04 & 38.59$\pm$0.04 & 38.80$\pm$0.05 & 38.30$\pm$0.06 & 39.21$\pm$0.04  & 0.47 & 39.92$\pm$0.06  \\
L(Pa$\alpha$) & erg~s$^{-1}$                 & 38.38$\pm$0.05 & 37.94$\pm$0.06  & 38.81$\pm$0.05 & 37.59$\pm$0.11 & 38.37$\pm$0.08  & 38.08$\pm$0.05 & 38.10$\pm$0.06 & 38.27$\pm$0.07 & 37.82$\pm$0.08 & 38.58$\pm$0.05  & 0.42  & 39.48$\pm$0.09  \\
E(B-V)$_{gas}$  & mag                     &  0.23$\pm$0.08  & 0.44$\pm$0.09   & 0.82$\pm$0.08    & 0.48$\pm$0.15  & 0.30$\pm$0.11    & 0.28$\pm$0.08   & 0.48$\pm$0.09   & 0.44$\pm$0.10   & 0.49$\pm$0.12   & 0.32$\pm$0.08    &             &  0.53$\pm$0.13  \\
A(H$\alpha$) &   mag                        & 0.58$\pm$0.20  & 1.12$\pm$0.23    & 2.08$\pm$0.20    & 1.22$\pm$0.38  & 0.76$\pm$0.28    &  0.71$\pm$0.20  &  1.22$\pm$0.23 & 1.12$\pm$0.25   & 1.24$\pm$0.30    & 0.81$\pm$0.20   &             &   1.34$\pm$0.33  \\
L(H$\alpha$)$_{corr}$ &erg~s$^{-1}$ &39.31$\pm$0.08 & 38.92$\pm$0.10 & 39.85$\pm$0.08 & 38.57$\pm$0.14 & 39.31$\pm$0.09  & 39.02$\pm$0.08 & 39.08$\pm$0.08  & 39.25$\pm$0.10 & 38.80$\pm$0.12 & 39.53$\pm$0.08 &             & 40.46$\pm$0.12\\
EW(H$\alpha$) & \AA                        & 225$\pm$27      &  165$\pm$20      & 324$\pm$40         & 450$\pm$77    & 83$\pm$10            & 32$\pm$4        & 47$\pm$6             & 183$\pm$27      & 375$\pm$56      &  155$\pm$19        &             &   25$\pm$5\\
EW(H$\alpha$)$_{dif}$ & \AA           &  291$\pm$35     &  258$\pm$31      & 482$\pm$59         & 234$\pm$40       & 109$\pm$13        & 55$\pm$7          & 108$\pm$13         & 291$\pm$44      & 584$\pm$88      &   214$\pm$26       &             &   45$\pm$8\\
\enddata
\tablecomments{Location, aperture  sizes  and photometry of the 10 regions along the starburst  ring of NGC~3351 (Figure~\ref{fig:Pictures}). Columns (1) and (2) list the property reported in each row and their units: right ascension, declination, radius of the 
photometric aperture (in arcseconds and parsec), luminosity density in each HST band, luminosity in the H$\alpha$ and Pa$\alpha$ lines, color excess, E(B--V)$_{gas}$, derived from these lines, the dust  attenuation at H$\alpha$ (A(H$\alpha$)=2.54$\times$E(B--V)$_{gas}$), the extinction--corrected H$\alpha$ luminosity, and the equivalent width of H$\alpha$ without (EW(H$\alpha$)) and with (EW(H$\alpha$)$_{dif}$) inclusion of differential extinction  correction between line and continuum. All listed values are corrected for foreground Milky Way dust absorption, and H$\alpha$ is also corrected for the [NII] contribution. Measurements for the 10 regions are listed in columns (3) through (12). Column (13) lists the fraction of the light in the Central Region which is in the 10 regions, while column (14) lists the photometry for the residual light in the Central Region, after subtraction of the contribution of  the 10 ring regions  (Res.).}
\end{deluxetable*}
\scriptsize{$^1$ The regions are circular on the sky, but  are ellipses in spatial coordinates with the major axis a factor 1.3 larger  than the minor axis, because of the 40$^o$ inclination angle of the galaxy. }
\end{longrotatetable}

\subsection{Stellar Population Models}\label{subsec:popmodels}

Spectral energy distributions (SEDs) from the UV to the NIR  are generated using the Starburst99 spectral synthesis models  \citep{Leitherer+1999}, using both instantaneous and constant star formation, 
with a \citet{Kroupa2001} IMF in the range 0.1--120~M$_{\odot}$ and metallicity Z=0.02 ($\sim$solar), which is the closest value to the measured oxygen abundance of NGC~3351 for which models are available.  
The extent of the regions, several tens to hundreds of pc, justifies the use of both instantaneous burst (SSP) and constant star formation (CSP) models. We can expect some of the regions to display a more complex star formation history 
than our simplistic dichotomic approach. However, we rely on the fact that the regions are relatively small in area (with the exception of the Residual Region), typically no larger than giant HII regions \citep{Kennicutt1984,Hunt+2009}, to justify our approach. 
We will discuss the Residual Region separately. 

We generate the models using Padova tracks with AGB treatment \citep{Girardi+2000, Vazquez+2005}. Since the regions under consideration are massive, M$\gtrsim$10$^5$~M$_{\odot}$, we expect negligible impact from 
stochastic sampling of the IMF \citep{Cervino+2002}, and hence use the default deterministic models, which imply full sampling of the stellar IMF. The Starburst99 models include nebular continuum, but not nebular emission lines. 
These are added by Yggdrasil \citep{Zackrisson+2011}, which uses Starburst99 stellar populations as an input for CLOUDY \citep{Ferland+2013}, with hydrogen number density $\rm n_H=10^2\ cm^{-3}$ and gas filling factor =0.01, typical 
of HII regions \citep{Croxall+2016}. Yggdrasil offers nebular emission models with 0\%, 50\%, and 100\% covering factor, implying that 0\%, 50\%, or 100\% of the ionizing  photons are used to ionize the gas in the nebula. 
The presence of at least some nebular emission in all the considered regions implies $>$0\% covering factor. Tests run on the remaining two available options, 50\% and 100\%, indicate that the latter generally produce poorer 
fits (larger $\chi^2_{red}$) than the former. Therefore, we adopt a 50\% covering factor for the ionized gas for all our regions, meaning that only 50\% of the nebular emission is spatially coincident with the region. While we 
expect the covering factor for most of our regions to be $<$100\%, a general 50\% covering factor is still an over--simplification, although it provides reasonable matches between the data and the models. The models used here  
assume non--rotating, single stars. There is, however, increasing evidence that stellar population models implementing rotating, binary (or multiple) stars are better fits for many data, especially at young ages and low metallicities \citep{Stanway+2020}. 
Differences in the ages and masses resulting from SED fits that use different input population models can be at the level of factors $\sim$2--3, but differences in colors excesses are decidedly more modest, $<$0.1~mag \citep{Wofford+2016}. 
Thus the specifics of the stellar population models are not expected to have an impact on this work, which mostly concentrates on attenuation effects. 
Yggdrasil  models are available in the age range 1~Myr--14~Gyr for instantaneous star formation (with nebular emission becoming negligible beyond $\approx$7~Myr) and 1--100~Myr for constant star formation. 
We extend the constant star formation models to 14~Gyr by combining the stellar inputs from Starburst99 with the nebular inputs made available by Yggdrasil. Age steps range from 1~Myr  below 15~Myr to 1~Gyr  above 1~Gyr.

\subsection{Dust Attenuation Models}\label{subsec:attmodels}

\begin{figure}[ht!]
\plotone{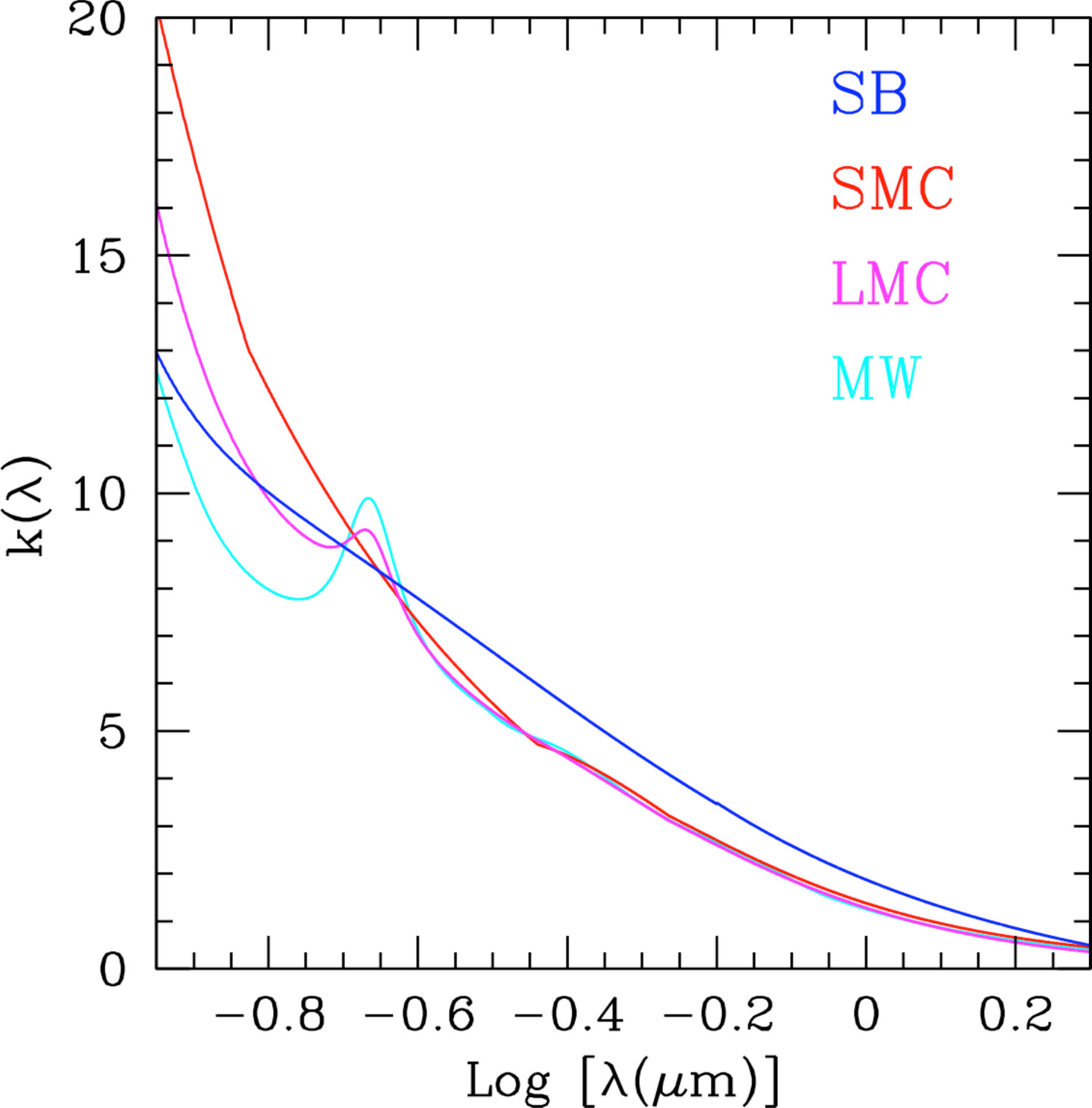}
\caption{The four extinction/attenuation curves used in this work, shown in the range 0.1--2~$\mu$m, relevant to the present analysis. The functional form for the three extinction curves: Milky Way (MW), Large Magellanic Cloud (LMC),
and Small Magellanic Cloud (SMC)  are  from \citet{Cardelli+1989} and \citet{Pei1992}, as parametrized by \citet{Fitzpatrick1999};  the functional form for the SB attenuation curve is from \citet{Calzetti+2000}. The only notable feature is present in the 
MW and LMC curves, at 0.2175~$\mu$m central wavelength; the other two curves do not have this feature. Although more recent versions of all curves 
exist \citep[e.g.,][]{Gordon+2003, Fitzpatrick+2019}, the expressions adopted cover sufficient parameter space for this work. \label{fig:ext_curves}}
\end{figure}

The stellar population model SEDs are attenuated with: the SB attenuation curve, and a Milky Way (MW), a Large Magellanic Cloud (LMC), and a Small Magellanic Cloud  (SMC) extinction curve \citep[as parametrized by][]{Fitzpatrick1999}. For the extinction curves, 
we adopt two dust geometries \citep{Calzetti2001}: foreground dust:
\begin{equation}
L(\lambda)_{obs} = L(\lambda)_{int} 10^{[-0.4 E(B-V) k(\lambda)]},
\end{equation}
and a homogeneous mixture of dust, stars, and gas:
\begin{equation}
L(\lambda)_{obs} = {L(\lambda)_{int} \{1- e^{[-0.921 E(B-V) k(\lambda)]}\} \over 0.921 E(B-V) k(\lambda)}.
\end{equation}
The latter is to account for scenarios where the emitting population(s) are buried in the environmental dust. 
The observed (output) and intrinsic (input) luminosity densities L($\lambda$) are linked by the attenuation: $E(B-V) k(\lambda)$, where E(B--V) is the color excess and k($\lambda$) provides the functional  form 
of the extinction curve, normalized at 0.55~$\mu$m to k(V)=3.1 (Figure~\ref{fig:ext_curves}). For the foreground geometry, we consider  both cases of equal and 
differential attenuation for the nebular gas and stellar continuum; for the differential attenuation, we assume that the stellar continuum is subject to half the attenuation of the 
nebular gas \citep{Calzetti+1994, Wild+2011, Kreckel+2013}. For the SB attenuation curve, we apply Eq.~3 with the functional form k($\lambda$) of \citet{Calzetti+2000}. In this case, the dust geometry and the differential attenuation between 
gas and stars are `built--in' into the functional form of the curve, as per results by \citet{Calzetti+1994}. \citet{Calzetti+2000} provides the normalization for  the attenuation curve: k(V)=4.05. In all cases, dust absorption and scattering are  
included in the expressions for k($\lambda$). We thus end up with 10 different models for the dust attenuation: one attenuation curve and nine geometry/extinction curve 
combinations (three curves times three different ways of attenuating gas and stars: foreground, foreground differential, and mixed). We generate the models in the color excess range E(B$-$V)=0--3~mag, with step size of 0.01.

As in the case of the star formation histories described in the previous section, the adoption of two basic geometries may appear to be an over--simplification of an otherwise 
complex distribution. However, these cases represent opposite extrema in terms of the effective reddening on the SED, with the foreground screen causing the most reddening and the homogenous mixture causing the least reddening \citep[][see their Figure 1.4]{Calzetti2013}. Our goal is to bracket possible geometries for the dust attenuation in the 50--150~pc regions, where we expect the stellar populations to be close to co--eval or at least 
to  display a relatively small range of ages. We do not expect the above descriptions for the attenuation to apply to the integrated light from the Central Region, as they are likely too simplistic for a complex region; this is already 
partially suggested by Figure~\ref{fig:IRX_beta} and discussed more at length in Section~\ref{sec:central}. 

Although we use all three extinction curves in fitting the FUV--to--NIR SEDs, we note that the observed SED of the central region of NGC~3351 does not show evidence for the presence of the 
0.2175~$\mu$m absorption feature. This is highlighted in the right panel of Figure~\ref{fig:central_SEDs}, which zooms into 
the FUV--to-NIR (0.15--1.9~$\mu$m) portion of the regions's SED and the UV--optical spectrum of \citet{StorchiBergmann+1995}. The spectrum is particularly constraining for the absence of the 
0.2175~$\mu$m feature in the attenuation, since photometry can sometimes give degenerate results between a feature--less and relatively grey extinction/attenuation curve and a curve with strong 0.2175~$\mu$m feature
and steep UV rise. The absence of an obvious broad absorption feature in the otherwise extremely red UV spectrum of the Central Region leads us to give preference in what follows to the SB and SMC curves, which do not contain the 
feature \citep{Pei1992, Calzetti+1994}, over either the MW or LMC curves, both carriers of the feature \citep{Gordon+2003}. Processing of dust carriers in the UV--intense and turbulent starburst 
environments may explain the absence of the 0.2175~$\mu$m absorption feature in these regions \citep{Gordon+1997, Fischera+2011}. 

For a fixed dust geometry/extinction curve choice, the free parameters in the fit of a dust--attenuated stellar population SED are: 1) the age (for an instantaneous burst) or star formation duration (for 
constant star formation); 2) the mass in stars, M$_{\star}$, and 3) the color excess E(B--V)$_{star}$ from the effects of dust.

\subsection{Dust Emission Models}\label{subsec:dustmodels}

Fits of the IR SED are used to calculate the total infrared luminosity, L(TIR), and the dust mass of the Central Region. The dust mass is used 
as a comparison term for the dust column density traced by the color excess E(B--V) at optical wavelengths. 
As already defined in Section~\ref{sec:galaxy}, L(TIR) is the emission in the range 3--1100~$\mu$m, which we adopt 
for uniformity with other authors. For the fits (Figure~\ref{fig:central_SEDs} , left), we employ the models of \citet{DraineLi2007}, as implemented in \citet{Draine+2007}\footnote{The models are publicly available at: 
https://www.astro.princeton.edu/~draine/dust/irem.html}. The models consist of a mixture of   
carbonaceous grains (including PAHs) and amorphous silicates, with size distributions that aim at reproducing the  MW,  
LMC, and SMC extinction curves. For each grain distribution (extinction curve), a range of PAH dust mass fractions is considered, 
between q$_{PAH}$=0.01\% and 4.6\%, in several discrete values. The highest q$_{PAH}$ value is consistent with Milky Way--type dust, while the SMC and LMC--type dust have lower q$_{PAH}$ values.

The dust mixture is heated by the combination of two starlight intensity 
components:  the diffuse starlight that permeates the interstellar medium, described by the energy density parameter {\em U$_{min}$}, and a range of regions with a power law 
distribution of intensities, $dM_{dust}/dU \propto U^{-\alpha}$, between  {\em U$_{min}$}, and {\em U$_{max}$}  and slope $\alpha$=$2$\footnote{The adoption of this value of  $\alpha$  is supported by 
the findings of \citet{Aniano+2020} for the center of NGC~3351.}. The two starlight  intensity components are added together in proportion to 
($1-\gamma$) and $\gamma$, where 0$\le\gamma\le$1. The parameter $\gamma$ is related to the fraction of starlight intensity due to 
current star formation. 
The emission in each band will also be proportional to the total  
dust mass, M$_{dust}$. \citet{Draine+2007} and \citet{Aniano+2020} established that the derived dust emission parameters are not sensitive to the choice of U$_{max}$, which we fix at a value of 10$^5$, borrowing  
from  the  study  of \citet{Calzetti+2018}.
Thus, for each given extinction curve, there are a total of four parameters for the models: q$_{PAH}$, U$_{min}$, $\gamma$,  and M$_{dust}$.

\subsection{Fitting Approach}\label{subsec:fitting}

The dust--attenuated stellar model SEDs and the dust emission model SEDs are convolved with the bandpasses of the relevant facility/filter combinations, in order to produce synthetic photometry  to be compared with 
the data.  We use $\chi^2$--minimization between the models and the data, taking into account the measurement uncertainties, to obtain the distribution of solutions and the reduced 
$\chi^2$ value for each. We then plot the distribution of solutions within the  90\% significance level for the appropriate number of degrees of freedom, and select the best values 
and the uncertainty for the parameters of each region based on the shape of the reduced $\chi^2$ probability distribution. 

In the UV--NIR range, our fits are performed with 10 or 12 datapoints\footnote{The SEDs of  the ring regions, R1, R2, etc., are each defined by 10 datapoints;  12 datapoints are available for the SEDs of the Central and Residual Regions, 
since, for these, we can add  the two GALEX measurements.} which, for three parameters (age, stellar mass, E(B--V)$_{star}$), imply 6 or 8 degrees of freedom in the fits. In comparison to recent literature, our 
UV--NIR fits use at least twice as many datapoints (bandpasses) than other similar analyses of the same galaxy \citep{Turner+2021}, thus yielding more stringently--determined ages, masses, and extinctions, but encompass regions 
that are at  least several tens of pc in size, as opposed to individual star clusters \citep{Adamo+2017}, which adds complications in the modeling of the SFH.  In the IR--sub-mm range, the dust emission is fit 
to data at wavelengths $\ge$8~$\mu$m, implying 6 datapoints and (for four parameters) one degree of freedom. 

For some comparisons (see Figure~\ref{fig:IRX_beta}), we use the slope of the UV spectrum, $\beta$, measured from the  GALEX bands as described in \citet{Kong+2004} and \citet{Calzetti+2005}. Comparisons with models 
require that the exact same bands are used for data and models. As an example, for the intrinsic SEDs of young  stellar populations, the relation between the UV slope measured in the GALEX bands and that measured from
spectra as described in \citet{Calzetti+1994} is: $\beta_{GALEX}\simeq 0.94 \beta_{Calzetti}$.  While not large, this difference, which is  entirely  due to the GALEX bands probing a UV spectral region that 
is $\sim$0.2~$\mu$m redward of the region used by \citet{Calzetti+1994}, can lead to  confusion if not properly included. Furthermore, differences from 
different ways of  measuring $\beta$ are magnified for dust--attenuated SEDs.  Due to the non--linearity of the FUV raise of extinction/attenuation curves, the UV slope $\beta$ 
derived from two photometric  bands depends on the filters' shapes and central wavelengths. This effect is exacerbated for highly non--linear curves, and at high E(B--V) values, because the attenuated UV
spectrum will increasingly deviate from a single power law shape. 
For example, a slope $\beta$ measured using a FUV filter just 100~\AA\ bluer than the GALEX FUV produces a flatter expected IRX--$\beta$  relation for the SMC curve in Figure~\ref{fig:IRX_beta}, with the 
IRX being $\sim$0.2~dex lower at $\beta=0$ (E(B$-$V=0.2)) than 
the fiducial SMC relation. This effect is less important in the case of the SB curve, which has a less steep FUV raise than the 
SMC extinction curve, which would  cause a change in the IRX by $\sim$0.09~dex at $\beta=0$ (E(B$-$V=0.5)) in the same experiment. This problem will generally be more pronounced when comparing model expectations 
with observations of galaxies at a range of redshifts, and will be  less pronounced if spectra, rather than sparse photometry, are used to measure $\beta$. In our study, the model 
spectra are convolved with the GALEX filter bandpasses, in order to derive expected trends that are consistent with  the way  the data are measured. 

\section{The Central Region}\label{sec:central}

The photometry listed in Table~\ref{tab:central} is  used  in this section to derive the characteristics of the Central 
Region and set the stage for the analysis of the ring regions. 

\subsection{Dust Emission  Properties}\label{sec:central_dust_emission}

The small number (5) of datapoints available at $\lambda\ge$8~$\mu$m provide a weak constraint on the physical  parameters that characterize the IR SED 
of the Central Region. Of the four free parameters (Section~\ref{subsec:fitting}) that define the shape and luminosity of the IR SED, q$_{PAH}$ is fairly slow--varying across 
a galaxy \citep{Aniano+2020}, and can be fit using a larger area than the Central Region. We thus fit the shape of the IR SED with the \citet{DraineLi2007} models 
described in Section~\ref{subsec:dustmodels}, using the $\sim$36$^{\prime\prime}$--aperture photometry from 8~$\mu$m to 500~$\mu$m of Table~\ref{tab:SPIRE500}, 
i.e., a total of eight datapoints. We obtain a best--fit q$_{PAH}$ value of 1.8\%, close to the 2\% value derived by \citet{Aniano+2020} for the central area of  the galaxy, 
and  U$_{min}$=12$\pm$3, which we also apply to the higher resolution data, since the Central Region provides over 70\% of the flux in the $\sim$36$^{\prime\prime}$ 
aperture. 

Adopting the above values of q$_{PAH}$ and U$_{min}$, we fit the IR SED of the Central Region in the 22$^{\prime\prime}$ aperture (Table~\ref{tab:central}),
obtaining  $\gamma$= 0.03$^{+0.03}_{-0.01}$, and a dust mass M$_{dust}$=(7.7$^{+1.5}_{-3.5}$)$\times$10$^5$~M$_{\odot}$. The 
post--Planck correction factor described in \citet{Aniano+2020} has been applied to the value of M$_{dust}$. The high value of U$_{min}$,  U$_{min}\simeq$12, 
is not uncommon in regions of strong star formation \citep[e.g.,][]{Calzetti+2018}. Using the formula of \citet{Draine+2014}, the 
characteristic dust temperature in the Central Region of NGC~3351 is T$_d\sim$27~K, consistent  with its starburst  nature, and in agreement  with the 
temperature determination of  \citet[][]{Nersesian+2020}. 

Integrating under the best--fit SED (Figure~\ref{fig:central_SEDs}, left), we derive a total IR luminosity L(TIR)$_{mod}$=10$^{(43.15\pm0.08})$~erg~s$^{-1}$ (Table~\ref{tab:linecentral}). 
This value is only 20\% higher than what is obtained by applying the empirical formula of \citet{DaleHelou2002} to the IR photometry of the central region of NGC~3351. We will use L(TIR)$_{mod}$
as our fiducial dust emission luminosity in the rest of the paper.

\subsection{Dust Attenuation and Stellar Population Properties}\label{sec:central_stellar_pops}

As a  starting point, we apply the standard approach used for unresolved galaxies to the FUV--to--NIR SED of the Central Region: a global fit using assumptions for the dust 
attenuation curve  and the star formation history. The assumptions are those described in Section~\ref{sec:models}. 

The best fit models for the dust attenuated stellar SEDs are shown in the right panel of Figure~\ref{fig:central_SEDs}.
The SB curve yields the best agreement between model and observations across all 12 datapoints (2 GALEX and 10 HST), with a reduced $\chi^2_{red}$=1.2 for instantaneous 
burst models (shown in Figure~\ref{fig:central_SEDs}) and about twice that value for constant star formation models. 
Conversely, the SMC curve produces an unacceptable fit to the data with $\chi^2_{red}$=9.1 for instantaneous burst models (shown in Figure~\ref{fig:central_SEDs}) 
and  a 50\% larger  $\chi^2_{red}$ value for constant star formation. These $\chi^2_{red}$ values are for the SMC curve used with  foreground dust;  mixed dust models yield even worse $\chi^2_{red}$  values. A close inspection of the models with the SMC extinction curve shows that they overproduce the U, B, and V band luminosities, while underproducing
the NIR luminosities. Use of the MW or the LMC extinction curves on the FUV--to--NIR SED produces even worse fits than the SMC one; this is explained by the absence of a 
0.2175~$\mu$m feature in the UV spectrum of the region \citep{Kinney+1993, StorchiBergmann+1995}. 

We use the best fits above to estimate  the expected TIR dust emission in both cases of the SB and SMC curves. The TIR emission is taken as the difference 
between the intrinsic SED and the attenuated SED, integrated from $\lambda$=0 to the NIR included. The calculations yield that the population model with the SB curve 
over--produces the observed IR dust luminosity by a factor $\sim$7.5, giving L(TIR)$=$10$^{44.02}$, while the model with the SMC extinction produces a closer value to the observed one, L(TIR)=10$^{43.26}$, 
being only $\sim$30\% larger. Thus, from the point of view of dust luminosity, the SMC curve 
 performs better, i.e., produces a closer value to the observed one, than the SB curve, a result we had already inferred from the location of  
 the Central Region data relative to both the SB and SMC curves in  the IRX--$\beta$  diagram (Figure~\ref{fig:IRX_beta}). 
 
In summary, the simplified star  formation histories adopted in our fitting approach of the FUV--to--NIR SED of the Central Region can either account for the shape of 
the SED, at the cost of over--predicting the dust luminosity (the case of the SB curve), or reproduce the dust luminosity at the cost of not fitting the shape of the 
FUV--to--NIR SED (the case of the SMC curve). This is a clear indication that the star formation history is far more complex than our simple assumptions, and a region--by--region 
analysis is required.

Incidentally, the average gas color excess, E(B--V)$_{gas}$, in the Central Region (Table~\ref{tab:linecentral}) corresponds to a dust mass M$_{dust}\sim$1.9$\times$10$^5$~M$_{\odot}$, 
assuming the relation between gas column density and E(B--V) of \citet{Bohlin+1978}, and a dust--to--hydrogen~mass ratio of 0.01, appropriate for a galaxy with solar  metallicity 
\citep{Draine+2007}. The color excess only accounts for the dust mass between the stellar populations and the observer, missing the dust behind the stars. If we 
assume a mid--plane geometry for the dust, the dust mass doubles to M$_{dust}\sim$3.8$\times$10$^5$~M$_{\odot}$, which is about a factor 2 lower than 
 the mean dust mass derived from the fit of the IR SED in the previous section, although the two numbers are consistent with each other within the uncertainties. 
 This consistency implies that there is little evidence for the Central Region to include subregions that are deeply buried in dust, and our analysis will not miss major dust heating components.

\section{Modeling the Individual Regions Along the Starburst Ring}\label{sec:regions}

The SEDs of the ring regions include data in the range $\lambda\sim$0.27--1.9~$\mu$m and do not have individual measurements 
at shorter wavelengths because of resolution  limitations in the GALEX images. As a consequence, we cannot univocally determine whether a 0.2175~$\mu$m 
feature is present in the SEDs of these regions. Thus, we take the SED of the Central Region as our clue that we should not expect a feature to be present, and only 
use the SMC and SB curves for our fits. We run tests using the MW and LMC curves for a subset of the regions, and find that, indeed, 
these fits cannot be uniquely separated from those using the SMC or SB curve; in general, the reduced  $\chi^2$ from the fits using the MW and LMC 
curves fall in--between the reduced $\chi^2$ values of the fits using the SMC and the SB curves. 

\begin{figure}
\plottwo{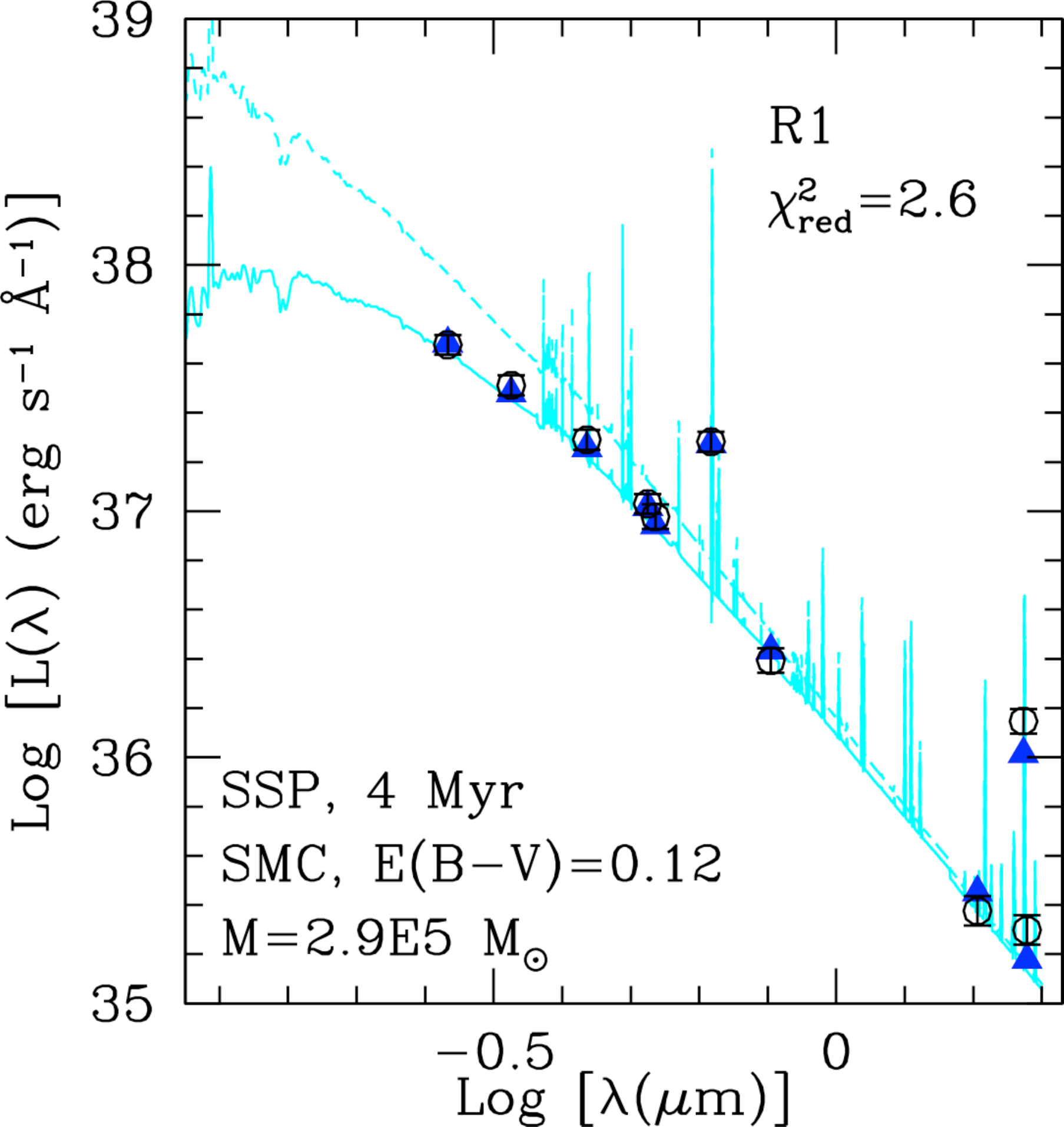}{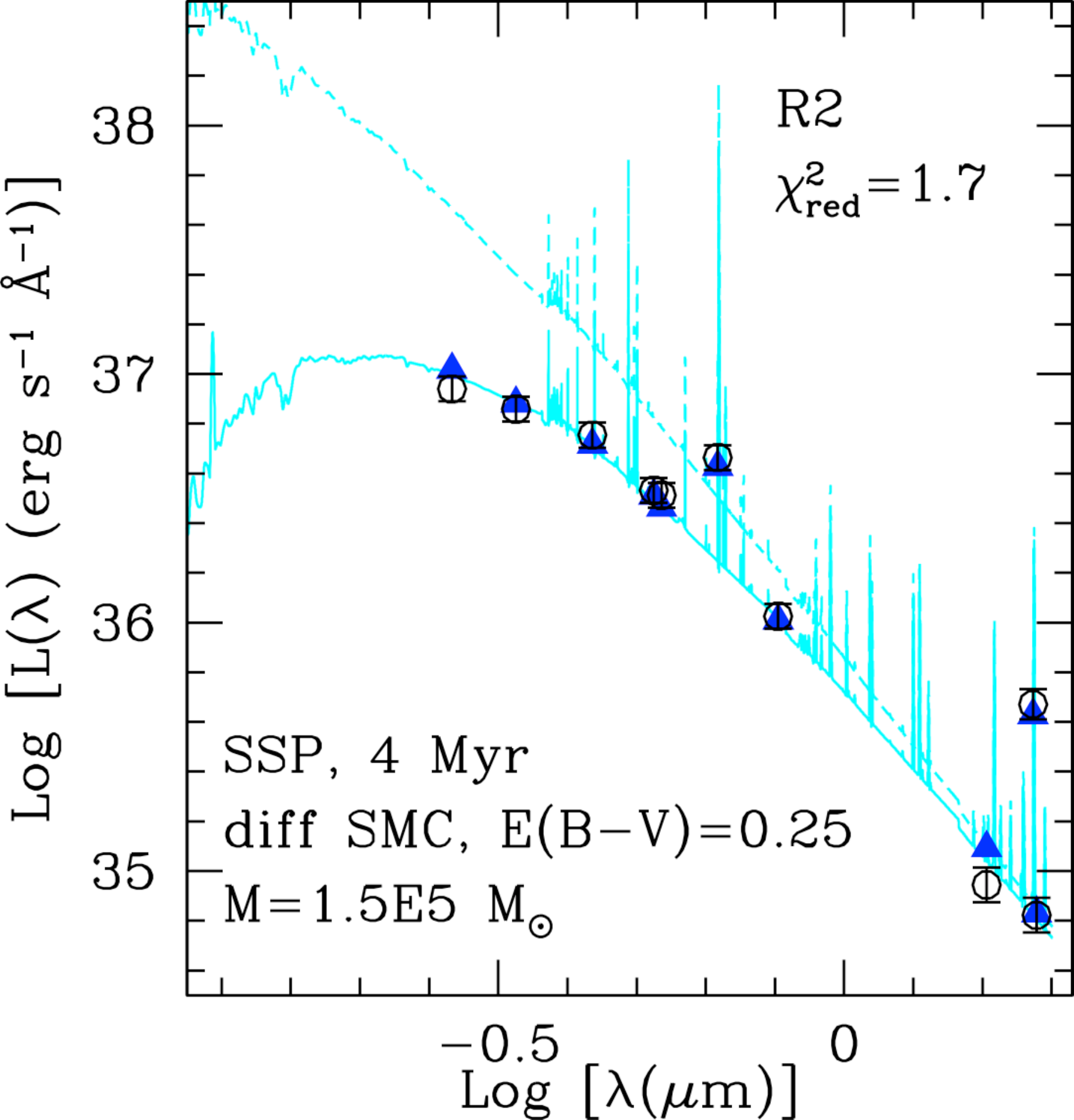}
\plottwo{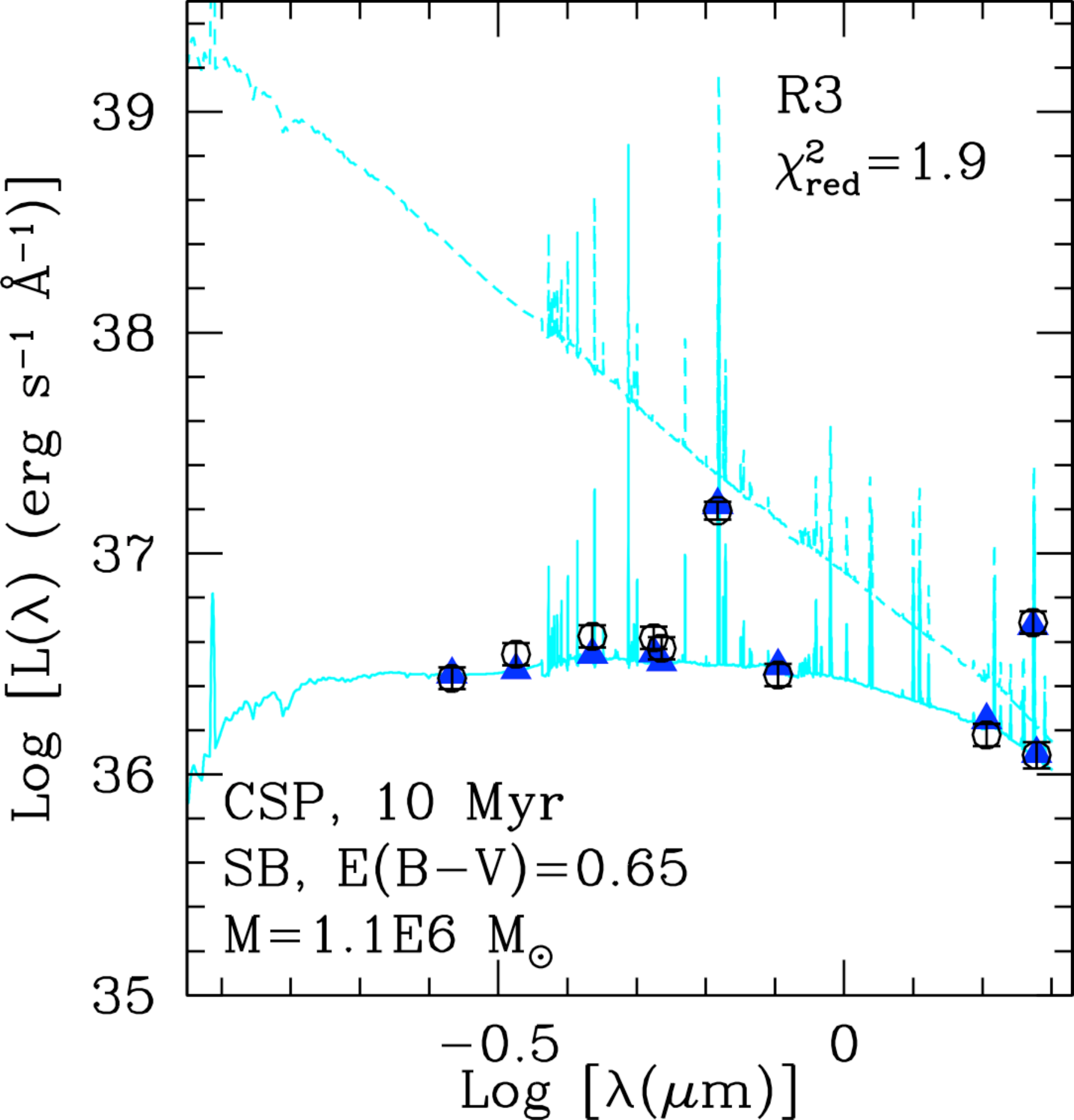}{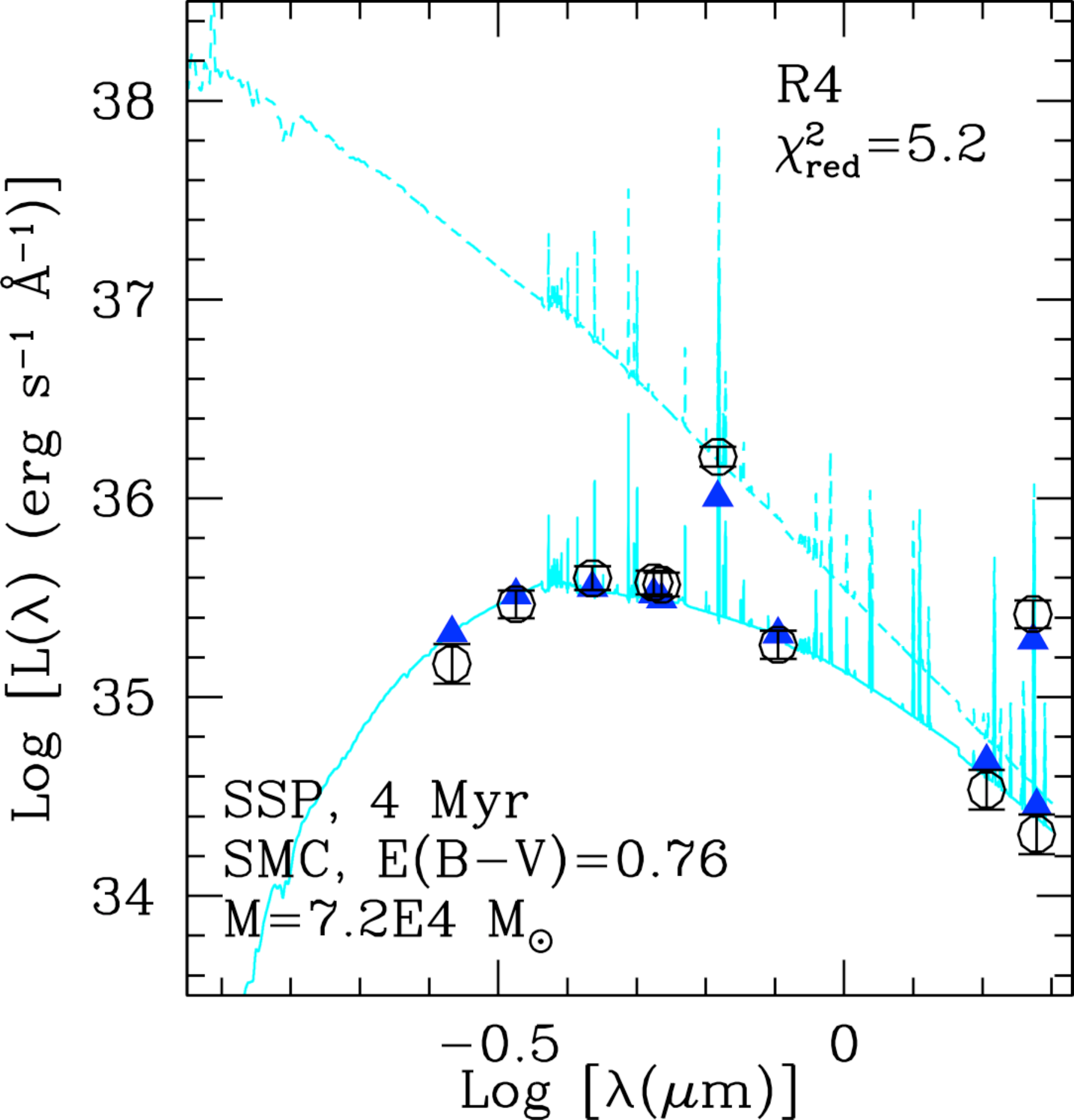}
\caption{Photometry (black circles with 1$\sigma$ error bars) for four of the ten ring regions, together with best fit models (cyan lines, dash=stellar population only; continuous=stellar population+dust attenuation) and synthetic photometry (blue triangles). The parameters listed in each panel are the central values from Table~\ref{tab:fits_regions}. The value of the reduced $\chi^2$ is also shown in each panel. \label{fig:region_bestfit_1}}
\end{figure}

\begin{figure}
\plottwo{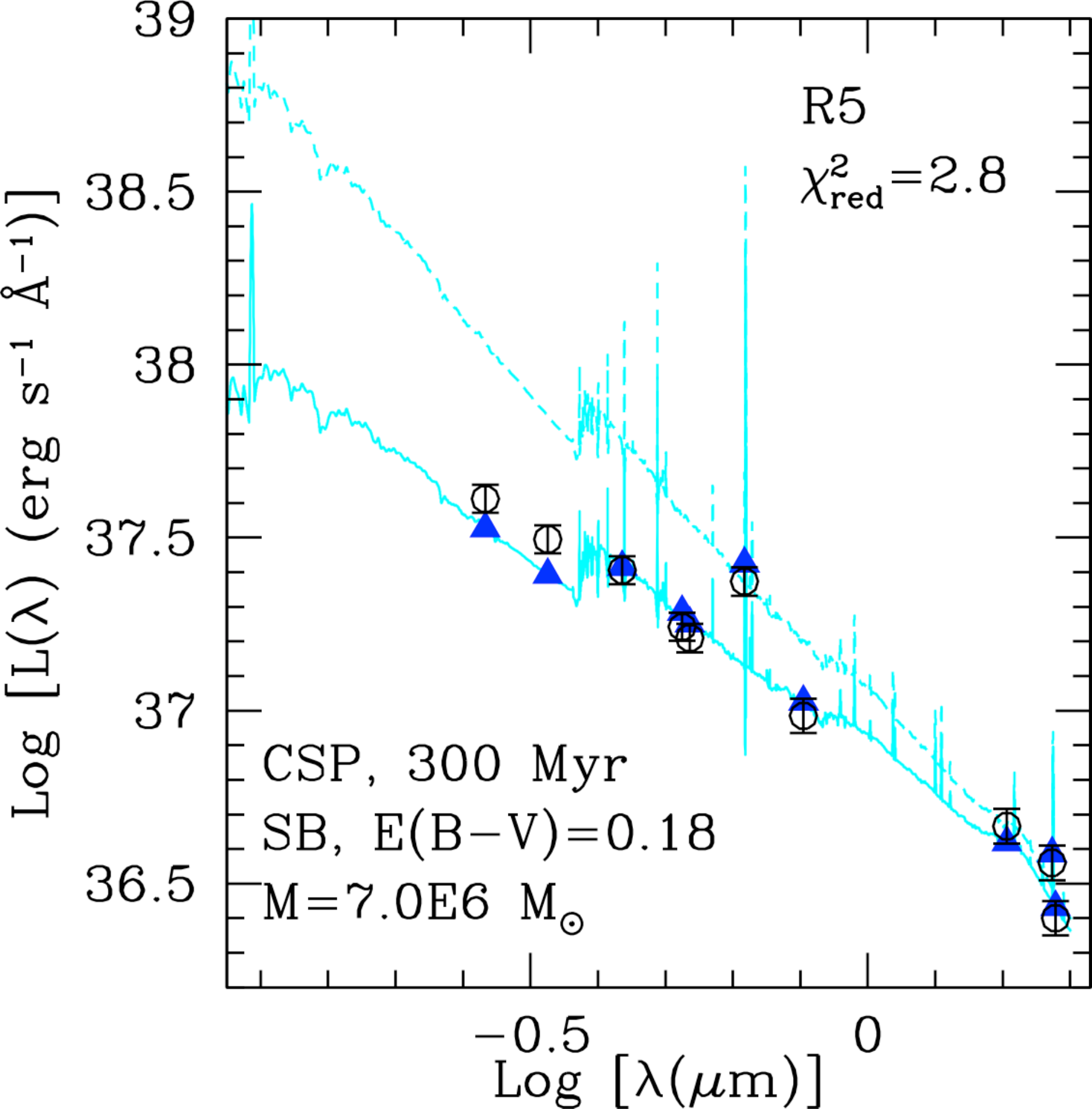}{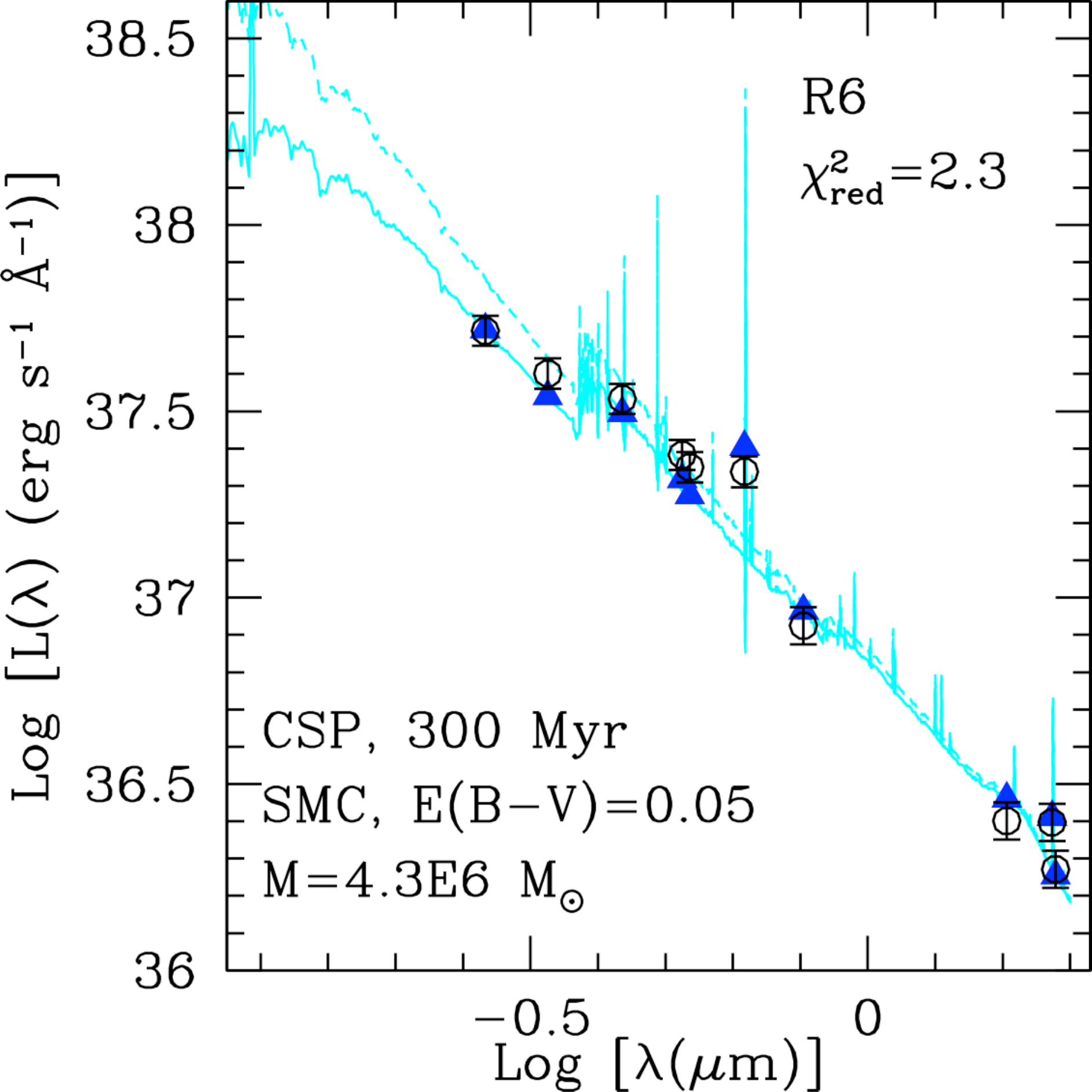}
\plottwo{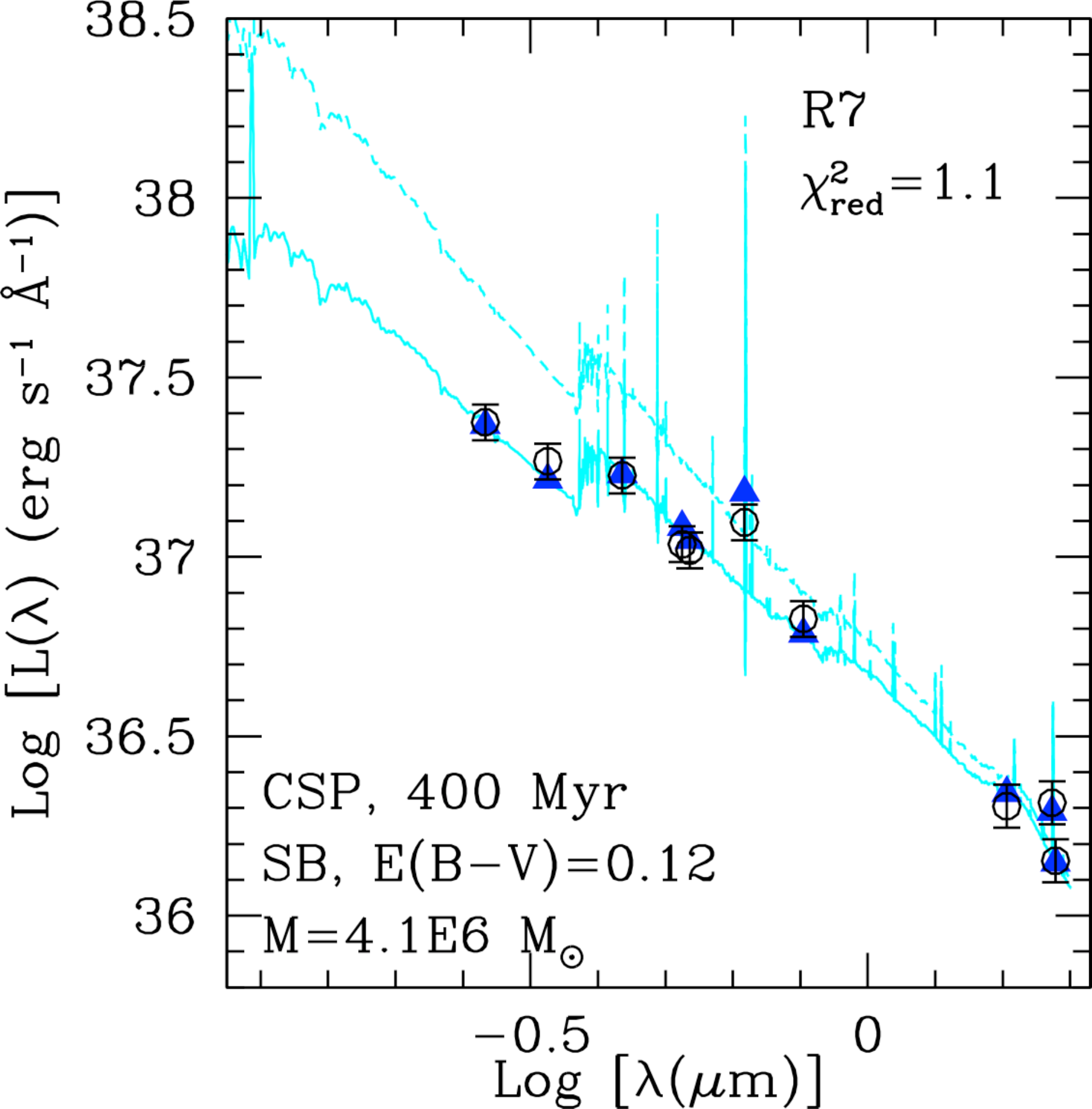}{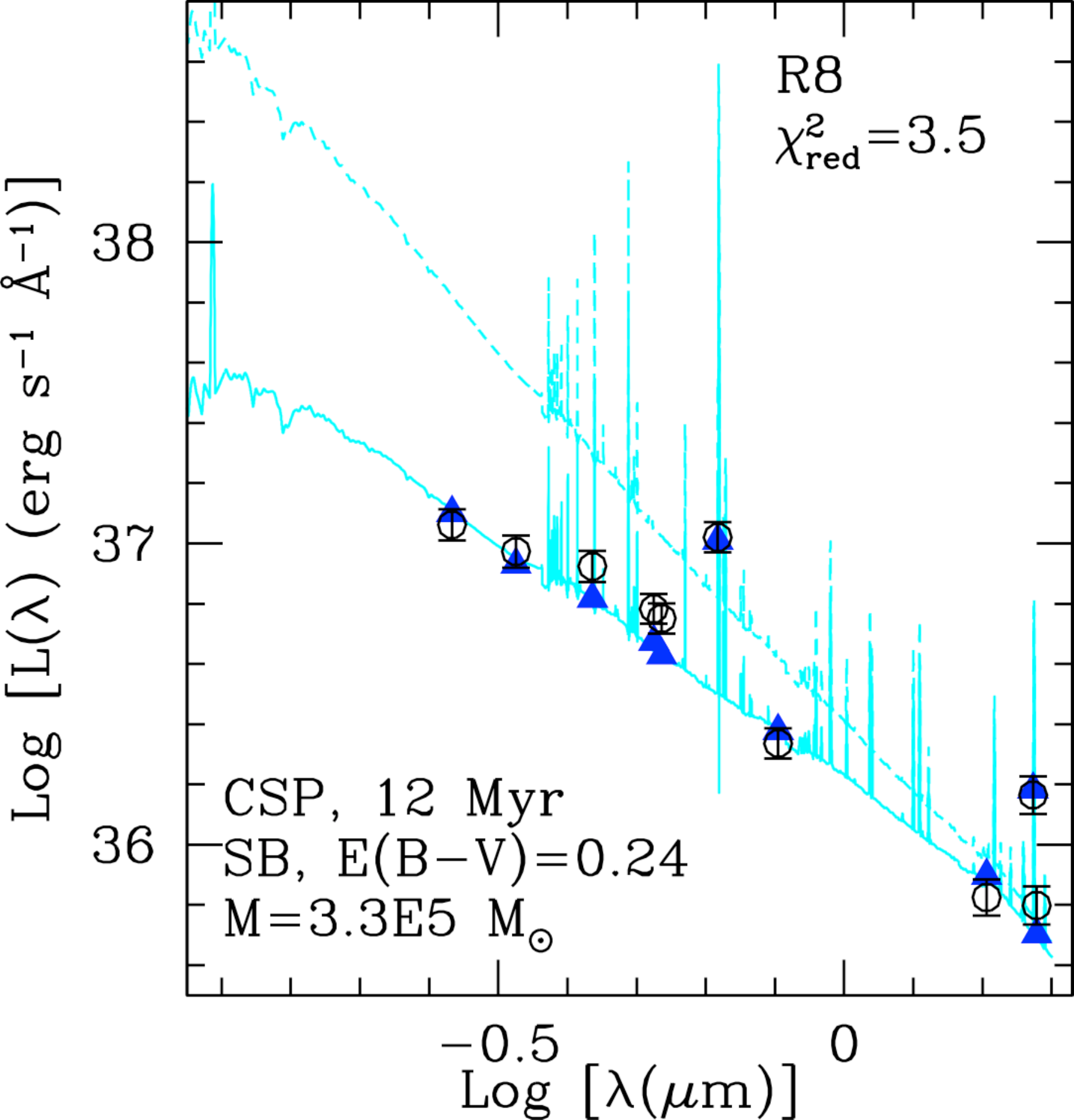}
\plottwo{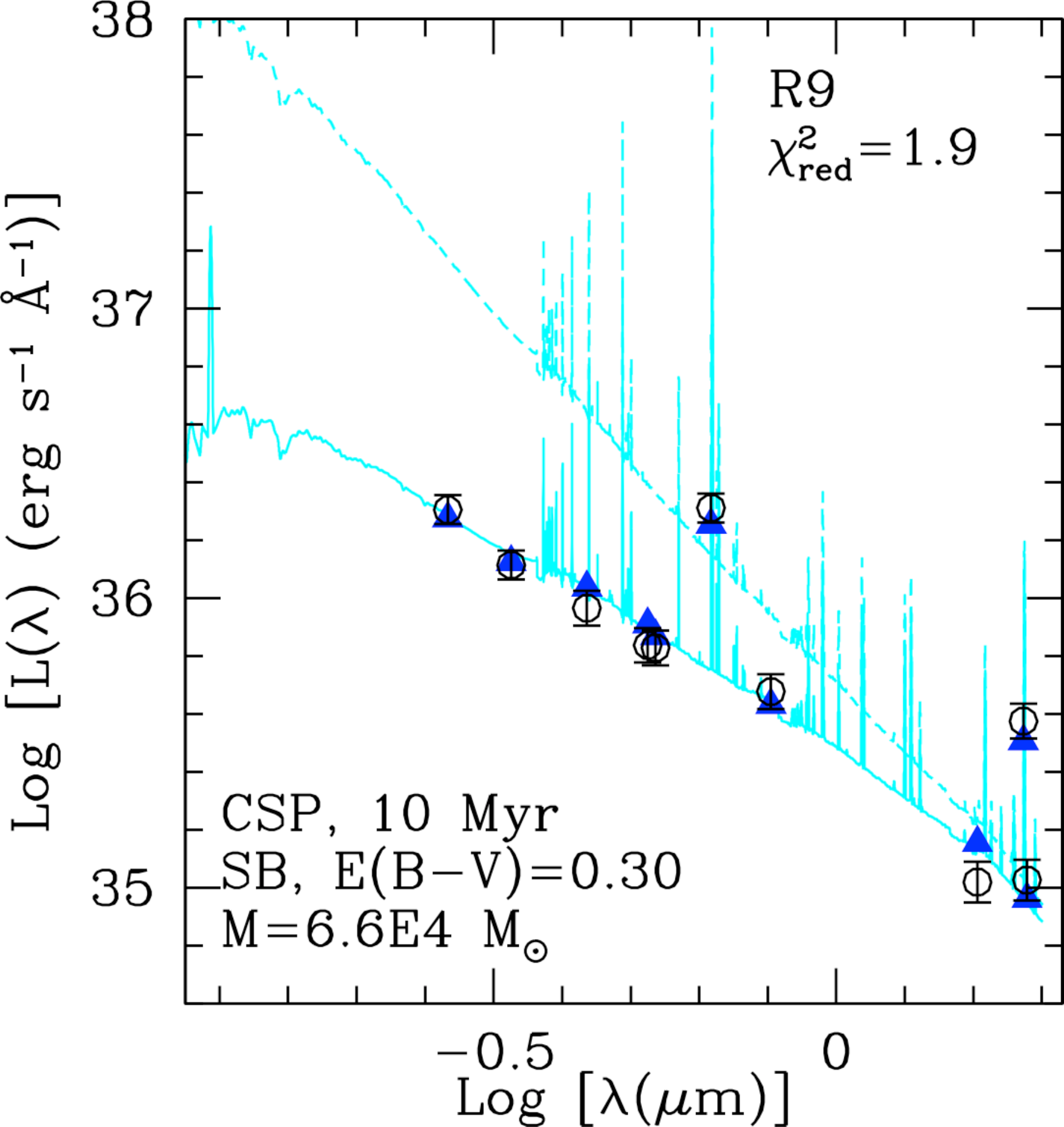}{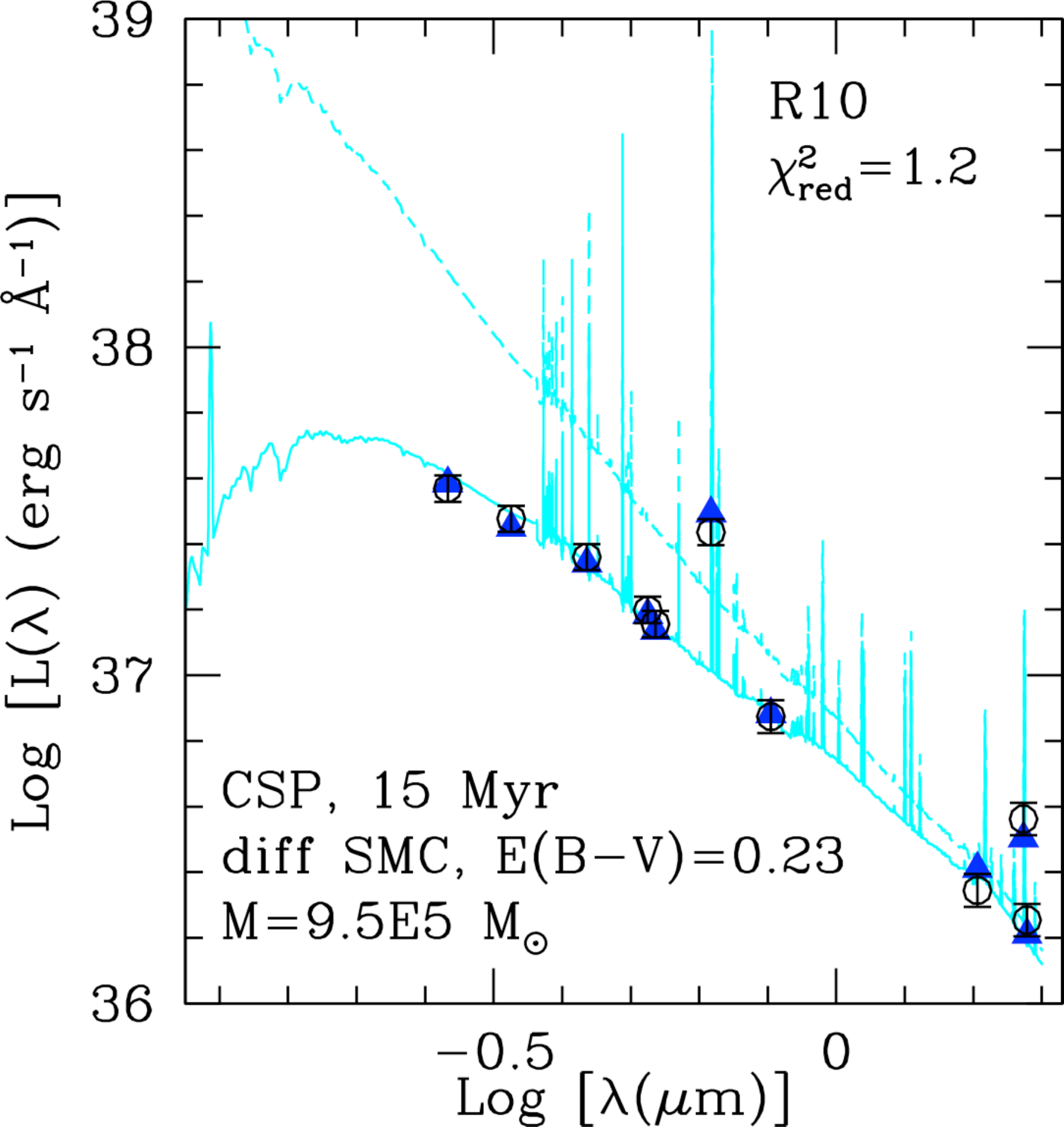}
\caption{As Figure~\ref{fig:region_bestfit_1} for the remaining six ring regions.\label{fig:region_bestfit_2}}
\end{figure}

The summary of our best fit parameters is listed in Table~\ref{tab:fits_regions}.  The small sizes of the regions justify the use of simplified star formation histories: instantaneous burst 
and continuous star formation, as discussed in Section~\ref{subsec:popmodels}. In fact,
most fits are reasonable, as inferred from the small values of the reduced $\chi^2$. The SEDs of the best fitting stellar population and dust attenuation models for the 10 ring  regions  are shown in Figures~\ref{fig:region_bestfit_1} and \ref{fig:region_bestfit_2}.
As a reminder, with 10 photometric datapoints and three parameters (age, stellar mass, and color excess), we end 
up with 6 degrees of freedom for the fits.  Examples of the distributions of the fitted parameters, from which their uncertainties 
are derived, are provided in Appendix~\ref{Distributions}. In general, if the stellar population is well fit by a SSP model, the CSP models are excluded at the level of several $\sigma$, and vice versa. Thus, from the point of 
view of the star formation history, the fits are unique.

For the extinction/attenuation, both the SB curve and the SMC curve provide reasonable fits  in most cases. Thus the attenuation curve is mostly degenerate in our  situation, mainly due to the absence  of UV data at wavelengths $\lambda<$0.27~$\mu$m, where differences between the curves become  more pronounced. Table~\ref{tab:fits_regions} lists the curve that provides the smallest reduced 
$\chi^2$ value, although in general the other curve provides a best fit $\chi^2$ value that is within 20\% of the one listed. 
As an example, the best fit for R4 is the SMC extinction curve with foreground geometry, color excess E(B--V)$_{star}$=0.76 and a 4~Myr old instantaneous burst population model, which yields  $\chi^2_{red}$=5.2; switching to the SB curve would only 
slightly worsen the goodness of fit ($\chi^2_{red}$=5.9) while still requiring a 4~Myr old instantaneous burst model with slightly lower color excess, E(B--V)$_{star}$=0.65. Exceptions to this almost `interchangeability' of the 
extinction/attenuation curves are R3, where the SB curves provides a best fit  that has a 40\% lower  $\chi^2_{red}$ value than the SMC curve, and R10, for which a differential 
SMC curve yields a 35\% lower $\chi^2_{red}$ value than the SB curve. In those cases that are better fit with the SMC curve, the data are in better agreement with foreground extinction models than with mixed geometry models, or have, in a few cases, comparable goodness of fit. The photometry of two regions, R2 and R10, gives strong preference 
to the differential attenuation model (Section~\ref{subsec:attmodels}). For color excess values E(B--V)$\lesssim$0.1-0.15~mag (e.g., R1, R6, and R7 in Table~\ref{tab:fits_regions}) the different dust geometries are virtually indistinguishable on the basis of the data; 
this is an obvious consequence of the minor role played by dust attenuation on the stellar SED at low values of the  color excess. For R5, mixed  geometry models with an SMC curve provide a reduced $\chi^2$ that is only slightly 
worse than the SB attenuation, but give the same solution for the star formation duration of the stellar population. Mixed dust geometry models, however, are strongly disfavored by our most strongly attenuated 
regions, R3 and R4.  In summary, foreground dust geometries are preferred for  the 10 ring regions,  with the SB curve and the SMC curve providing the best fit each for half of the regions. 

The three regions better fit with instantaneous burst models (R1, R2, and R4) have uniformly young ages, around 4~Myr, with relatively small uncertainties (Table~\ref{tab:fits_regions}). They are 
also all better fit with the SMC extinction curve, with foreground geometry;  in all cases the reduced $\chi^2$ does not change if the extinction is implemented with and without differential extinction between nebular gas and stellar continuum. 
For all other regions, which are better fit with constant star formation models (Figures~\ref{fig:region_bestfit_1} and \ref{fig:region_bestfit_2}), the duration of the star formation spans a large range, from 10~Myr to $\sim$400~Myr. The large uncertainties 
on the durations reflect the fact that for constant star formation the UV--NIR SED changes slowly with age. 

As a sanity check, we derive independent  ages using the EW(H$\alpha$), including and excluding: (1)  leakage of ionizing photons from the  region, which corresponds to assuming both 50\% and  100\% 
covering fraction, and (2) differential extinction between 
emission lines and underlying stellar continuum. For  the calculation, we adopt the same SFH (instantaneous  burst or constant star  formation) as the best--fits for the SEDs. 
The values derived with this range of assumptions are listed in the row called Age$_{lines}$ in Table~\ref{tab:fits_regions}. 
Adopting a 50\%  level of leakage from HII regions is justified in light of previous results for nearby galaxies 
\citep{Ferguson+1996,Oey+2007} and of the preference for  our SEDs to  be better fit with models that use gas covering fraction 50\%  (as opposed to 100\%, see Section~\ref{subsec:popmodels}). In most cases, the age ranges inferred from the  EW(H$\alpha$) are
consistent with those derived from the SED fits. Discrepancies can be observed for R6, R8, and R10, with older ages predicted by the EW(H$\alpha$) than by the SED fits. We attribute this discrepancy to  the  possibility that in these regions 
the ionizing photon leakage is higher than 50\%.  Again, the large range in ages from EW(H$\alpha$) for the constant star formation cases are a direct consequence of the slow SED variations of these models with time, with slow build up of light in the stellar continuum and no change in the H$\alpha$  luminosity. 

\begin{figure}[t]
\plotone{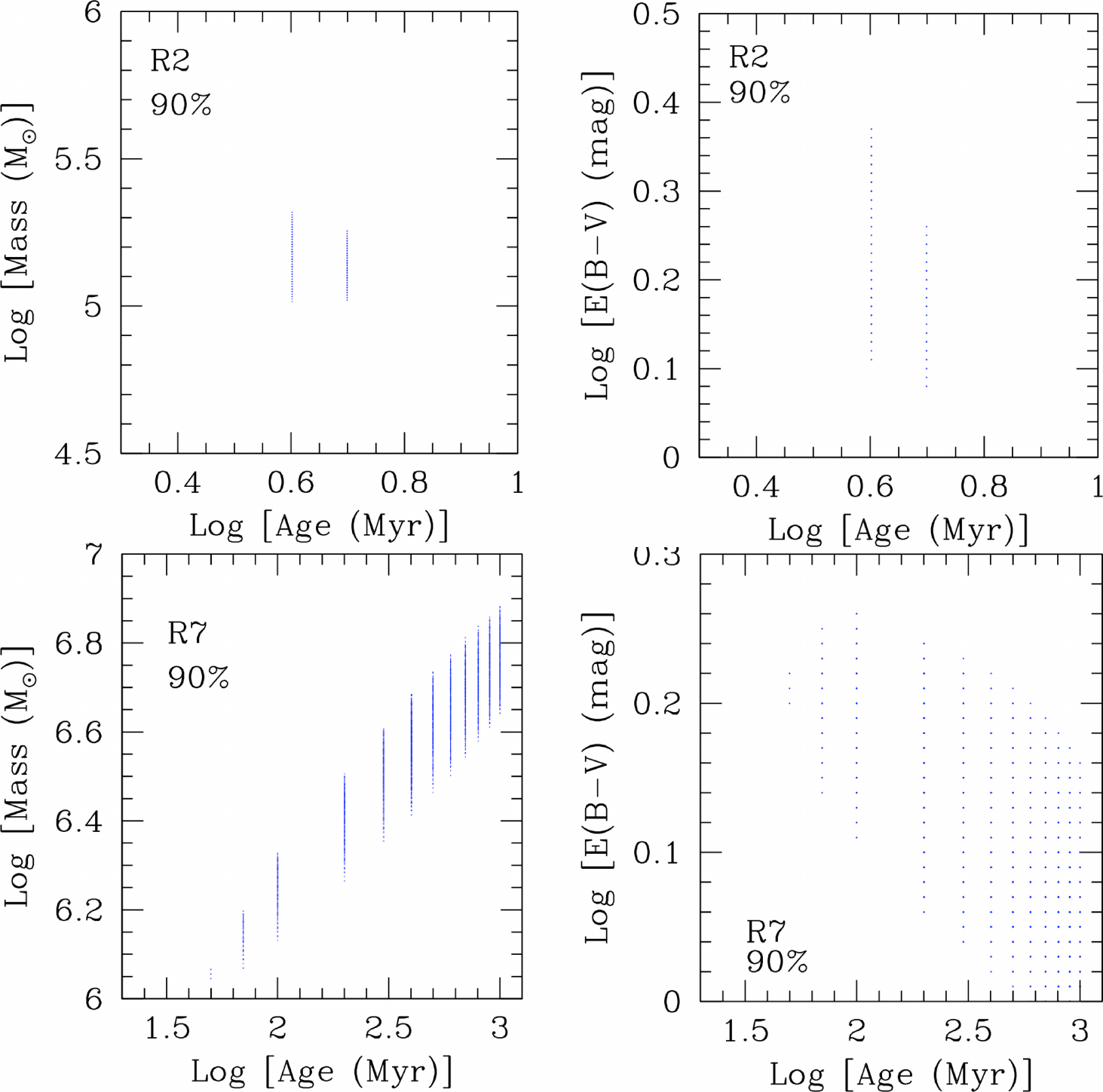}
\caption{Example of covariance between age, stellar mass, and color excess, shown for R2 and R7. The points show the location of the 90\% confidence level outputs from the $\chi^2$  minimization between data and models. R2 is chosen as the example for the outputs that have  SSP models for the best  fit stellar population. These cases are characterized by a small range of acceptable ages,  which makes covariances not easily distinguishable from random trends. For R7, the best fit population models are the CSP ones; this specific example yields the largest range of acceptable ages among our regions, which magnifies the trends for the covariances. \label{fig:covariance}}
\end{figure}

The uncertainties in the values of the three parameters: age, stellar mass, and color excess, are co--variant. An increase in age is generally accompanied by a decrease of color excess and an increase in mass. Examples are shown in  
Figure~\ref{fig:covariance} for both instantaneous burst and constant star formation populations. In our case, the regions that are better fit by instantaneous burst models only show weak--to--no covariance among parameters, which is simply due to the small range of acceptable ages. Typically, for fits that  accept a wider range of ages than our cases, covariances among the best fit parameters of an instantaneous burst population are easy to understand. When age increases, the stellar population SED becomes intrinsically 
redder and smaller values of  E(B--V) are required to fit the observed photometry; in addition, an aging stellar population becomes progressively fainter,  implying that a larger mass is required to increase the intrinsic luminosity of the population in order to account 
for the observations. For constant star formation, the reasoning is similar: increasing durations accumulate more mass, with minimal or no increase in the luminosity of the red portion of the SED (which is where the normalization, 
hence the mass, of the region is constrained). This implies that, when fixing one of the three parameters, the ranges of allowed values for the remaining two parameters are smaller, roughly by a factor 1.3--2, than the formal uncertainties listed in Table~\ref{tab:fits_regions}.

\begin{longrotatetable}
\begin{deluxetable*}{lrrrrrrrrrrrr}
\tablecaption{Physical Characteristics of Individual Regions\label{tab:fits_regions}}
\tablewidth{1050pt}
\tabletypesize{\scriptsize}
\tablehead{
\colhead{Property} &\colhead{Units} &\colhead{R1}& 
\colhead{R2} & \colhead{R3} & 
\colhead{R4} & \colhead{R5} & 
\colhead{R6} & \colhead{R7} & 
\colhead{R8} & \colhead{R9} & \colhead{R10} & \colhead{Res.} \\ 
\colhead{(1)} & \colhead{(2)} & \colhead{(3)} & \colhead{(4)} & 
\colhead{(5)} & \colhead{(6)} & \colhead{(7)} &
\colhead{(8)} & \colhead{(9)} & \colhead{(10)} & \colhead{(11)} & \colhead{(12)}
& \colhead{(13)} 
} 
\startdata
Pop. Model &                       & SSP                     & SSP                   & CSP                   & SSP                 & CSP                  & CSP                  & CSP                 & CSP                                & CSP                      & CSP                  &  Diff--CSP$^1$ \\
Age             & Myr                & 4$^{+1}_{-0}$  &4$^{+1}_{-0}$  &10$^{+6}_{-3}$ &4$\pm$1   &300$^{+600}_{-100}$ &300$^{+300}_{-100}$ &400$^{+600}_{-200}$ &12$^{+40}_{-5}$     & 10$^{+10}_{-3}$    & 15$\pm$5            &  1000--(400$^{+300}_{-100}$)\\
Atten/Ext.    &                       & SMC                    & diff SMC            & SB                     & SMC                & SB                   &  SMC                 & SB                    & SB                                    & SB                        & diff SMC             &   SB\\
E(B-V)$_{star}$ & mag       & 0.12$\pm$0.05    & 0.25$\pm$0.10  &  0.65$\pm$0.08 & 0.76$\pm$0.12 & 0.18$\pm$0.07 & 0.05$\pm$0.05 & 0.12$\pm$0.08  & 0.24$\pm$0.08            & 0.30$\pm$0.10      & 0.23$\pm$0.07      &  0.28$\pm$0.15\\
Mass           & M$_{\odot}$   & (2.9$\pm$0.6)E5 &(1.5$\pm$0.4)E5&(1.1$\pm$0.3)E6& (7.2$\pm$1.9)E4 &(7.0$\pm$3.4)E6&(4.3$\pm$1.8)E6 &(4.1$\pm$1.5)E6 & (3.3$\pm$1.4)E5 & (6.6$\pm$2.7)E4      & (9.7$\pm$4.8)E5   & (1.2$\pm$0.5)E9\\
$\chi^2_{red}$&                    & 2.6                       & 1.7                    & 1.9                    & 5.2                    & 2.8                   & 2.3                    & 1.1                     & 3.5                               & 1.9                           &  1.2                 &    2.1$^2$       \\
Age$_{lines}$ & Myr             & 4--5                     & 4--5                   & 2--90                 & 2--5                   & 300--5000        & $>$1000           & $>$300            & 20--500                         & 6--70                        & 50--800             &$>$4000\\
\enddata
\tablecomments{Physical characteristics of the 10 regions along the starburst  ring of NGC~3351 (Figure~\ref{fig:Pictures}), as derived from the fits to the photometry of Table~\ref{tab:regions}. Columns (1) and (2) list the property reported in each row and their units (as appropriate), as obtained from the best--fit models: stellar population model (SSP=instantaneous; CSP=constant star formation; Diff--CSP=difference between the two constant star formation models of listed ages), age/duration, attenuation/extinction curve (SB or SMC, with [`diff'] or without differential attenuation), color excess E(B--V)$_{star}$ of the stellar continuum, stellar mass, reduced $\chi^2$ value for the best fit model, and age inferred from the EW(H$\alpha$). The range in EW(H$\alpha$) ages accounts from presence/absence of differential attenuation, and for presence/absence of ionizing photon leakage from the aperture.  Best--fit for the 10 regions are listed in columns (3) through (12). Column (13) lists the best--fit result for the residual light in the central region. Reported uncertainties are 1~$\sigma$.}
\end{deluxetable*}
\scriptsize{$^1$ The best fit model for the Res. Region is indicated as Diff--CSP, which is the difference between two constant star formation (CSP)  models, with durations 1~Gyr and $\sim$0.4~Gyr, respectively. See Section~\ref{sec:residual}.\\
$^2$ The value of the reduced $\chi^2$ for the Res. Region is from the 10 HST photometric datapoints between 0.27~$\mu$m and 1.9~$\mu$m (Table~\ref{tab:regions}). Adding the extrapolated GALEX NUV luminosity density, L(0.23)=10$^{37.80 \pm 0.35}$~erg~s$^{-1}$~\AA$^{-1}$, to the fit yields $\chi^2_{red}=1.9$. All fits of the Res. SED are constrained to a FUV luminosity density upper limit L(0.15)$\lesssim$10$^{36.3}$~erg~s$^{-1}$~\AA$^{-1}$. See text. }
\end{longrotatetable}

\begin{figure}
\plottwo{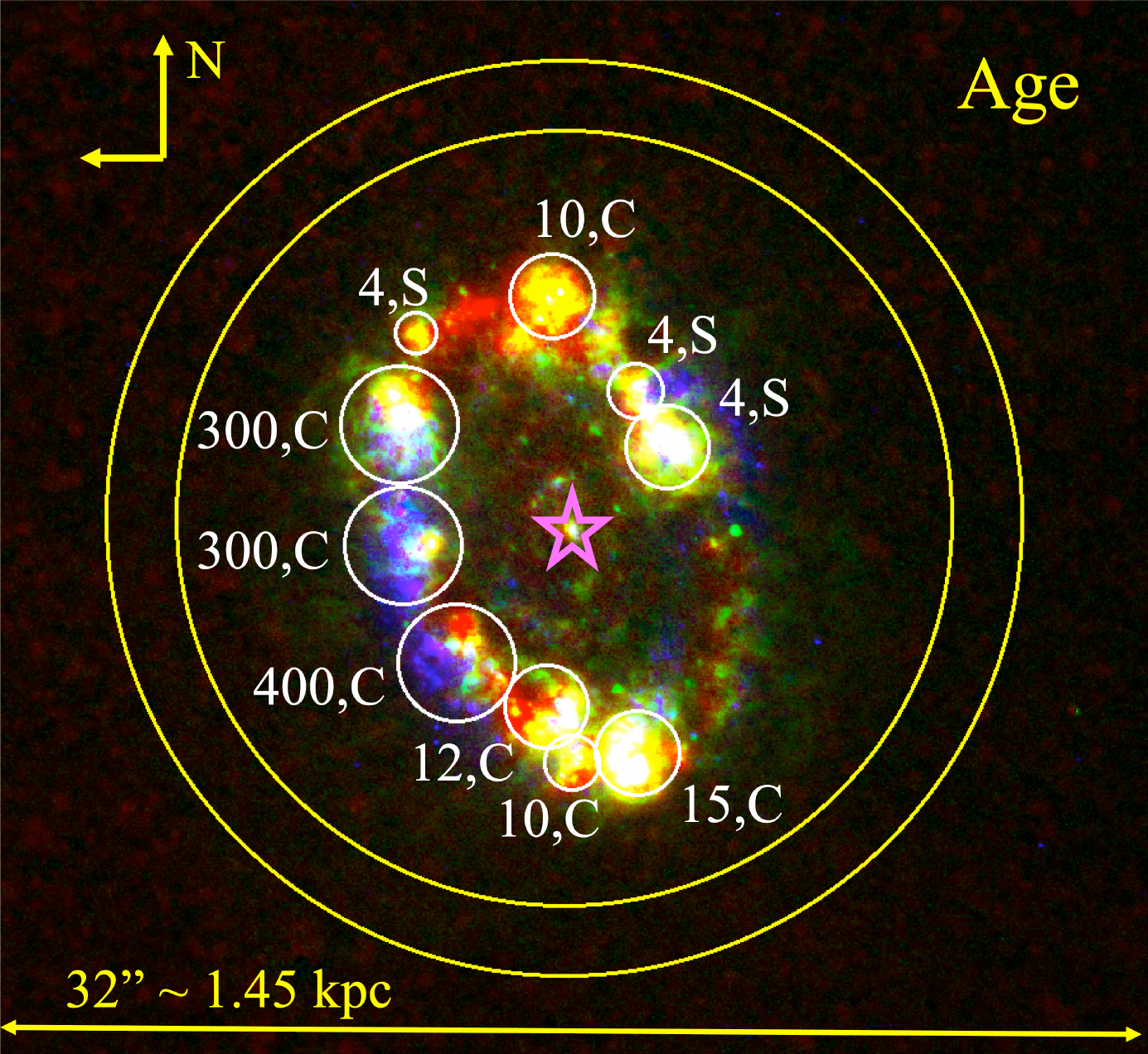}{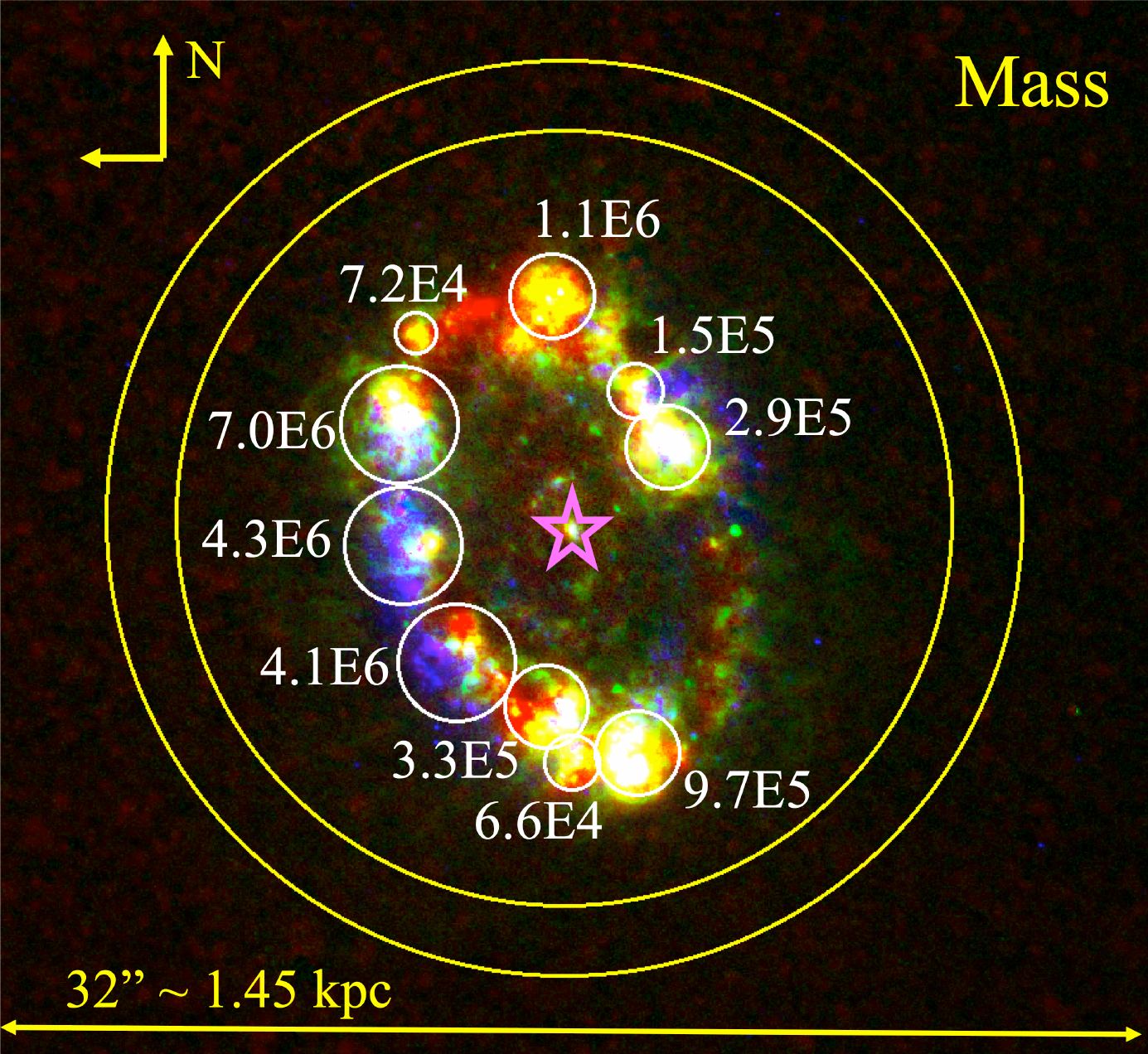}
\caption{The best--fit values of the ages and masses for the 10 ring regions shown on the central panel of Figure~\ref{fig:Pictures}. In the left hand--side panel, the ages are given in Myr, with a letter following the numerical value of the age to indicate whether the best  fit model is an instantaneous burst population (S) or constant star formation (C). The masses in the right hand--side panel are given in units of M$_{\odot}$. \label{fig:region_ages_masses}}
\end{figure}

The distribution of ages of the 10 ring regions indicates that ages/durations decrease along the ring moving from the East side to the West side (Figure~\ref{fig:region_ages_masses}, left), with the youngest regions (R1, R2, R3, R4) located to the North-West  side, as if star formation has been propagating along the ring over the past $\sim$400 Myr. The young regions are also located in the area of peak 24~$\mu$m emission, which coincides with the position of R3 and R4 (Figure~\ref{fig:Pictures}); in fact, these two regions are the most extincted among the ten. The secondary peak of  24~$\mu$m emission coincides with the other ring location hosting young regions, R8, R9, and R10. The two 24~$\mu$m peaks are approximately co--located with the two peaks of CO emission in the  galaxy and identify where the gas and dust streams feed the ring, to support the star formation in this region \citep{Kenney+1992, Regan+2001, Leaman+2019}. They also coincide with two emission peaks at 3--33~GHz \citep{Linden+2020}; these authors find that star formation at the location of those radio peaks has been roughly constant for the past $\sim$10~Myr, which supports our findings from the best--fit stellar populations in R3 and in R8/R9/R10.  The most prominent GALEX FUV peak is  co--located with R5, R6, R7, all three with low E(B--V)$_{star}$ and long star  formation durations (Table~\ref{tab:fits_regions}), indicative that the star formation has been on--going long enough in the area to  clear most of the dust. Extended star formation timescales, up to a few hundreds Myr, have been advocated to explain several characteristics, including multi--band colors and the EW(H$\alpha$), of ringlike star forming structures located at the Inner Lindblad Resonance of galaxies \citep{Kennicutt+1989, Kormendy+2004}. Our results for the ring of star formation in NGC~3351, which is near the galaxy's Inner Lindblad Resonance \citep{Kenney+1992}, are in agreement with these earlier findings. The existence of regions with progressively older ages/longer durations along the ring supports a scenario of ``string--of--pearls'' star formation, similar to what inferred for other galaxies  \citep{Boeker+2008, vanderLaan+2013}. 

With the exception of R4, the color excess measured from the hydrogen recombination lines, E(B--V)$_{gas}$, is  larger, by a median factor 1.8$\pm$0.2, than the color excess derived from the best  fits of the stellar continuum, E(B--V)$_{star}$, confirming results from previous studies of both galaxies and regions within galaxies  \citep{Calzetti+1994, Kreckel+2013}. The deviation of R4 from this trend may be explained by geometrical effects, if the ionizing photons leaving this heavily extincted cluster reach regions that contain gas with lower dust optical depth. The extrapolation of our best fit models to the GALEX wavelengths gives that the 10 ring regions contribute $\sim$80\% of  the NUV light of the Central Region, but overestimates the FUV emission by  a factor 1.8$^{+0.6}_{-0.4}$. Different choices of ages/masses/extinction values within our uncertainty ellipses do not decrease the discrepancy in the FUV, which we carry as a limitation of our modeling approach. The only way  to circumvent this  limitation would be the availability of a high angular resolution FUV image to constrain the models. 

The sum across the 10 ring regions of the extinction corrected luminosity in H$\alpha$ is listed in Table~\ref{tab:properties_central}. This value is $\sim$40\% of the total in the central region, with a few percent uncertainty depending on how the extinction correction for the Central Region is performed. This fraction is consistent with our assumption above that about  50\% of  the ionizing photons have leaked out of the ring regions. 
Depending on whether the ionized gas emission outside of the 10 ring regions can be considered part of these regions or not, the SFR in the ring, from the  extinction corrected H$\alpha$ luminosity, is in the range $\sim$0.1-0.3~M$_{\odot}$~yr$^{-1}$, while the sum of the masses gives M$\sim$1.8$\times$10$^7$~M$_{\odot}$ (Figure~\ref{fig:region_ages_masses}, right, and Table~\ref{tab:properties_central}). With these  values, the central ring is located between one and two  orders of magnitude above the Main Sequence of Star Formation for its stellar mass. The ring in NGC~3351 is therefore a very active region of current star formation.  Although this argument is  an oversimplification, as it assumes that the central ring of star formation is a  stand--alone galaxy (which it is not), it gives a sense for the level of activity in the region.  

When the results above are placed in the context of the information listed in column 13 of Table~\ref{tab:regions}, the trend is for the 10 ring regions to provide a decreasing fraction of the total light in the central region for increasing wavelength, from $\approx$100\% in the FUV down to 7\% in the H--band. Beyond the B--band, the ring regions represent less than 1/3 of the total light. The sum of the stellar mass in these regions represents about 1.5\% of the total in the Central Region, well below the mass uncertainty. Because of these characteristics, modeling of the Residual Region is a key piece of information for understanding the stellar population and dust attenuation in the center of NGC~3351.

\begin{deluxetable*}{lcl}
\tablecaption{Central Region's Intrinsic Properties\label{tab:properties_central}}
\tablewidth{450pt}
\tablehead{
\colhead{Parameter} & \colhead{Units} & \colhead{Value}}
\startdata
Log[L(H$\alpha$)$_{corr, \Sigma(R1-R10)}$]  $^1$  & erg~s${-1}$ &   40.30$\pm$0.04 \\
SFR(H$\alpha$)$_{\Sigma(R1-R10)}$ & M$_{\odot}$~yr$^{-1}$ & 0.11$\pm$0.01\\
M$_{\Sigma(R1-R10)}$ & M$_{\odot}$ & (1.8$\pm$0.4)E7\\
Log[L(H$\alpha$)$_{corr, sum}$] $^1$ & erg~s${-1}$  & 40.69$\pm$0.08\\
Log[L(TIR)$_{sum}$] & erg~s${-1}$  & 43.17$\pm$0.07\\
SFR(H$\alpha$)$_{sum}$ & M$_{\odot}$~yr$^{-1}$ & 0.27$\pm$0.05\\
SFR(TIR)$_{sum}$& M$_{\odot}$~yr$^{-1}$ & 0.41$\pm$0.05\\
SFR(H$\alpha$)$_{Central}$ & M$_{\odot}$~yr$^{-1}$ & 0.30$\pm$0.06\\
SFR(H$\alpha$+24)$_{Central}$ & M$_{\odot}$~yr$^{-1}$ & 0.38$\pm$0.05, 0.28$\pm$0.05\\
SFR(TIR)$_{Central}$ & M$_{\odot}$~yr$^{-1}$ & 0.40$\pm$0.06\\
SFR(FUV+24)$_{Central}$ & M$_{\odot}$~yr$^{-1}$ & 0.50$\pm$0.06, 0.33$\pm$0.05\\
M$_{\star, Central}$ & M$_{\odot}$ & (1.2$\pm$0.5)E9\\
\enddata
$^1$ The H$\alpha$ luminosity obtained from the sum of the individually extinction corrected ($corr$) regions from Table~\ref{tab:regions}.  L(H$\alpha$)$_{corr, sum}$  is within 10\% of the H$\alpha$ luminosity  derived using the global measurements of the Central Region from Table~\ref{tab:linecentral}; this indicates that the color excess of Table~\ref{tab:linecentral},  obtained from integrated photometry, is representative of the entire area, and region--to--region variations of the color excess do  not impact the final luminosity values. \\
\tablecomments{The SFRs  are derived using the formulae of \citet{Calzetti2013}. For the mixed--wavelength SFRs, each with two  values, the second value is derived using the proportionality constants of \citet{KennicuttEvans2012} between the two luminosities. The meaning of  each subscript is explained in Table~\ref{tab:nomenclature}.}
\end{deluxetable*}

\section{Modeling the Residual Region\label{sec:residual}}

\begin{figure}
\plottwo{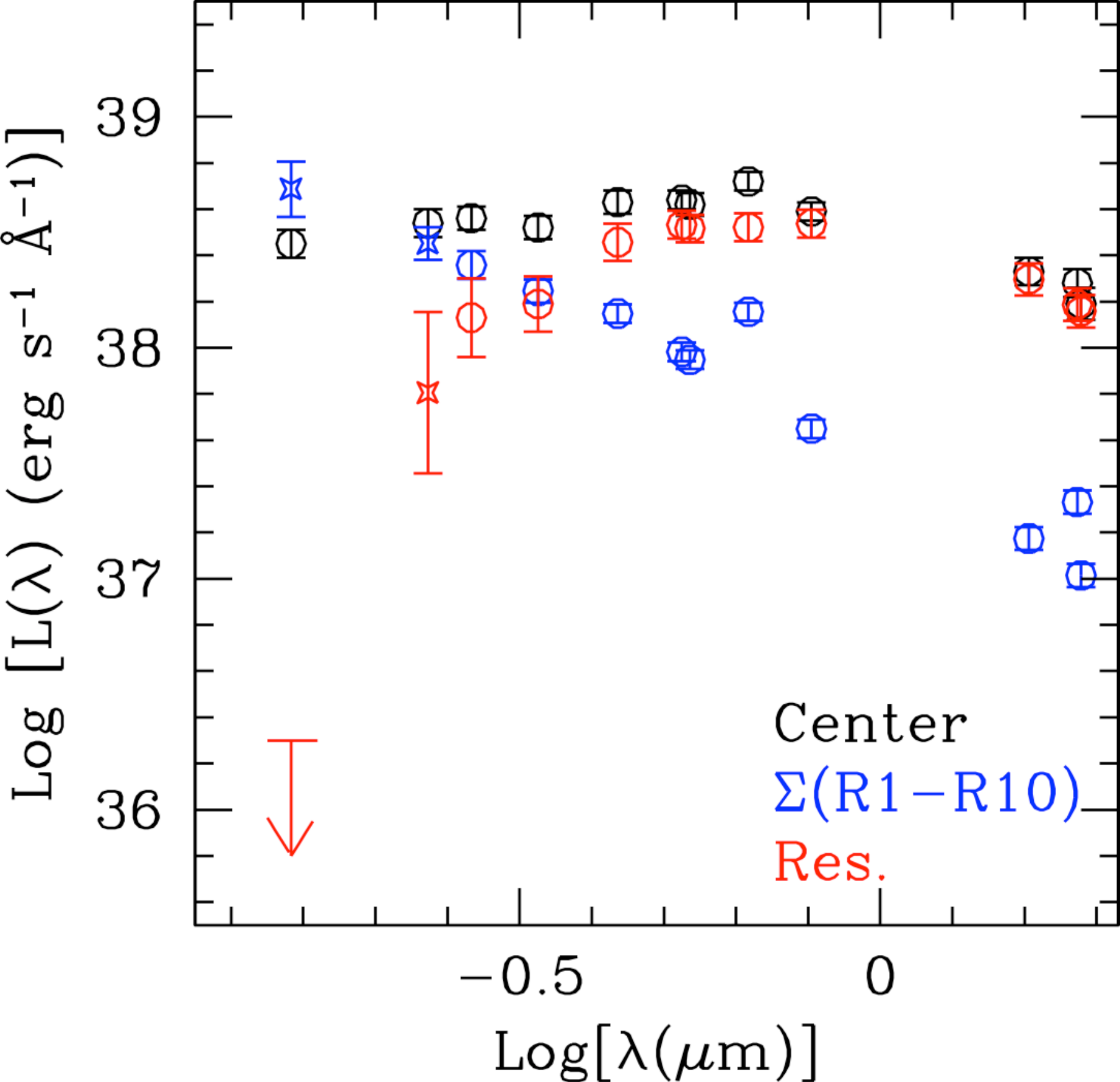}{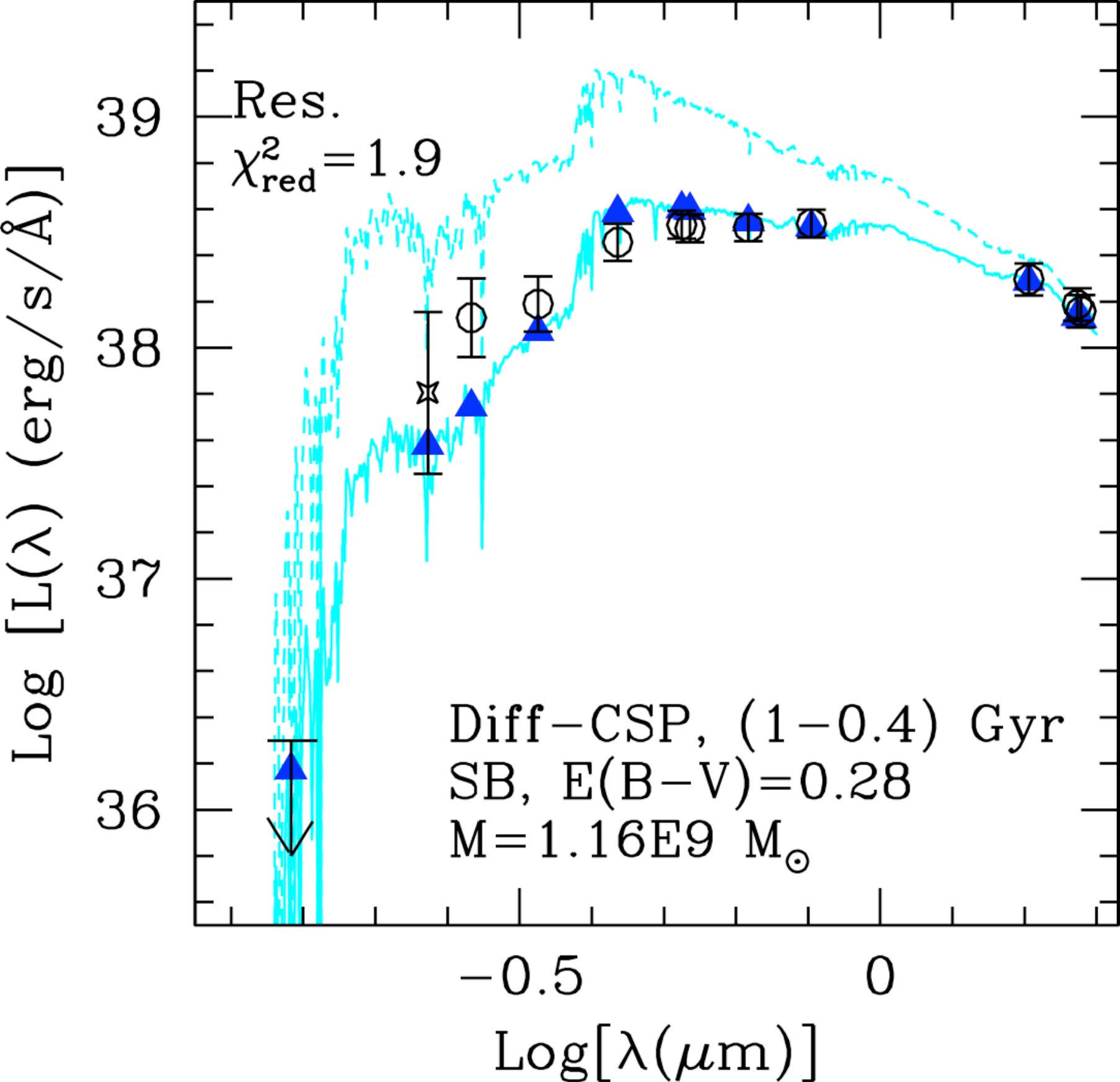}
\caption{{\bf LEFT:} The comparison between the photometry of  the Central Region (black circles, same photometry as Figure~\ref{fig:central_SEDs}, right), the sum of the 10 ring regions (blue circles and crosses), and the Res. Region (red circles, cross, and upper limit). All data are shown with 1~$\sigma$  error bars, from Table~\ref{tab:regions}. The crosses are the extrapolation to the GALEX FUV and NUV bands of the best fit attenuated models to the ring regions. The upper limit for the FUV photometry of  the Res. Region is due to the excess FUV flux predicted by the sum of  the ring region models. {\bf RIGHT:} The Res. observed/extrapolated photometry (black circles, cross, and upper limit) with the best  fit stellar population model (cyan) both attenuated (continuous line) and intrinsic  (dashed line). The photometry predicted by the best-fit model  is shown as blue triangles. The parameters listed in the panel are the central values from Table~\ref{tab:fits_regions}. \label{fig:region_residual}}
\end{figure}

The photometry of the FUV--NIR SED of the Central Region, of the sum  of the ring regions, and of the Res. Region is shown in Figure~\ref{fig:region_residual} (left). The Res. Region  photometry is in the last column of Table~\ref{tab:regions}, and is the 
difference between the photometry of  the Central Region and  the sum of the photometry  of the 10 ring regions (Figure~\ref{fig:Pictures} and Table~\ref{tab:nomenclature}). Any of the ring regions only contributes a few \% of the total light at long wavelengths, as already discussed in Section~\ref{sec:photometry}. 

About half of  the ionized gas is located in the Res. Region, but unlike the case of the ring regions, the gas in the Res. Region is generally diffuse; most of it is associated with the ring of star formation, an indication that the gas 
is ionized by the UV photons from the ring clusters. An exception is the nuclear cluster, which shows concentrated H$\alpha$ and Pa$\alpha$. A fit  of the nuclear cluster photometry reveals that this is a relatively young, $\sim$7~Myr old, instantaneous burst 
population, with non--negligible color excess, E(B--V)$\sim$0.3--0.5, for both the  stellar continuum and the ionized gas. The stellar mass enclosed in a region with 0.4$^{\prime\prime}$ diameter ($\sim$20~pc) is M$\sim$1.8$\times$10$^5$~M$_{\odot}$, comparable to 
the mass of R2. The extrapolation of the best fit model down to the GALEX wavelengths yields a luminosity density L($\lambda$)$\sim$10$^{36}$~erg~s$^{-1}$~\AA$^{-1}$ for both FUV and NUV. We use this result to help guide us in setting an upper limit flux in the FUV for the entire Res. Region, since the ring regions already account for the entirety of the FUV emission (with limitations, as discussed in Section~\ref{sec:regions}). A visual inspection of the GALEX/FUV image suggests  that the Res. Region may include about twice as much FUV light as the nuclear cluster. We thus set the upper limit  for the Res.  Region in the GALEX/FUV band to L(0.15)$\lesssim$10$^{36.3}$~erg~s$^{-1}$~\AA$^{-1}$. The accuracy of this estimate will have negligible impact on our results, since the upper limit is less than 1\%  of  the FUV luminosity of the Central Region. The extrapolated value for the GALEX NUV band, which is simply the difference between the photometry of the Central Region and the sum of the extrapolated model NUV photometry of the 10 Ring Regions, is L(0.23)=10$^{37.80\pm 0.35}$~erg~s$^{-1}$~\AA$^{-1}$. 

From the results of the previous section, we have seen that the ring is hosting an extended episode of  star formation, a few hundred Myrs in duration. This extended star formation sits on top of the Res. Region which forms a stellar background characterized by a red SED (Figure~\ref{fig:region_residual},  left). Thus the Res. Region may contain the accumulated  stellar populations of  past episodes of star formation, possibly prior to those in the ring. Our simple model SEDs, that only include instantaneous bursts and constant star formation, cannot capture this scenario. To address this possibility, we construct a new set of SED models by subtracting from the 1--14~Gyr CSP models younger CSP models, in the duration range 100 to 900  Myr. Each of these models exemplifies the case of  past star formation that  was constant over the period of time YY$-$XX, with XX=100,...,900~Myr and YY=1,2,...,14~Gyr. We draw the separating point at 1~Gyr, which is about 2--3 times older than the oldest stellar population in the ring. In other words, we are assuming that star formation in the central region has been on--going continuously for at  least 1~Gyr, but regions younger than $\ge$100 Myr are located along the ring. These new stellar SED models are called Diff--CSP in Table~\ref{tab:fits_regions}, last 
column.  

Because the SED of the Res. Region has relatively large error bars and very little UV emission, several models fit the observed photometry with comparable goodness--of--fit, including models with instantaneous bursts that are heavily dust obscured.  As discussed in Section~\ref{sec:central}, the Central Region does not appear to include areas with large dust obscuration values, since the dust  column density retrieved from the mean value of the color excess E(B--V)$_{gas}$ is consistent, within a factor of 2, with the value of the dust column density obtained from the IR SED fit. In addition, we have seen in Section~\ref{sec:photometry} that the value of  the color excess we derive from the H$\alpha$/Pa$\alpha$  ratio agrees with the value derived  from H$\alpha$/H$\beta$ by \citet{StorchiBergmann+1995} and \citet{Moustakas+2010}; constant E(B--V)$_{gas}$ from nebular emission at different wavelengths is indicative of foreground dust \citep{Calzetti+1994}. Finally, an inspection of the stellar--continuum--subtracted IRAC/8~$\mu$m image, which is the dust emission image with the highest angular resolution in our collection, indicates that 80\% of the emission at this wavelength is associated with the ring of star formation; this is consistent with the absence of  a local peak of 24~$\mu$m emission in correspondence of the nucleus. These three lines of evidence lead us to  exclude best  fits that  imply large values of the color excess for the Res. Region and only consider those cases that have E(B--V)$\lesssim$1~mag (Table~\ref{tab:linecentral}). This condition is imposed alongside the first one: that  the best fit solution has age/duration longer than those of the regions along the ring ($\gtrsim$100~Myr).

With the above constraints, only the Diff--CSP models with the SB attenuation curve provide satisfactory fits to the data (Figure~\ref{fig:region_residual}, right panel). Cases that subtract a CSP  model with duration $\ge$300--400~Myr from the  1--14~Gyr CSP models minimize the difference between data  and models, and are generally acceptable. The minimum value $\chi^2_{red}$=1.9 is obtained for the (1~Gyr-400~Myr) model (Figure~\ref{fig:region_residual}, right, and Table~\ref{tab:fits_regions}, last column). Longer star formation durations, $\ge$2~Gyr, require progressively lower values of the color excess, E(B--V)=0.3 to 0.1, but also produce increasingly slightly worse fits, $\chi^2_{red}$=2.1--2.4. The (1~Gyr--400~Myr) model not only gives the lowest $\chi^2_{red}$ value, but also  predicts a FUV photometric point that is close to the upper limit of the SED. Furthermore, our photometric approach supports shorter SFH timescales: the luminosities  of both the Central Region and Residual Region  are measured with {\em local} background subtraction, which removes the contribution from the long--timescale (older)  stellar populations that provide most  of the diffuse light from the galaxy. Thus, we will be using the (1~Gyr--400~Myr) model solution in the following sections, keeping in mind that  durations $>$1~Gyr do not change the quantitative results that will be presented.

\section{The Attenuation in the Central Region of NGC~3351}\label{sec:attenuation}

In this section we aim at  accomplishing two  goals. The first  is to leverage the region--by--region SED fitting performed in the previous sections to derive the intrinsic SED of the Central  Region by  combining the intrinsic SEDs of individual regions. The second goal is to compare the intrinsic  SED  with  the  observed SED in order to derive a net attenuation curve for the Central Region of  NGC\,3351. The two  goals are separated and detailed in two of the sub--sections that follow.

\subsection{Avoiding a `Circular' Analysis}\label{sec:circularity}

A common concern when performing analyses that involve SED fits  is whether the 
results are an effect of circular arguments. Specifically, we use  photometry measured 
in the individual ring regions to derive  the intrinsic  SEDs of the populations in those regions, with assumptions 
for  the attenuation curves. Then we combine the intrinsic SEDs of 
the individual ring regions to construct the intrinsic SED of the more extended ($\sim$1~kpc) Central  
Region (Section~\ref{sec:intrinsic}), in order to derive its attenuation curve (Section~\ref{sec:fit_attenuation}). How is this not circular?

We argue that we avoid circularity by modeling areas of emission with physical sizes that are significantly 
smaller than the size of the region we desire to recover the attenuation curve for.  
The ring regions have physical sizes $\le$180~pc (deprojected) and are sufficiently small that they can host a limited number of stellar populations of 
separate ages; in fact, their SEDs can be modeled reasonably well with simple assumptions for the 
SFH of the region (instantaneous and constant star  formation), and for the dust  geometry (foreground or mixed), 
as seen in Section~\ref{sec:regions}. 

\begin{figure}
\plotone{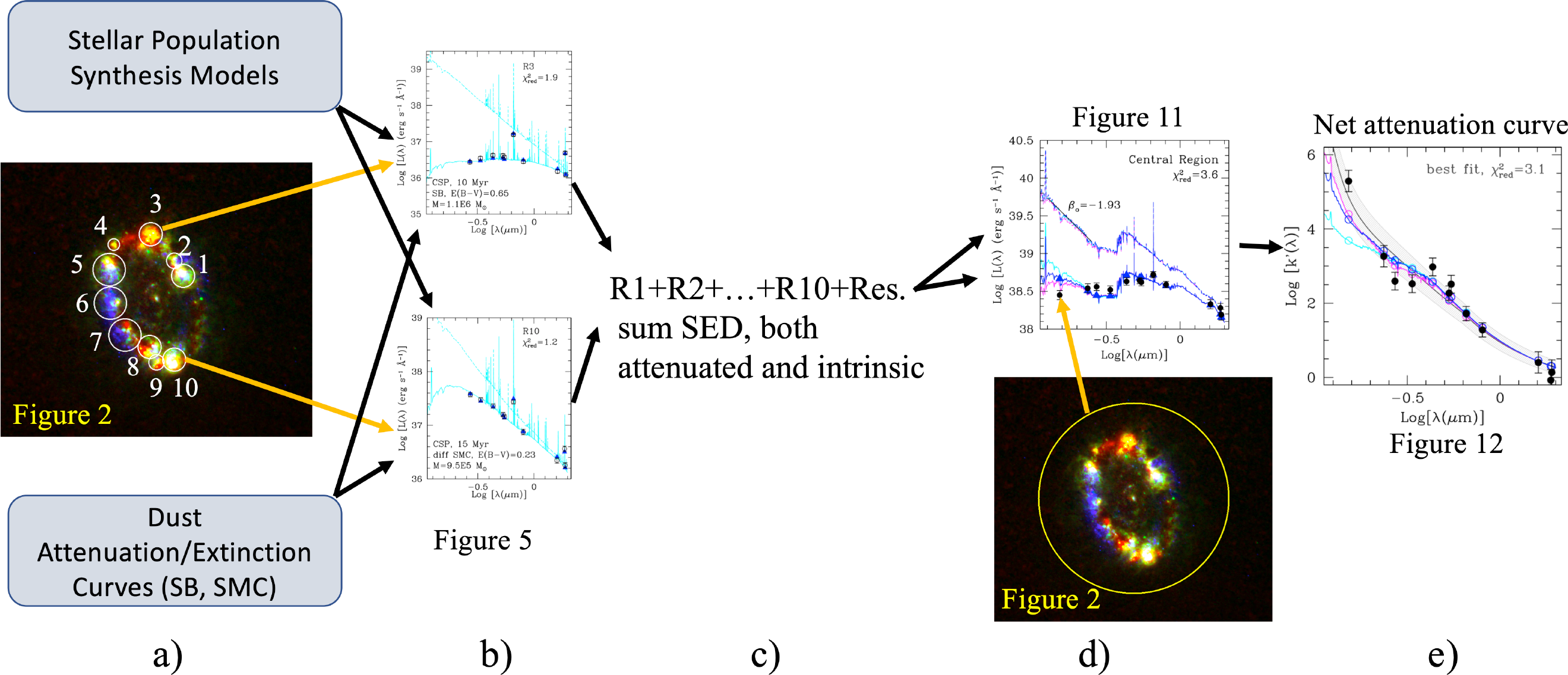}
\caption{The steps to derive the net attenuation  curve of the Central Region of NGC\,3351 are schematically shown from left to right. The photometry of physically small, 
$\sim$180~pc or less (de--projected), regions is modeled with simplified assumptions for the star formation history and the attenuation or extinction curve (panels a and b). The models, both intrinsic and attenuated are  then summed together (panel c) to create  the Central Region SEDs. The attenuated SED is compared with the measured photometry  for verification (panel d), while the infrared SED (not shown) provides an energy--balance check on the  intrinsic SED. The ratio between the so--constructed intrinsic and attenuated SEDs yields the  net attenuation curve of the Central Region (panel e). Individual panels are identified with the number corresponding to the Figure where they appear.\label{fig:circularity}}
\end{figure}

Conversely, large, $\gtrsim$1~kpc, physical  regions, such as the Central Region, generally 
contain several stellar populations components, with  separate ages and different masses and dust optical depths. 
In fact, we cannot reproduce {\em both} the UV--to--NIR SED and the IRX--$\beta$ value of  the Central 
Region with simple assumptions for the SFH and the dust geometry, as seen in 
Section~\ref{sec:central}. Only by dividing the Central Region into physically smaller (and simpler) 
sub--regions we are able to reconstruct both its intrinsic stellar SED and the appropriate attenuation 
curve, as shown in the next sub--sections, and schematically represented  in  Figure~\ref{fig:circularity}.

\subsection{The Intrinsic SED of the Central Region}\label{sec:intrinsic}

\begin{figure}
\plotone{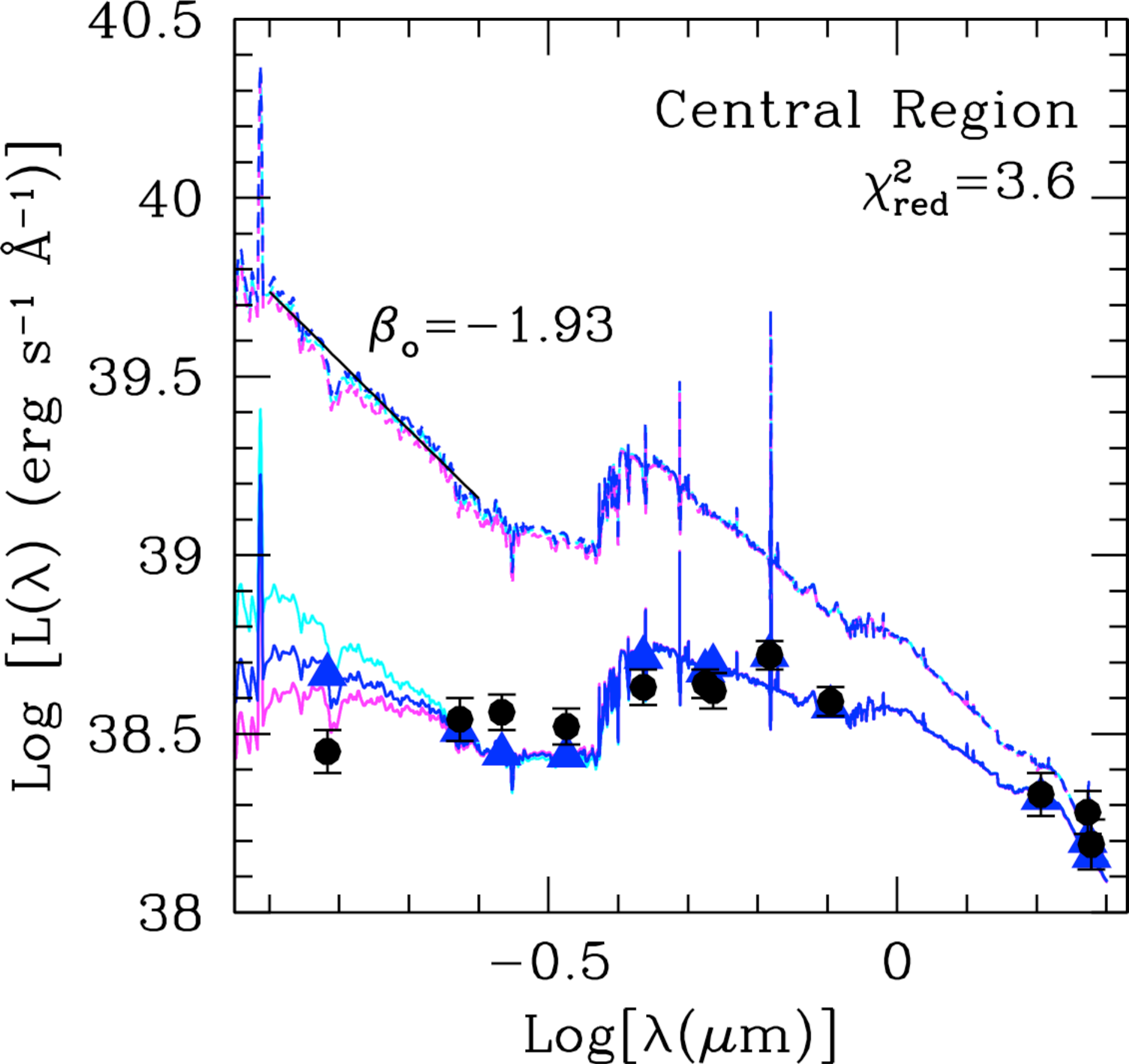}
\caption{The sum SED of all the modeled regions: R1, R2, ..., R10, and Res., is shown for both the attenuated (continuum lines) and intrinsic (dashed lines) cases. The observed photometry (black circles) is shown with 1~$\sigma$  error bars. The three model spectra correspond to: (blue line) the sum of the best fits for each region, from Table~\ref{tab:fits_regions}; (magenta line) the sum of the fits obtained by using the SMC extinction curve for each region, with the exclusion of Res. for which only the SB curve provides an acceptable fit (Section~\ref{sec:residual}); (cyan line) the sum  of the fits obtained by using the SB attenuation for each region. The photometric values for the best fits of the attenuated spectrum are shown as blue  triangles. The $\chi^2_{red}$ values reported at  the top right of  the figure refers to the comparison between the best fit photometry and the data.\label{fig:best_fits}}
\end{figure}

The  sum of the attenuated model SEDs of all the regions discussed so  far, R1, R2, ...., R10, and Res., should reproduce the observed SED of the Central Region. This is shown in  Figure~\ref{fig:best_fits}, where the observed FUV--to--NIR photometry of  the Central Region is compared with the synthetic photometry of the sum of the reddened SEDs from the individual regions' best fits (Table~\ref{tab:fits_regions}). The main discrepancy between data and models is in the FUV photometry. This is from the over--prediction of the FUV luminosity from  the 10 ring regions, discussed in Section~\ref{sec:regions}. 
The second-largest discrepancy is  for the photometry at 0.27~$\mu$m  (the third datapoint from left in Figure~\ref{fig:best_fits}), which simply reflects the fact that  this point is under--predicted in the fit of the Res. Region. This is a consequence of the upper limit in the FUV for the Res. Region SED, which pushes all models for this region down to low NUV luminosities;  the availability of a high--resolution FUV image will likely help improve the fits across the entire UV wavelength range. Despite these discrepancies, the reduced $\chi^2$ from all 12 datapoints, $\chi^2_{red}$=3.6, is sufficiently small to yield an acceptable fit\footnote{This value of $\chi^2_{red}$ is calculated assuming three degrees of freedom in the comparison between data and model. This  is done to make the reduced $\chi^2$ value  comparable to those in previous sections. However, there are zero degrees of freedom in the comparison between the sum model SED and the observed photometry, since there are no scalings or other adjustments to the individual best--fit models to produce the sum SED.}; when limited to the 10 datapoints at $\lambda\ge$0.27~$\mu$m, $\chi^2_{red}$ improves to a value of 2.7. These values are in--between those of the best fits attempted on the global photometry of the Central Region (Section~\ref{sec:central}): they are about a factor  3 worse than the $\chi^2_{red}$ obtained using the SB curve and a factor 2.5 better than using the SMC curve (Figure~\ref{fig:central_SEDs}, right).

In addition to the sum of the  attenuated  model spectra and photometry from Table~\ref{tab:fits_regions}, Figure~\ref{fig:best_fits} shows the two extreme cases of only using the SB  curve for the attenuation in all regions (cyan lines) or only 
the SMC extinction curve, with the exception of the Res. Region, which is only fit by the SB  curve (magenta lines). These two extreme  cases bracket the optimal fit case by design. The change from one  choice of attenuation curve to the  other 
affects significantly only the GALEX FUV and NUV photometric points. This is because the longer wavelengths become increasingly dominated by the light from the Res. Region, for which the best-fit model does not change in the three representations shown 
in Figure~\ref{fig:best_fits}.  

While there are significant differences between the model SEDs of the attenuated light, the intrinsic sum model SED is well constrained, with little difference between the adopted attenuation curves for fitting the observed  photometry 
of the individual regions. This is due to the best fit stellar population SEDs  being robust against the choice of attenuation curve; the important nuance to appreciate is that it is the SEDs that are similar to each other, not necessarily the 
characteristics of the stellar populations, which have small age variations between one choice of attenuation curve and another. The availability of two hydrogen recombination lines, in addition to multi--wavelength coverage from NUV to NIR, helps with the  
recovery of the intrinsic stellar population SED, especially at blue wavelengths.  

The comparison of the sum of the individually dust--corrected luminosities at H$\alpha$, L(H$\alpha$)$_{corr, sum}$ (Table~\ref{tab:properties_central}), with the dust--corrected L(H$\alpha$)$_{Central}$=10$^{40.73\pm0.10}$~erg~s$^{-1}$ from the integrated photometry of  the Central Region  (Table~\ref{tab:linecentral}) offers an additional check to our approach of modeling the SED of the central region as a sum of independently fitted regions. The two values are within 10\% of  each other, i.e., they are consistent with  each other within the 1~$\sigma$ uncertainty, despite the fact that the individual ring regions span a significant range of line extinctions, from E(B--V)$_{gas}$$\sim$0.2~mag to 0.8~mag. In addition, the sum of the dust corrected H$\alpha$ luminosity  of the ring regions, L(H$\alpha$)$_{corr, \Sigma  (R1-R10)}$ is about 37\%--41\% of the total luminosity in the central region, again comparable to the fraction in the observed luminosities. Thus, extinction corrections do not alter the ratio of the ring regions  line luminosities relative to the total.  The SFR calculated from the dust--corrected H$\alpha$ is systematically lower, by 30\%  to 70\%, than the values derived from the infrared luminosity or from composite tracers that include a UV band or optical line emission and an infrared band. For the composite tracers, the discrepancy depends on the calibration used for the ratio between the  two luminosities. The largest  discrepancy is for  the SFR(FUV$+$24) which uses the formula of \citet{Liu+2011} as calibrated by \citet{Calzetti2013}; this formula is  derived using young star--forming regions. If we use the  formula in \citet{KennicuttEvans2012} \citep[again, with the calibration of][]{Calzetti2013}, which  is derived for whole galaxies, the resulting  SFR(FUV$+$24) is consistent with the extinction--corrected H$\alpha$ luminosity one within 1~$\sigma$. The same is true for the composite indicator SFR(H$\alpha+$24). Thus, the \citet{KennicuttEvans2012} composite SFR indicators may better describe the complex stellar populations that are present in the Central Region. 

Adopting the sum model SED of Figure~\ref{fig:best_fits} as the intrinsic stellar population SED of the central region, the expected IR emission from dust can be calculated from the integral of the difference between the intrinsic stellar population and 
the dust reddened stellar population. For the latter, we adopt the best fitting model to the observed photometry for the Central Region from Figure~\ref{fig:central_SEDs} (right), extrapolated to wavelengths  $<$0.15~$\mu$m  in order to account for the light blueward of the GALEX FUV band. The result of the integration is the expected IR  luminosity, and we find L(IR)$_{sum}$=10$^{43.17\pm0.07}$~erg~s$^{-1}$ (Table~\ref{tab:properties_central}). If instead we use the models of Figure~\ref{fig:best_fits}  for the reddened stellar populations, we obtain L(IR)$_{sum}$=10$^{43.08\pm0.03}$~erg~s$^{-1}$. Both values agree with each other and with the measured IR dust  emission values of the central region listed in Table~\ref{tab:linecentral}. Thus, our intrinsic  stellar population, reconstructed by  simply  summing together the dust--free SEDs of the individual regions, is able to predict the correct amount of IR emission from the central region of NGC\,3351. 

We can use our best  fit models to analyze the contribution of  the different populations to the TIR emission. We find that the Res. Region and the sum of the 10 ring regions each contribute about 50\% of the TIR emission in the Central Region. This lends  more support to the use  of  the composite SFR indicators from \citet{KennicuttEvans2012} for the Central Region. In fact, a significant portion of the TIR  emission is  contributed by the Res. Region, which contains stellar populations that are old enough to no  longer contribute to the current  star formation,  but still heat  the dust. 

\subsection{The Attenuation Curve}\label{sec:fit_attenuation}

\begin{figure}
\plottwo{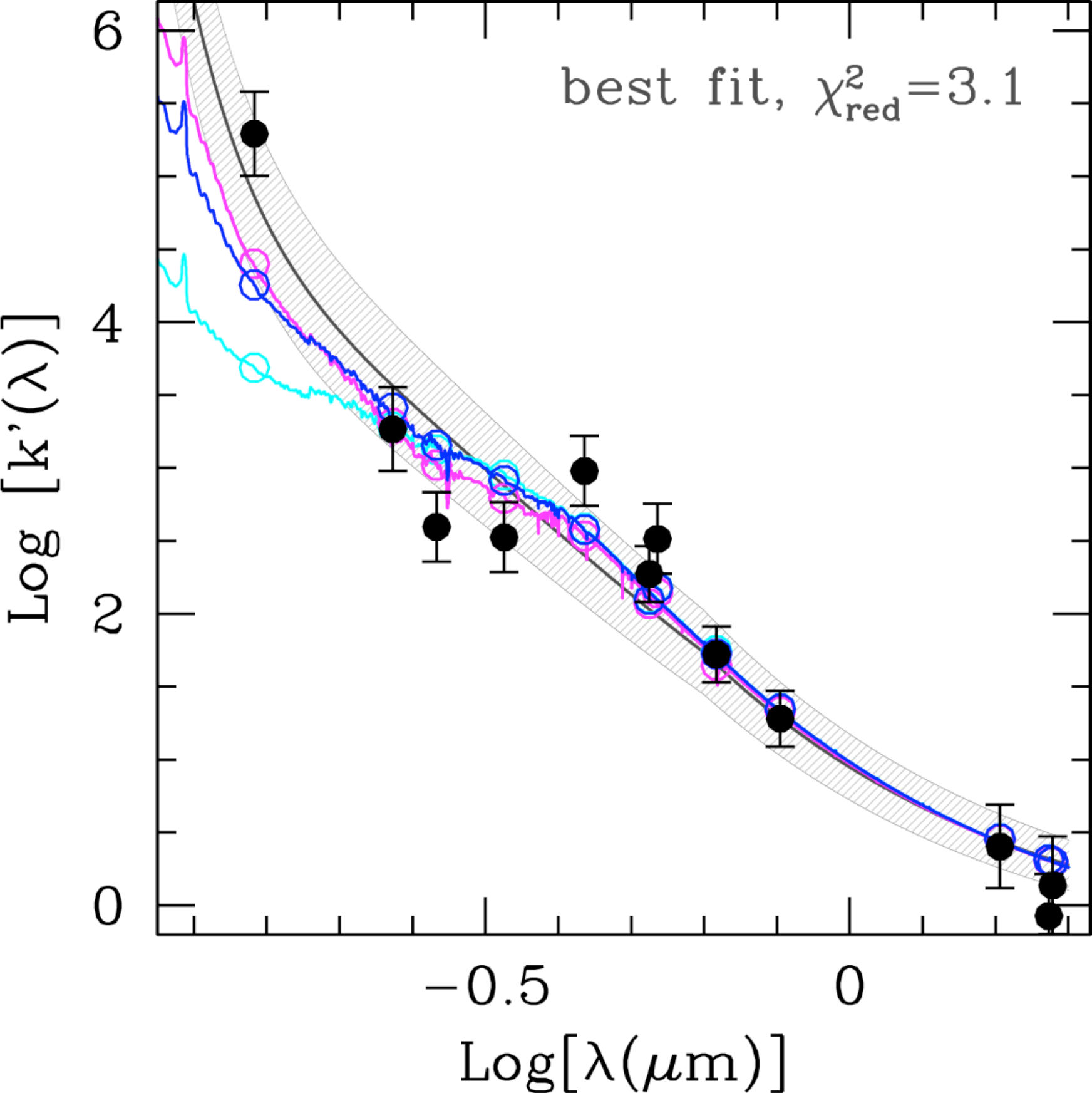}{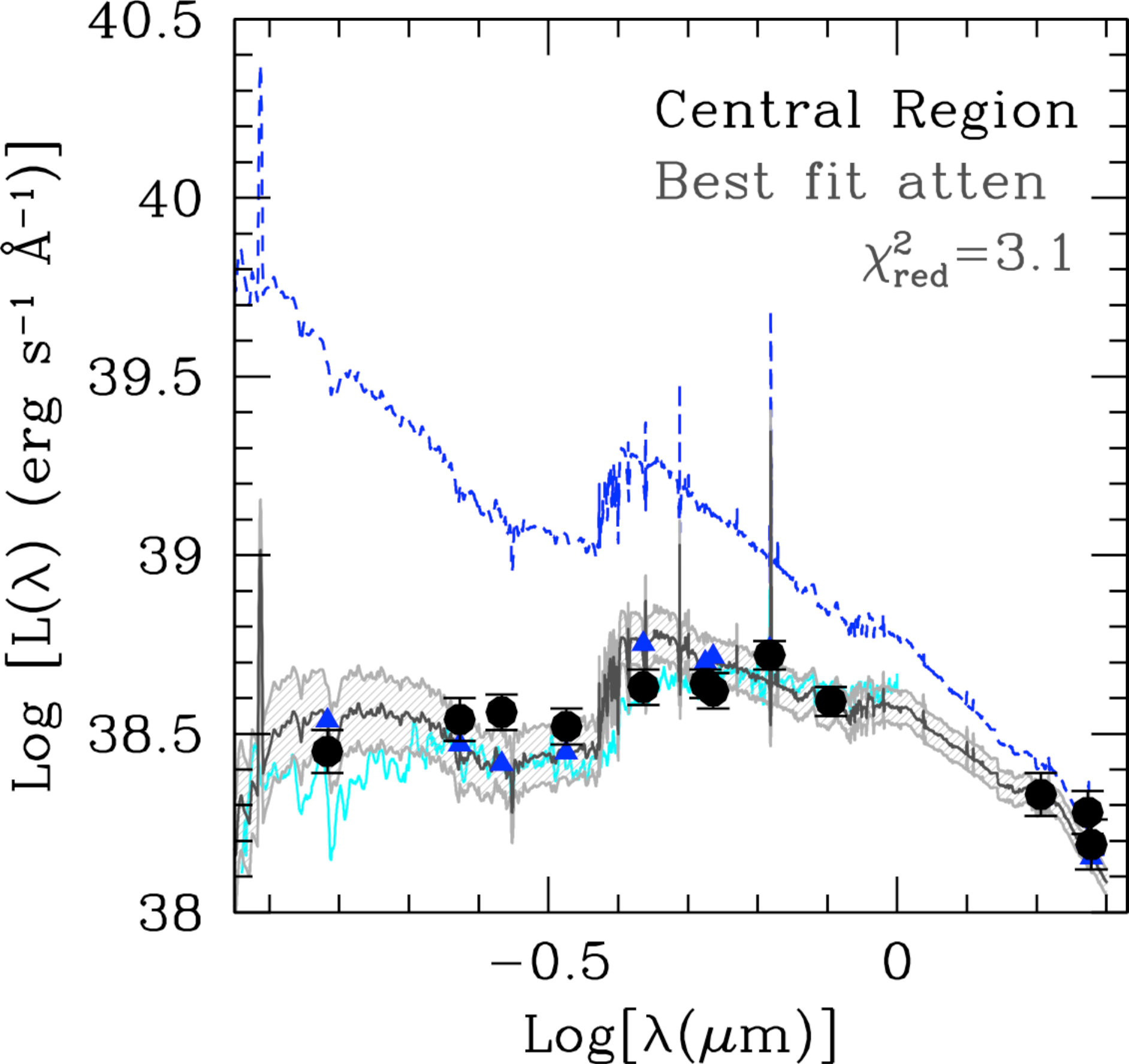}
\caption{{\bf LEFT:} The net attenuation curve k$^{\prime}(\lambda)$ required to account for the  observations (black filled circles), and for the sum of the reddened model spectra and photometry of ring$+$Res. regions (lines and empty circles: blue for sum of optimal fits; cyan for SB attenuation curve; magenta for SMC curve; see Figure~\ref{fig:best_fits}), adopting the intrinsic stellar population's SED of  Section~\ref{sec:intrinsic} as ground truth. The best fit to the attenuation of the observed photometry is shown as a dark--grey line; the grey hatched region marks the 1~$\sigma$ uncertainty range for the curve. The reduced $\chi^2$ is reported at the top right of the figure. {\bf RIGHT:}  The reddened spectrum (grey continuous line, with the grey hatched region representing the 1~$\sigma$ confidence level) and photometry (blue triangles) resulting from the attenuation curve to the left applied to the intrinsic stellar populations' SED (blue dashed line; see Figure~\ref{fig:best_fits}). The observed photometry (black circles) is shown with 1~$\sigma$ error bars, together with the observed spectrum of \citet{StorchiBergmann+1995} (cyan continuous line). The $\chi^2_{red}$ between model and observed photometry is reported at  the top right of  the figure; it is identical to the one to the  left by construction. \label{fig:attenuation}}
\end{figure}

The ratio of the intrinsic stellar population's SED to its attenuated (observed)  counterpart provides a measure of the wavelength--dependence of the attenuation in a region or galaxy. We adopt the formulation of \citet{Calzetti+1994} and \citet{Calzetti+2000}, 
which borrows directly from the practice used for deriving extinction curves from  stellar SEDs (see equation~3):
\begin{equation}
k^{\prime}(\lambda) = {Log_{10} [L(\lambda)_{int} / L(\lambda)_{obs}] \over 0.4\ E(B-V)_{gas}},
\end{equation}
where L($\lambda$)$_{int}$ and L($\lambda$)$_{obs}$ are the intrinsic and  observed luminosities at wavelength $\lambda$, and E(B--V)$_{gas}$ is the color excess of the ionized gas. We specify the color excess of the ionized gas, in order to be able to compare 
our attenuation curve  $k^{\prime}(\lambda)$ with previous derivations. Figure~\ref{fig:attenuation}, left panel, shows the results for $k^{\prime}(\lambda)$, with the discrete photometric values (black filled circles) obtained from the ratio of the intrinsic population SED to the observed photometry, and the color lines and circles from the ratio of intrinsic to attenuated model spectra and photometry. We fit the observed photometry with a polynomial curve similar to the SB one, resulting in the curve of Figure~\ref{fig:attenuation} (left; grey line with the 1~$\sigma$ uncertainty marked by the hatched light-grey region). The best--fit expression is:
\begin{equation}
\begin{split}
k^{\prime}(\lambda) & =(1.00\pm 0.08) (-2.365  + 1.345/\lambda) + (1.97\pm 0.15), \  \ \ \  \  \ \  \ \ \ \ \  \ \  \  \ \ \ \  \ \ \ \ \  \ \ \ \ \  \  \ \ \  \ \ \  for \ \ \  0.63~\mu m \le \lambda \le 2.20~\mu m;\\
                                 &=(1.00\pm 0.08) (-2.450  + 1.809/\lambda  - 0.293/\lambda^2 + 0.0216/\lambda^3) + (1.97\pm 0.15), \  \ \ \  for \ \ \ 0.12~\mu m \le \lambda < 0.63~\mu m.
\end{split}
\end{equation}
The uncertainties above indicate the non--covariant region of the two variables: scaling (steepness)  and normalization (vertical offset). We elect not  to model the higher order curvatures of the data, as these could be an effect of inaccuracies in the modeling of the intrinsic SED, as opposed to actual features. Eq.~5 yields k$^{\prime}$(B)--k$^{\prime}$(V)=0.40, implying that, when using the ionized gas color excess as reference, the steepness of the curve has to be reduced by more than a factor 2 relative to a standard extinction curve, in order to fit the observed photometry. This means  that the stellar continuum is less attenuated than the ionized gas, a result that had already been recovered for both starburst and star--forming galaxies at a range of redshifts in previous investigations \citep[][; see, however, \citet{Shivaei+2015}]{Calzetti+1994, Calzetti+2000, Wild+2011, Kreckel+2013, Reddy+2015, Shivaei+2020a}. 

As a sanity check, we apply Eq.~6 to the intrinsic stellar population  SED to verify that we recover the observed photometry of  the Central Region. The result of this experiment is shown in the right panel of Figure~\ref{fig:attenuation}, with the same symbols as the left panel. The few discrepancies observed are the same that can be inferred from the left  panel, with the attenuated model spectrum showing more structure than the observed photometry or spectrum, as already discussed in previous sections. Nevertheless, the agreement between data and models is acceptable, with a $\chi^2_{red}$=3.1\footnote{The meaning of $\chi^2_{red}$ in this context is the same as the one in Section~\ref{sec:intrinsic}}.

\section{Results and Discussion}\label{sec:results}

\subsection{The Star Formation History in the Center of NGC\,3351}

\begin{figure}
\plotone{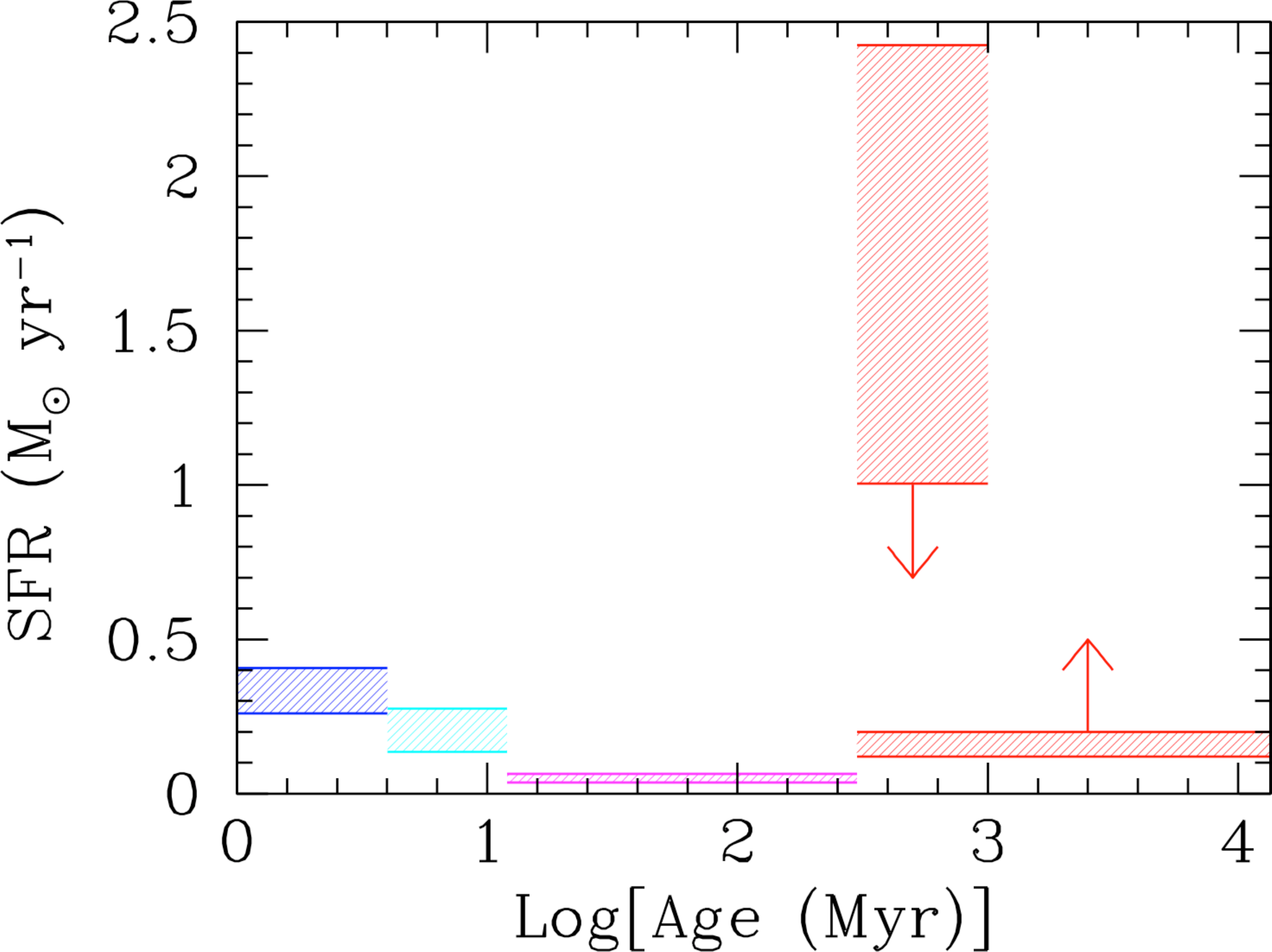}
\caption{The star formation history of the Central Region of NGC\,3351, as calculated from the masses and durations of Table~\ref{tab:fits_regions}, and from the addition of the best fit constant star formation model in the age range 400~Myr--14~Gyr (see text). The hatched horizontal  regions mark the $\pm$1~$\sigma$ uncertainty  region for each age bin. Age bins are: 0--4~Myr (blue hatched region), 4--12~Myr (cyan hatched region), 12--300~Myr (magenta hatched region), and $>$300~Myr (red hatched region). The oldest bin spans different ages ranges, with an average SFR that is no lower than $\sim$0.1~M$_{\odot}$~yr$^{-1}$. The complex SFH of the Central Region of  the  galaxy is apparent from the data, with a non--monotonic trend as a function of time. \label{fig:SFH}}
\end{figure}

The ages/durations and masses of each individual region, as listed in Table~\ref{tab:fits_regions}, are used here to produce a coarse--step SFH for the Central Region. Based on the ages and durations at our disposal, we select: 0--4~Myr, 4--12~Myr, 12--300~Myr and $>$300~Myr as our age bins. A finer--step age binning would require smaller uncertainties, which are currently not available to us. 
Our results are shown in Figure~\ref{fig:SFH}. In addition to the best fit values listed in Table~\ref{tab:fits_regions}, we include for the oldest age bin the average SFR expected from the best fit solution of the (14~Gyr$-$400~Myr) Diff-CSP model  
(Section~\ref{sec:residual}).  While we do not expect the SFR in the center of NGC~3351 to have remained constant over the past  $\sim$14~Gyr, this model offers a convenient lower limit to the average SFR for ages older than $\sim$300--400~Myr. This is likely a 
hard lower limit, as  bars are transient features in galaxies, with lifetimes of 1--2~Gyr in a barred Sb galaxy like  NGC~3351 \citep{Bournaud+2005}. These lifetimes, ultimately, suggest    that the  average SFR  derived from the (1~Gyr$-$400~Myr) model is preferable over the longer timescale ones. The SFR calculated for the oldest age bin spans a wide range, from $\sim$0.1~M$_{\odot}$~yr$^{-1}$ to $\sim$2.4~M$_{\odot}$~yr$^{-1}$, depending on the model selected. The large uncertainty reflects the double difficulty of recovering the light from stars older than a few hundred Myr when younger populations dominate the light output, and of discriminating populations with ages $\gtrsim$1~Gyr when multiple such populations are potentially present. 
 Irrespective of such limitations, the SFR in the $>$300~Myr age bin is at least as high as the SFR in the 4--12~Myr bin, at  least on average. Conversely, the current SFR is the highest over the most recent $\sim$300~Myr. Thus, the age range between $\sim$10--15~Myr and $\sim$300--400~Myr marks the period of lowest SFR for the Central Region over the most recent $\sim$1--2~Gyr. Most of the stellar mass was accumulated before the most recent 300--400~Myr.  

\begin{figure}
\plottwo{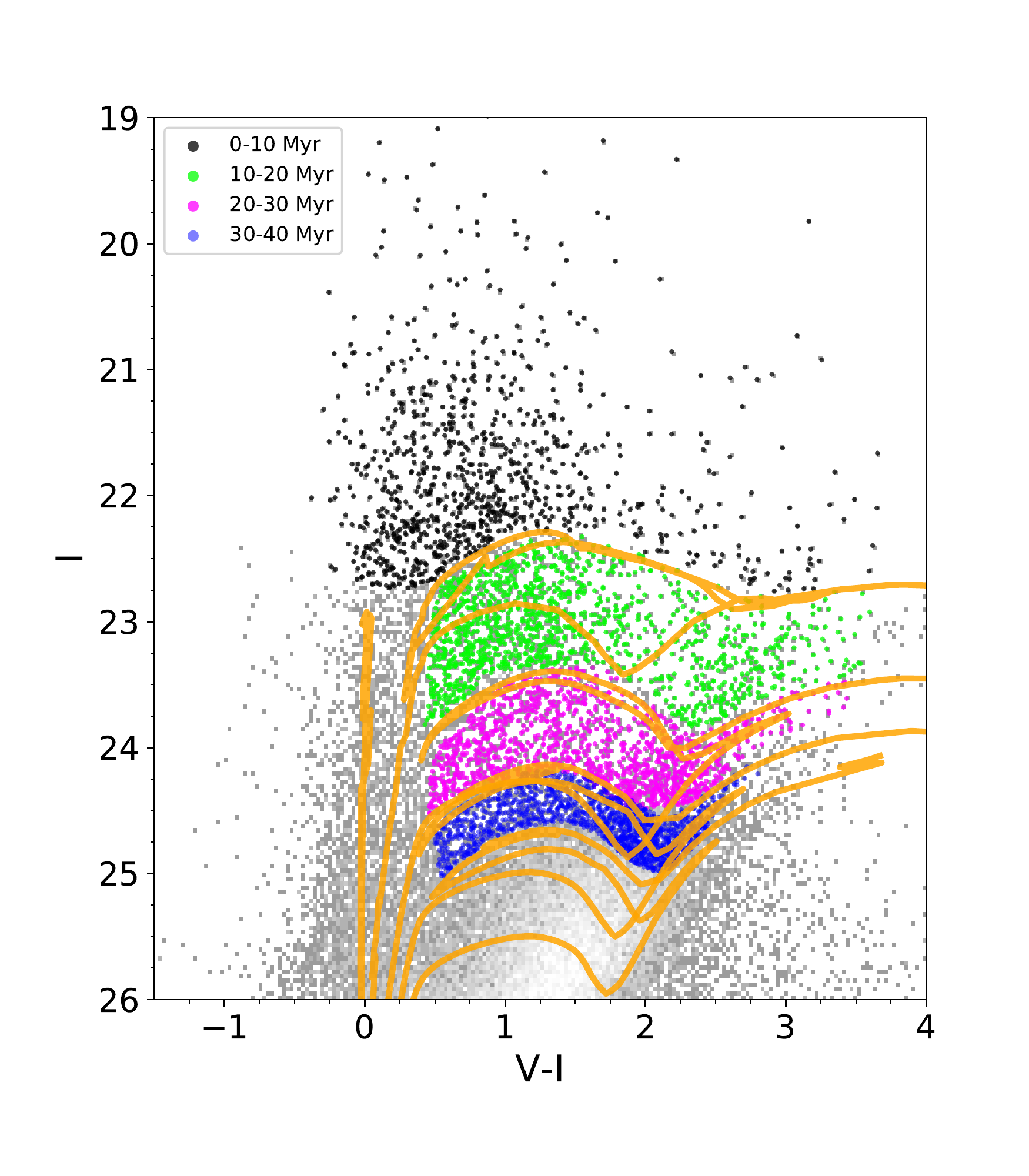}{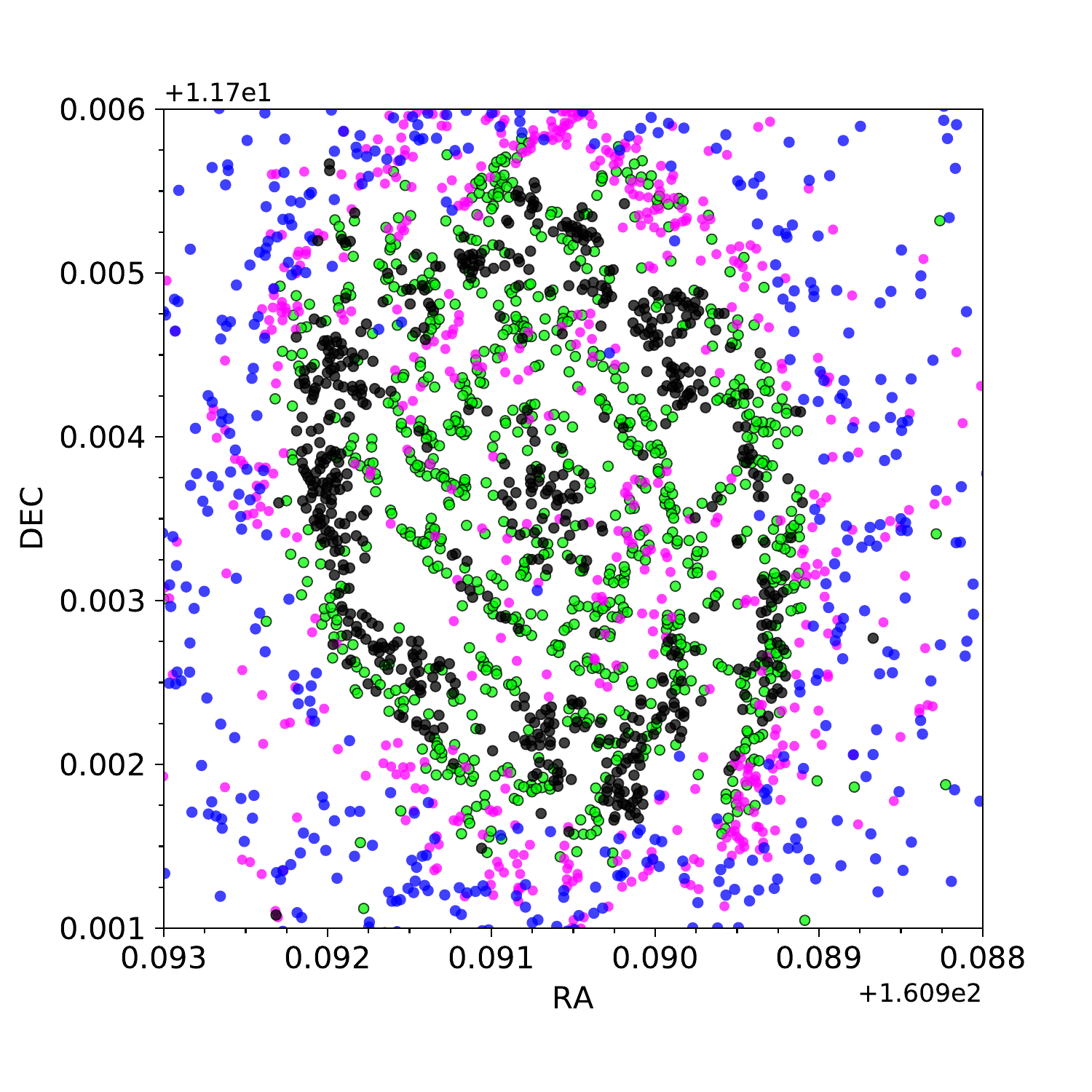}
\caption{{\bf LEFT:} The I--versus--(V$-$I) color--magnitude (CMD) diagram for the resolved stars in the Central Region, with tracks from the solar metallicity PARSEC--COLIBRI stellar models overlaid \citep[orange tracks][]{Bressan+2012, Marigo+2017}. Different colors identify  areas with  stars of different ages in the range 0--40~Myr, using the Blue-Loop region of the CMD \citep{Cignoni+2019}. {\bf RIGHT:}  The stars with identified ages from the CMD are re--mapped onto the Central region. The youngest stars, younger than 10~Myr, are concentrated along the ring and in the nucleus, in  agreement with the results from the  SED fitting. Stars with  ages in the range 10--20~Myr populate the region along and inside the ring, but are not present outside the ring region. Stars older than 20 Myr are more widely distributed, although their dearth inside the ring is likely due to incompleteness because of confusion. \label{fig:CMD}}
\end{figure}

We check some of our results on the age distributions in the Central Region with  color--magnitude diagrams (CMDs) from resolved stars,  
leveraging the stellar catalogs of the LEGUS project \citep{Sabbi+2018}. Of the five bands observed by LEGUS, we choose the two reddest ones, V and I, in order to  minimize the effects of dust  extinction (Figure~\ref{fig:CMD}). We employ the methodology of \citet[][]{Cignoni+2019} to identify age bins for the stars from the Blue Loop region of the CMD, using the PARSEC--COLIBRI stellar models \citep{Bressan+2012, Marigo+2017} with solar metallicity and adopting a reference color excess E(B--V)=0.3. Due to confusion, we can only recover ages up to $\sim$40~Myr from the CMD, which limits the lookback time we can explore with this  method. The stars are divided in four equal bins of age, which are given different colors (Figure~\ref{fig:CMD}, left), whose location is mapped back onto the Central Region (Figure~\ref{fig:CMD}, right). From the distribution of ages in the right panel of  Figure~\ref{fig:CMD}, we can infer that the ring and the nucleus contain the youngest stars, $<$10~Myr, in agreement with  our results and those of \citet{Turner+2021}. Stars in the age range 10--20~Myr are distributed more uniformly inside the region enclosed by the ring, but  are not present  outside of this region. Stars older than 20~Myr are more uniformly distributed, and their dearth inside the ring region is likely due to incompleteness caused by confusion.  While the derivation of a SFH from resolved stars would require the use of artificial star tests and is beyond the scope of this work, we note the general consistency with our earlier result that stellar populations younger 
than 10~Myr are located only along the ring and in the nucleus. This also strengthens our argument that the ionized gas found outside of the ring is due to ionizing photons that have leaked out of the ring region, with a small contribution from the nuclear cluster.

The general shape of the SFH of the Central Region from Figure~\ref{fig:SFH} suggests that star formation was quenched about 300--400~Myr ago, and resumed recently, about 15~Myr ago. This is consistent with recent results for the ages of  the stellar 
populations along the bar that feeds the Central Region: \citet{George+2019} find that the populations in the bar have ages $>$350~Myr, in addition to be devoid of  both molecular and atomic  gas. Thus, although the bar has quenched star formation 
\citep{Masters+2010, Masters+2011}, it has also driven gas into the center to ignite star  formation there  \citep{Combes+1985, Spinoso+2017}, but with a `lull' between $\sim$300~Myr and $\sim$15~Myr ago. The SFR also appears to have  increased over the past few Myrs, and it may still keep increasing in the future, fueled by the significant concentration of both molecular and atomic gas in the Central region \citep{Kenney+1992, Regan+2006, Schruba+2011, Leaman+2019}. 

 With a non--monotonic shape, the SFH of the Central Region does not fit either the instantaneous burst or the constant star formation models. This explains the failure of our initial simplistic approach, embodied by Figures~\ref{fig:IRX_beta} and ~\ref{fig:central_SEDs} (right), to account for all observed characteristics of this region. This is also a cautionary tale for standard approaches to the modeling of starburst or star--forming galaxies, where star formation histories are generally simplified to a few template behaviors: constant, increasing, decreasing, exponentially decreasing, etc. These smooth behaviors may  not apply to real situations  \citep[e.g.,][]{Leja+2019, Iyer+2019, Lower+2020}, and can affect conclusions on, e.g., the attenuation properties of the galaxy.

A consequence of the non--monotonic SFH in the Central  Region is that the UV slope, $\beta_o$, of the intrinsic stellar SED  is less negative, i.e., is redder, than the UV slope expected in a young or constantly star forming region. Using the FUV and NUV photometry values of the intrinsic stellar population SED of the Central region, we find:
\begin{equation}
\beta_o = -1.9\pm 0.1.
\end{equation}
For comparison, we obtain  $\beta_o= -2.4\pm 0.1$ for the intrinsic stellar population of the sum of the 10 ring regions, $\Sigma(R1-R10)$. As reference, at solar metallicity,  a stellar population formed at constant rate over 100~Myr  has $\beta_o= -2.3$, whilst  stellar populations formed over time intervals shorter than 10~Myr and instantaneous burst populations with age $<$10~Myr have bluer $\beta_o$ values \citep{Calzetti2001}. While the cumulative intrinsic  SED of the 10 ring regions is consistent with on--going star formation,  the red UV slope of the  Central Region reflects its complex SFH, including the build up of  stars during the star  formation that occurred prior to $\sim$300~Myr ago; these stars significantly contribute to the NUV luminosity, but not to the  FUV luminosity, of  the Central Region.

Redder--than-model--expectations UV slopes at zero attenuation are not new. For instance, using the Balmer ratio as an indicator of zero attenuation, \citet{Calzetti+1994}  and \citet{Battisti+2016} obtain $\beta_o\sim -1.7 - -1.6$, similar to what \citet{Reddy+2015}  and \citet{Shivaei+2020a} obtain at redshift z$\sim$2. Other authors obtain similar results for $\beta_o$ when extrapolating low L(TIR) values to zero attenuation in the IRX--$\beta$ diagram \citep{Calzetti+2005, Boquien+2012, Casey+2014}. In most cases, the observational result relies on the assumption that the galaxies/regions in the zero attenuation location are the parent galaxies/regions of  those at higher  attenuations; in a few cases, the result is the output of SED fits of kpc--size or larger regions, which carries a high degree of degeneracy between SFH and attenuation.  In the present case,  the red UV slope of the Central Region is the result  of modeling stellar populations at sub--kpc scales. This UV slope is a crucial `ingredient' for explaining the location of the Central Region on the IRX--$\beta$ diagram of Figure~\ref{fig:IRX_beta}, as we will see in the next  section. 

\subsection{The  `Net'  Attenuation Curve in the Center of NGC\,3351}

\begin{figure}
\plottwo{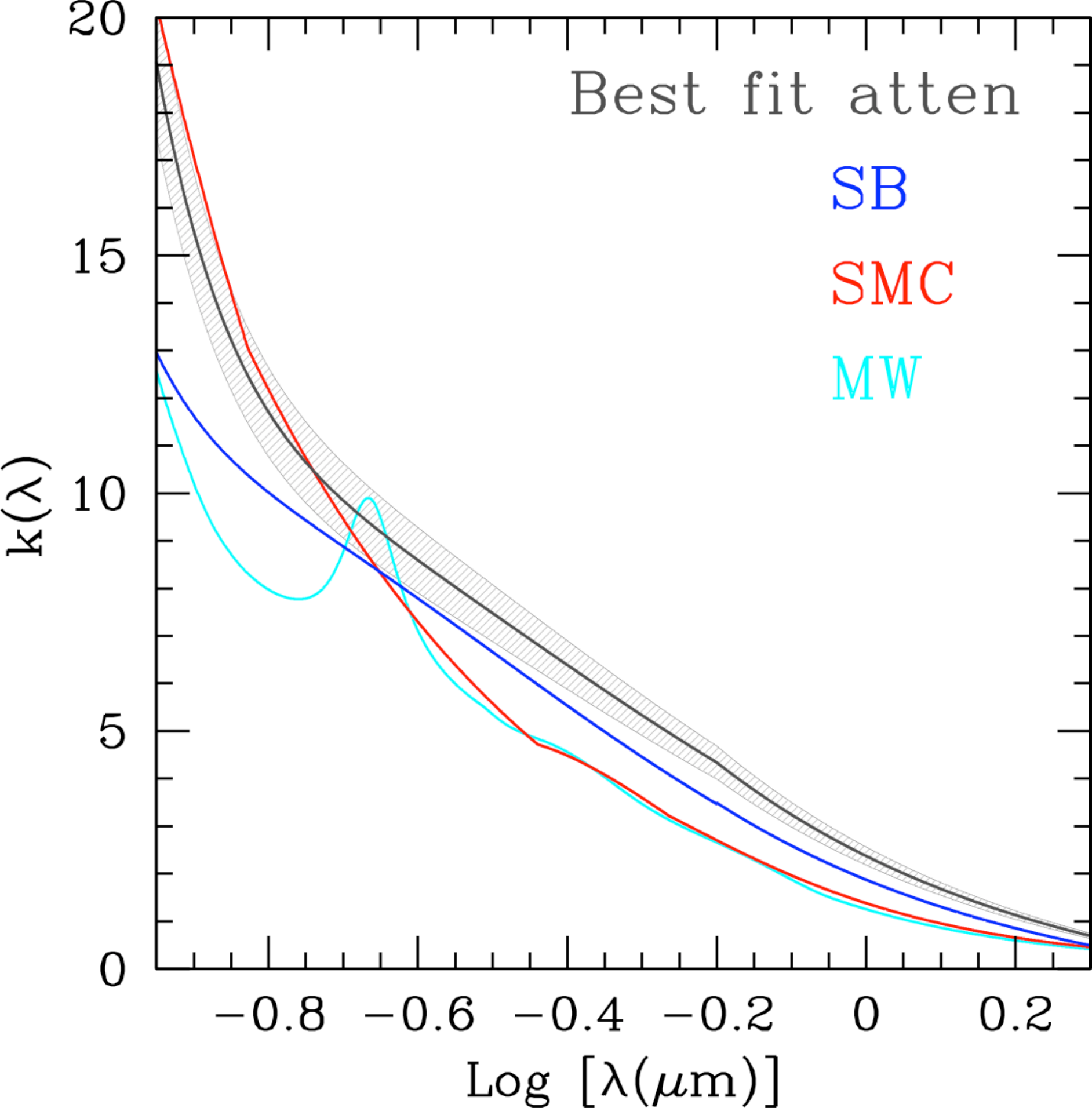}{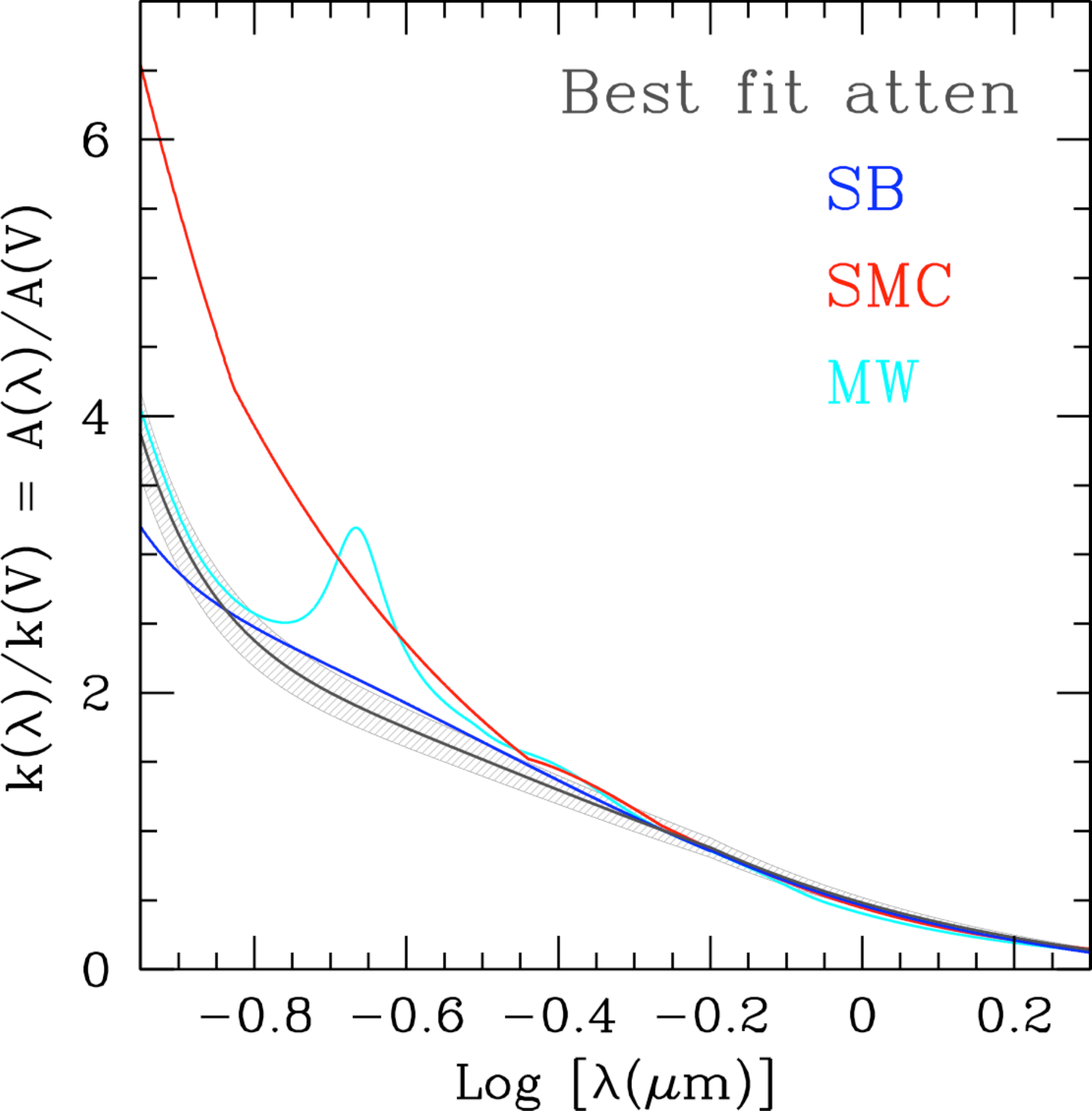}
\caption{{\bf LEFT:} The best fit attenuation curve from the left panel of Figure~\ref{fig:attenuation} is shown (grey line and hatched region), scaled to k(B)--k(V)=1, and compared to the SB attenuation and existing extinction curves (from Figure~\ref{fig:ext_curves}). All curves are at native R(V) or R$^{\prime}$(V) value.  {\bf RIGHT:} The same as the left panel, normalized to k($\lambda$)/k(V)=1. \label{fig:atten_comp}}
\end{figure}

The net attenuation required to recover the observed photometry from the intrinsic stellar population SED of the Central Region can be difficult to relate to previous results without some rescaling. Specifically, extinction 
curves are generally compared after rescaling them to a common value k(B)--k(V)=1. Figure~\ref{fig:atten_comp} compares the attenuation curve of equation~6 with the SB attenuation and existing extinction curves, each with their own normalization (R(V), left panel) and all normalized to k($\lambda$)/k(V)=1 (right panel).  

Our best--fit  net attenuation from equation~6, rescaled to k(B)--k(V)=1, becomes:
\begin{equation}
\begin{split}
k(\lambda) & =(2.49\pm 0.20) (-2.365  + 1.345/\lambda) + (4.93\pm 0.37), \  \ \ \  \  \ \  \ \ \ \ \  \ \  \  \ \ \ \  \ \ \ \ \  \ \ \ \ \  \  \ \ \  \ \ \  for \ \ \  0.63~\mu m \le \lambda \le 2.20~\mu m;\\
                                 &=(2.49\pm 0.20) (-2.450  + 1.809/\lambda  - 0.293/\lambda^2 + 0.0216/\lambda^3) + (4.93\pm 0.37), \  \ \ \  for \ \ \ 0.12~\mu m \le \lambda < 0.63~\mu m. 
\end{split}
\end{equation}
The normalization\footnote{The symbol R$^{\prime}$(V)  is preferred over R(V) in the case of  attenuation curves, since R(V) refers to the total extinction at V in the case of extinction curves, while R$^{\prime}$(V) combines extinction, scattering and geometrical effects \citep{Calzetti+2000}.}, R$^{\prime}$(V)=$4.93\pm0.37$, is much larger than most extinction curves (typically with R(V)=3.1 or smaller, with few exceptions), and larger than the value found for many attenuation curves, including the SB one \citep[R$^{\prime}$(V)=4.05$\pm$0.80][]{Calzetti+2000}. However, this is consistent with the R$^{\prime}$(V) values found by \citet{Battisti+2020} for  z$>$1 galaxies. When normalized to k($\lambda$)/k(V)=1, the FUV slope in equation~8 is close to that of the MW  extinction curve, and in--between those of the SB attenuation and SMC extinction curves (Figure~\ref{fig:atten_comp}, right). The  small metallicity range in the Central Region, which is  mostly solar, indicates that these two characteristics are unlikely to be due to effects of varying dust composition, but rather are a direct consequence of geometry, i.e., of a range of dust column densities distributed in front of stellar populations of different ages/durations. 

\begin{figure}
\plottwo{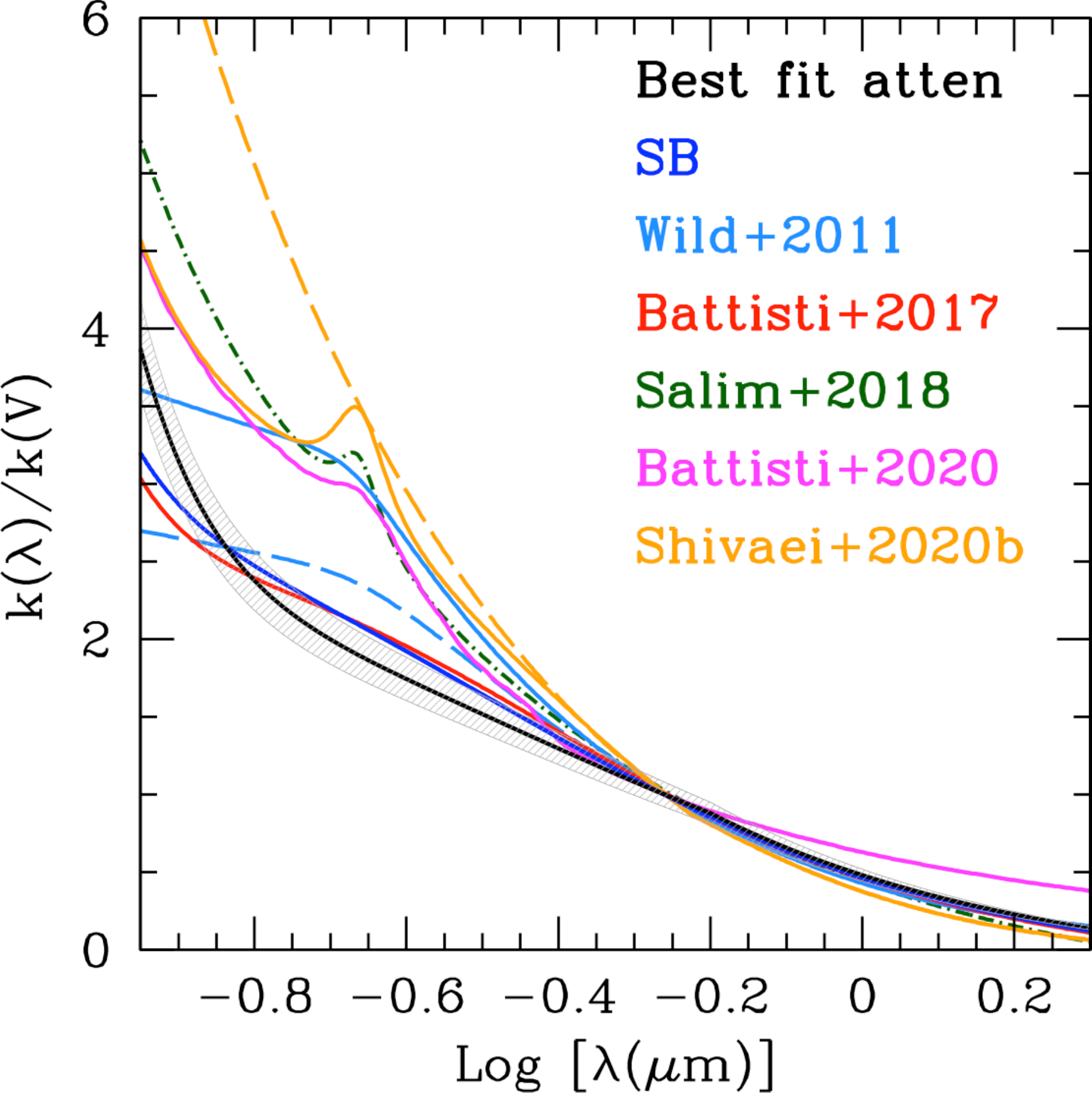}{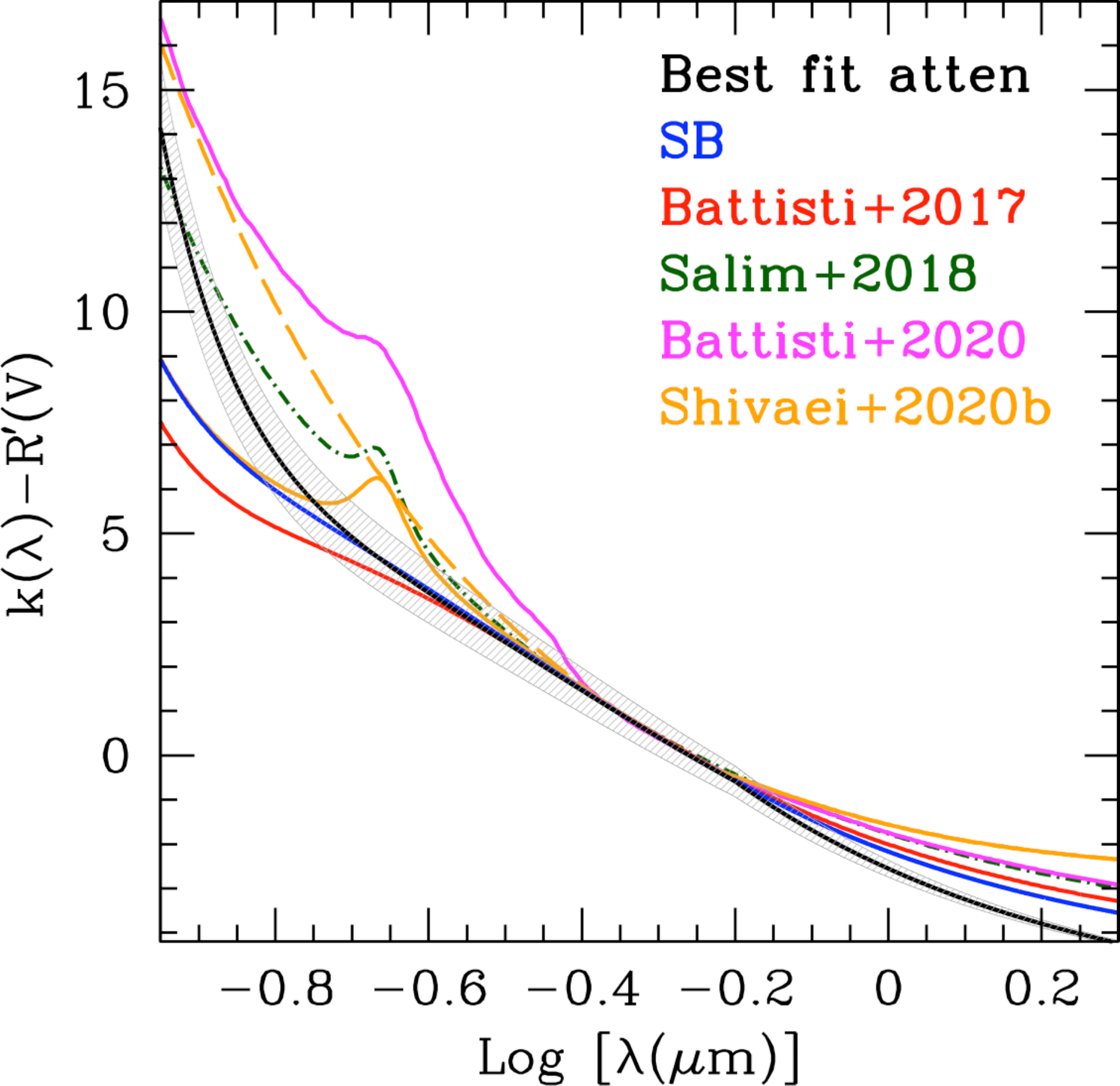}
\caption{{\bf LEFT:} The best fit attenuation curve from Figure~\ref{fig:atten_comp} is shown (black line and grey hatched region) normalized to k($\lambda$)/k(V)=1 and compared to attenuation curves at redshift z$\sim$0 as well as examples at high redshift, from the literature. In addition to  the SB curve \citep[blue line,]{Calzetti+2000}, the attenuation curves reported are from: \citet{Wild+2011}, for both low (dash light blue line) and high (continuous light blue line) stellar surface mass density, where the boundary between low and high is at 3$\times$10$^8$~M$_{\odot}$~kpc$^{-2}$; \citet{Battisti+2017a} (red line); \citet{Salim+2018} (dark green dash--dot line); \citet{Battisti+2020} for  an attenuation curve derived for the redshift range 1$\lesssim$z$\lesssim$3 (magenta line);  and \citet{Shivaei+2020a} for  high metallicity (continuous orange line) and low metallicity (dash  orange line) attenuation curves at z$\sim$2. {\bf RIGHT:} The same attenuation curves as in the left panel, plotted with a vertical shift  to R$^{\prime}$(V)=0. The curves of \citet{Wild+2011} are not shown as these authors do  not publish their curves' R$^{\prime}$(V). \label{fig:atten_comp_2}}
\end{figure}

Comparison with published attenuation curves at z$\sim$0 \citep{Wild+2011, Battisti+2017a, Salim+2018} and with selected examples at z$>$0 \citep{Battisti+2020, Shivaei+2020a} shows that  the general shape of the attenuation in the center of NGC~3351 is consistent with several existing ones, both when normalized to k($\lambda$)/k(V)=1 and when shifted to R$^{\prime}$(V)=0 (Figure~\ref{fig:atten_comp_2}). These two ways of  presenting comparisons  among attenuation curves are  both used in the literature, with one emphasizing the  selective attenuation (left panel of Figure~\ref{fig:atten_comp_2}) and the other emphasizing the impact of R$^{\prime}$(V) (right panel of Figure~\ref{fig:atten_comp_2}). In both cases, the attenuation curve in the center of NGC~3351 is shallower than those of \citet{Salim+2018}, \citet{Battisti+2020}, and of the low metallicity case in \citet{Shivaei+2020a}. \citet{Nersesian+2020} derived an attenuation curve for the galaxy  NGC~3351, finding a significantly steeper curve than ours, with a prominent 0.2175~$\mu$m feature. We attribute the differences between our results and theirs to differences in the two analyses: ours  targets the starburst center of the galaxy while \citet{Nersesian+2020} derive the attenuation curve for the entire galaxy, which include regions far more quiescent than the Central Region.

\begin{figure}
\plotone{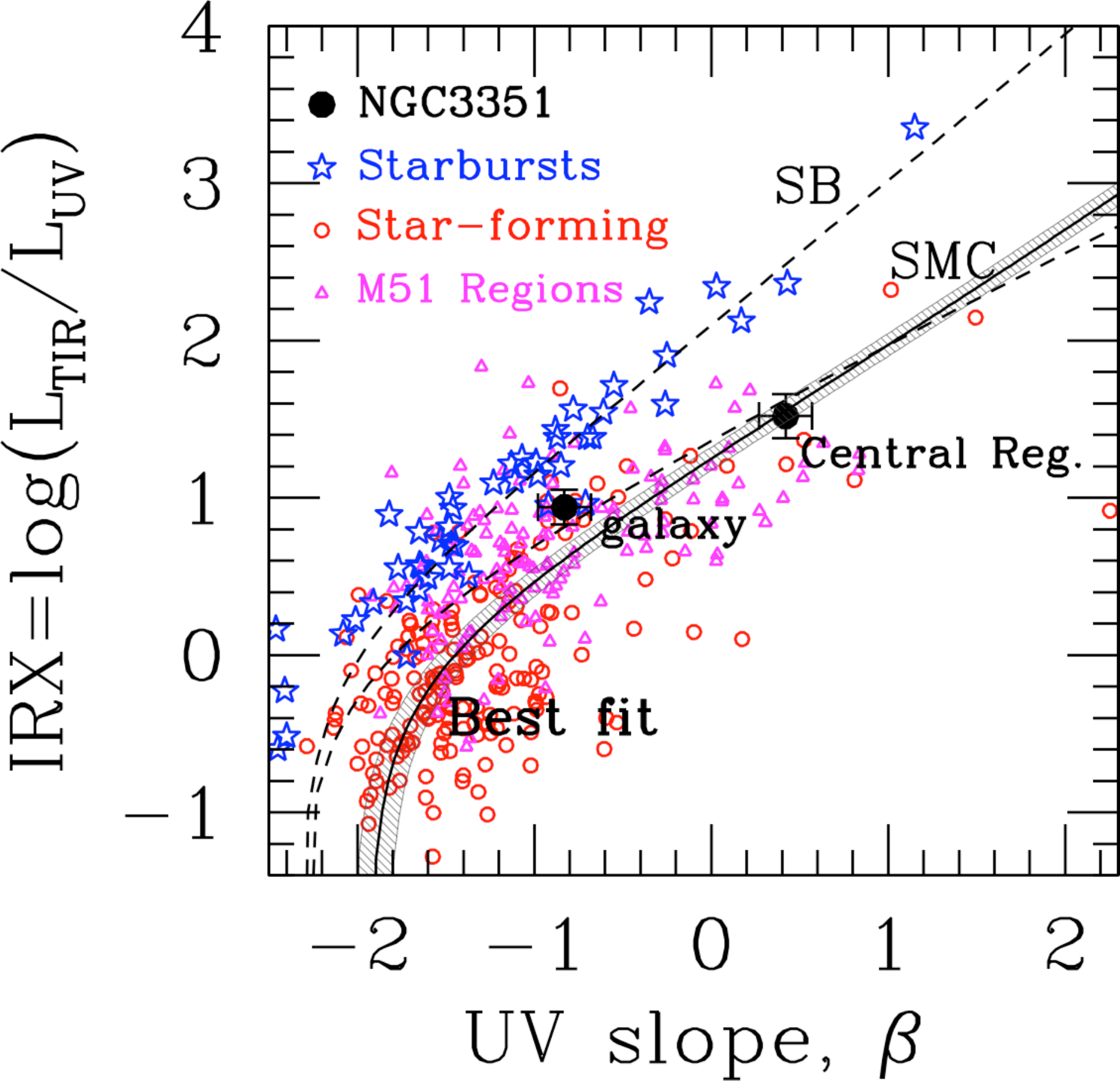}
\caption{The IRX--$\beta$  relation of Figure~\ref{fig:IRX_beta}, with the best fit attenuation curve of  Figure~\ref{fig:attenuation} added to the plot (black continuous line). The relation is anchored at $\beta_o=-1.9\pm0.1$, for the case of zero dust (zero IR emission). The grey hatched region includes the 1~$\sigma$ uncertainties both of the parameters of the attenuation curve and of  the intrinsic stellar population's UV slope $\beta_o$. The Central Region of  NGC\,3351 is well represented by this new attenuation curve. \label{fig:IRX_beta_new}}
\end{figure}

The shape and normalization of the net attenuation curve of equation~8 would not  be sufficient in itself to account for the location of the Central Region on the IRX--$\beta$ plot, since the k(FUV) values, which are the main determinant of the IRX--$\beta$ trend, are not as high as those of the SMC extinction curve, as shown in  the right panel of  Figure~\ref{fig:atten_comp}, where the two curves are normalized to k($\lambda$)/k(V)=1. 
However, if we anchor the attenuation curve of equation~8 to the value of $\beta_o$ in equation~7 and calculate the expected L(TIR)/L(FUV)  and $\beta$ for increasing amounts of dust, we recover the IRX--$\beta$ trend shown in Figure~\ref{fig:IRX_beta_new}. The net attenuation curve of NGC\,3351 accounts for the observed IRX and $\beta$ combination of the region, confirming the earlier calculation showing that the IR emission from the intrinsic population SED is consistent with observations. The slope of the net attenuation at high IRX values is  intermediate between those of the SB curve and the SMC curve, as expected  from the behavior in Figure~\ref{fig:atten_comp}, right. Thus, as already argued, the relatively shallow UV slope $\beta_o$ of the intrinsic stellar population  is the main reason for the location of the Central region of NGC\,3351 on the IRX--$\beta$  diagram below the locus of  central starbursts.

The above results, taken all together, indicate that most previous conclusions on the attenuation curves of galaxies are correct. (1) Net attenuations often indicate steeper curves in the UV than obtained for the starbursts of \citet{Calzetti+1994} and \citet{Calzetti+2000}  \citep[e.g.,][]{Reddy+2015, Salim+2018, Battisti+2020}. Of course, shallower curves are also  possible, e.g., as in the case of ULIRGs and other star--forming galaxies \citep{Calzetti2001, Goldader+2002, Battisti+2016, LoFaro+2017}. (2) The normalization R$^{\prime}$(V) of attenuation curves  is  also not fixed; at fixed E(B--V), a high normalization increases the infrared dust emission without reddening of the observed SEDs \citep[e.g.,][]{Battisti+2020}. In the case of the complex geometries of galaxies, the normalization is a short--hand for the mixed dust--stars optical depth or more complex geometries. (3) SFHs dictate what the intrinsic stellar population SED looks like. In the case of complex histories, the intrinsic  UV slope needs to be calculated `a posteriori', as the contribution of aged stellar populations may alter its value \citep[e.g.][]{Calzetti+2005, Overzier+2011, Casey+2014, Popping+2017, Narayanan+2018, Liang+2020}. Similarly, extremely young and/or extremely metal--poor stellar populations may display significantly bluer intrinsic SEDs than older or metal--rich galaxies \citep[e.g.][]{Reddy+2018, Schulz+2020}. The main conclusion is that, outside of the `central starburst' regime, where the star  formation is spatially concentrated in the center of the galaxy \citep{Calzetti+1994, Meurer+1999, Heckman2000}, the relative geometry of dust and different--aged stellar populations determine what the net attenuation curve looks like. This conclusion had been reached already using simulations \citep{Popping+2017,  Narayanan+2018, Liang+2020},  and we have confirmed it using observational  data.

One result that holds steady is the presence of a  differential attenuation between stars and ionized gas. From our fits, the median E(B--V)$_{star}$, is  smaller, by a factor 0.56$\pm$0.06, than the median color excess derived from the ionized gas, E(B--V)$_{gas}$. This should be  compared with the scaling of the net  attenuation curve we derive from the photometry  of the Central Region, which yields:
\begin{equation}
 E(B-V)_{star}=(0.40 \pm 0.03) E(B-V)_{gas}.
 \end{equation} 
 The two numbers are 2.4~$\sigma$ away from each other, which is not unexpected since the first value  is a median of ratios, while the second value, equation~9, is the ratio of sum luminosities. Yet both numbers point  to the fact that stars are systematically 
 less dust reddened than the ionized gas. 
Indeed, a consistent finding  among many authors is that the attenuation of the stellar continuum is lower than that of the ionized gas \citep{Calzetti+1994, Calzetti+2000, Calzetti2001, Kreckel+2013, Reddy+2015, Battisti+2016, Theios+2019, Shivaei+2020a};  this is perhaps not surprising, as stars and star clusters tend to blow away their natal dust cocoons or migrate away  from their dusty birthplaces as they age \citep{Grasha+2018, Grasha+2019, Krumholz+2019}. A common toy model that accounts for this general result is one where the youngest stars, and their associated ionized gas, reside in dustier regions than their more evolved counterparts \citep{Calzetti+1994, CharlotFall2000, Zaritsky+2004, Liang+2020}. The  value we find for NGC\,3351 is not very different from the value $0.44\pm0.04$ \citet{Calzetti+1994} find for local starburst galaxies. The median value we  derive from the individual fits of subregions, 0.56$\pm$0.06, is not inconsistent with the value 0.47$\pm$0.006 derived by \citet{Kreckel+2013}. These authors use a region--by--region approach for the optical SED fits of nearby galaxies, similar to our approach, but with much larger, kpc--sized regions, which may account for residual discrepancies.

\subsection{Caveats}\label{sec:caveats}

While the results  presented so far shed light on the dust attenuation trends that several authors have observed in and within galaxies, at least  two caveats are in order. 

First and foremost, this is one single example. Although the availability of high resolution multi--wavelength images from the HST has enabled us to study in  detail the stellar content and attenuation properties of the Central Region of NGC\,3351, it is not clear how representative of other galaxies the metal--rich circumnuclear starburst in a nearby spiral is. Thus, although a convenient analytical expression is given in equation~8, the reader should be cautioned about using it before  testing its applicability to their case. Specifically, 
the assumption that  goes into applying equation~8 to any other galaxy is that this galaxy contains similar distributions of both stellar populations and dust column densities as  those observed in the center of NGC\,3351, and variations in the dust column densities occur in a `lockstep' fashion across all  regions. This  is clearly an unlikely  scenario, which can only hold as an approximation.  

Second, our dataset lacks  high--resolution  FUV ($\sim$0.15~$\mu$m) images, which prevents us from constraining the stellar population fits of individual regions at this wavelength. As a result, the inferred FUV photometry from the sum of our best--fit SEDs is  larger, by a factor $\lesssim$2 than the observed FUV photometry (Section~\ref{sec:regions}).  This discrepancy  is currently fairly  resilient  to attempts to decrease it: regions that contain sufficient dust attenuation to lower the  difference between  models and observed FUV luminosity tend to be poor fits at  longer wavelengths. The FUV discrepancy may also drive the (smaller) discrepancies in the NUV and U bands: these result  from over--compensation of the models to attempts to  fit the data. Clearly a high--resolution FUV image will be required to clarify these issues. 

\section{Summary and Conclusions}\label{sec:conclusions}

We use a combination of archival low-- and high--angular  resolution images from the FUV ($\sim$0.15~$\mu$m) to the sub--mm (500$\mu$m) from the HST, GALEX, SST, and HSO, 
in order to  investigate the stellar populations and dust attenuation properties of the central $\sim$1~kpc of the nearby galaxy NGC\,3351. The Central Region contains a circumnuclear ring of star formation, sufficiently active to place the region on the starburst locus in the SFR--mass sequence of galaxies. We fit the multi--wavelength photometry of each component with a combination of stellar population and dust attenuation models. The results show that:
\begin{enumerate}
\item The star formation history of the central  $\sim$1~kpc in NGC\,3351 is not monotonic with time. It has a minimum in SFR around 15--300~Myr ago, and is several times higher both at earlier and later times (Figure~\ref{fig:SFH}). We find evidence that  the star formation has been propagating along the ring for the past  $\sim$400~Myr. The non--monotonic star formation history, with most of the stellar mass accumulating at times earlier than 300~Myr ago, causes the intrinsic UV slope of the dust--free stellar populations  to  have value $\beta_o=1.9\pm 0.1$ (equation~7).  This is significantly redder than what measured for stellar populations at solar metallicity produced by constant star formation or by a young instantaneous burst. 
\item The net dust attenuation  in the region is consistent with an attenuation curve that is as steep in the UV as the MW extinction curve, steeper  than the attenuation curve of \citet{Calzetti+1994,Calzetti+2000}, but significantly shallower than the SMC curve (Figure~\ref{fig:atten_comp}). The normalization of the net attenuation (the value of the curve at V) is significantly larger than those of existing extinction and attenuation curves, with R$^{\prime}$(V)=4.93$\pm$0.37 (equation~8). 
\item The combination of points 1 and 2 above provides an explanation for the location of the Central Region on the IRX--$\beta$ diagram (Figure~\ref{fig:IRX_beta_new}). Its location is {\em below} the locus occupied by starburst galaxies, which is unexpected given the active nature of the region, but explainable in light of the results above, especially in light of the red UV slope $\beta_o$.
\item The mean dust color excess for the stellar continuum is lower than the color excess measured for the ionized gas. When using the net attenuation curve of the Central Region, the two color excesses satisfy the following relation: E(B--V)$_{star}$=(0.40$\pm$0.03) E(B--V)$_{gas}$ (equation~9), comparable to what was derived for starburst galaxies \citep{Calzetti+1994}. When measured on a region--by--region basis, the relation between the two is: E(B--V)$_{star}$=(0.56$\pm$0.06) E(B--V)$_{gas}$, higher than the ratio derived from integral photometry. In either case, for one magnitude of reddening between B and V in the ionized gas, we only measure 0.4 (or  0.56) magnitudes of reddening in the stellar continuum.
\end{enumerate}

We do  not attempt to fit the 0.2175~$\mu$m bump typical  of extinction curves, as existing UV spectroscopy indicates this bump to be insignificant in the center of NGC\,3351. 

Despite all the consistencies listed above, the case presented here still represents one single realization, at $\sim$fixed metallicity, of the many possible geometrical distributions of different--age stellar populations and different--optical--depth dust. However, it helps understand the deviations of star--forming galaxies from the standard locus of starbursts as a mostly geometrical effect combined with a complex star formation history.  

\acknowledgments

The authors thank  the referee for comments that  have improved the manuscript.

Based on observations made with the NASA/ESA Hubble Space Telescope, obtained  at the Space Telescope Science Institute, which is operated by the 
Association of Universities for Research in Astronomy, Inc., under NASA contract NAS 5--26555. These observations are associated with program \# 13364. 

Based also on archival data from the following facilities: the NASA/ESA Hubble Space Telescope, and obtained from the Hubble Legacy Archive, 
which is a collaboration between the Space Telescope Science Institute (STScI/NASA), the Space Telescope European Coordinating 
Facility (ST-ECF/ESA) and the Canadian Astronomy Data Centre (CADC/NRC/CSA); the Spitzer Space Telescope, which was operated by the 
Jet Propulsion Laboratory, California Institute of Technology under a contract with NASA, and retrieved from the NASA/IPAC Infrared Science Archive, which is 
operated by the Jet Propulsion Laboratory, California Institute of Technology,  under  contract  with NASA; and the 
Herschel Space Observatory, an ESA space observatory with science instruments provided by European-led Principal Investigator consortia and with important participation from NASA.

This research has made use of the NASA/IPAC Extragalactic Database (NED) which is operated by the Jet
Propulsion Laboratory, California Institute of Technology, under contract with the National Aeronautics and Space
Administration.

D.C. acknowledges partial support from the National Aeronautics and Space Administration (NASA), via the JetPropulsion Laboratory Euclid Project Office, as part of 
the ``Science Investigations as Members of the Euclid Consortium and Euclid Science Team'' program.  M.M. acknowledges the support of the Swedish Research Council, Vetenskapsr{\aa}det (internationell postdok grant 2019-00502\_3). M. C. acknowledges the support of INFN ``Iniziativa specifica TAsP". 
A.A. acknowledges the support of the Swedish Research Council, Vetenskapsr{\aa}det, and the Swedish National Space Agency (SNSA).

\vspace{5mm}
\facilities{GALEX, HST(WFC3,NICMOS), SST(IRAC,MIPS), HSO(PACS)}

\software{FORTRAN, IRAF, SuperMongo}

\appendix

\section{Best Fit Parameters Distributions\label{Distributions}}

Examples of 90\% confidence level distributions of best fit parameters for the  regions in NGC~3351.

\begin{figure}[ht!]
\plotone{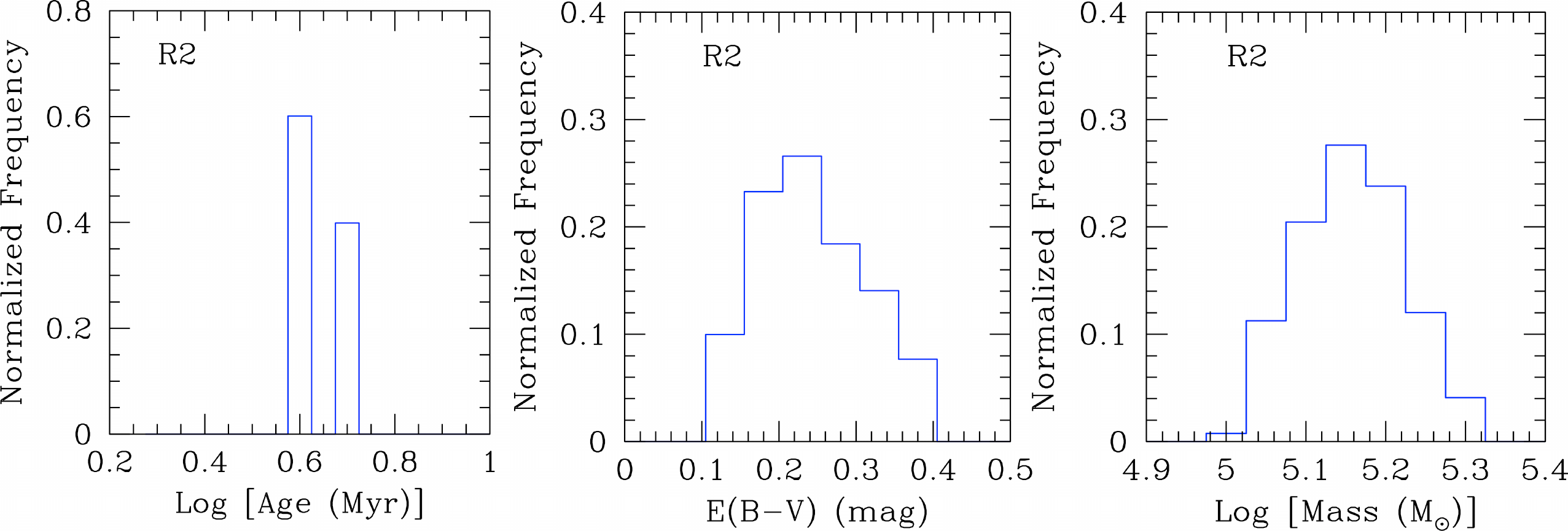}
\caption{The 90th percentile distribution of the best fit age, color excess E(B--V), and stellar mass for R2. This region is an example of instantaneous burst population with a small value of $\chi^2_{red}$=1.7.  The discreteness in the age distribution plot reflects the sampling of the stellar population ages in our models. Axis ranges change from plot  to plot and from Figure to Figure.\label{fig:distrib_1}}
\end{figure}

\begin{figure}[ht!]
\plotone{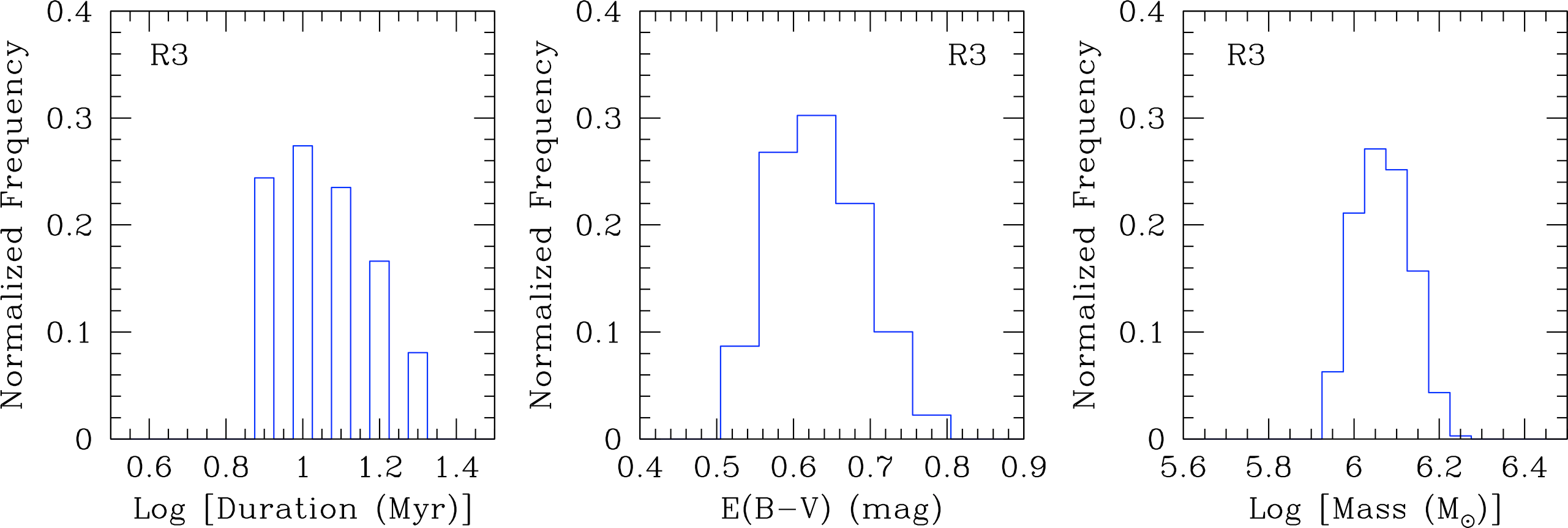}
\caption{The same as Figure~\ref{fig:distrib_1}, but for R3. This is an example of constant star formation with relatively short duration, and a small value of $\chi^2_{red}$=1.9.\label{fig:distrib_2}}
\end{figure}

\begin{figure}[ht!]
\plotone{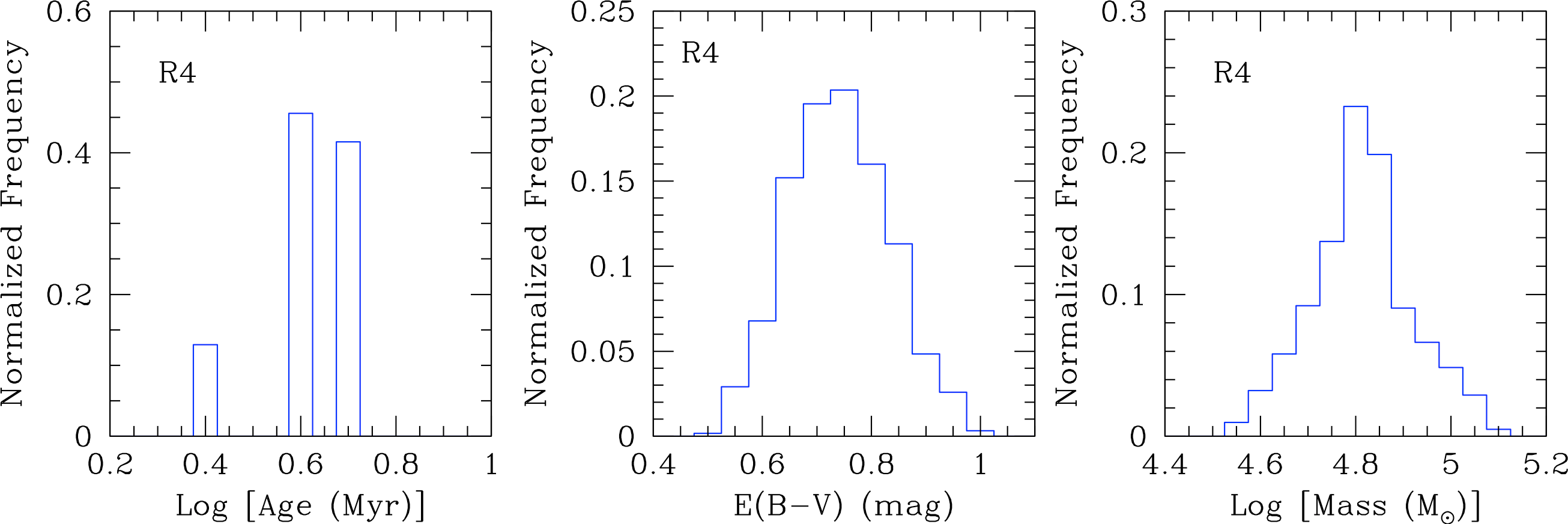}
\caption{The same as Figure~\ref{fig:distrib_1}, but for R4. This is an example of instantaneous burst with the poorest reduced $\chi^2$ value in our sample, $\chi^2_{red}$=5.2, but also with the largest attenuation of the stellar continuum, with E(B--V)=0.76.\label{fig:distrib_3}}
\end{figure}

\begin{figure}[ht!]
\plotone{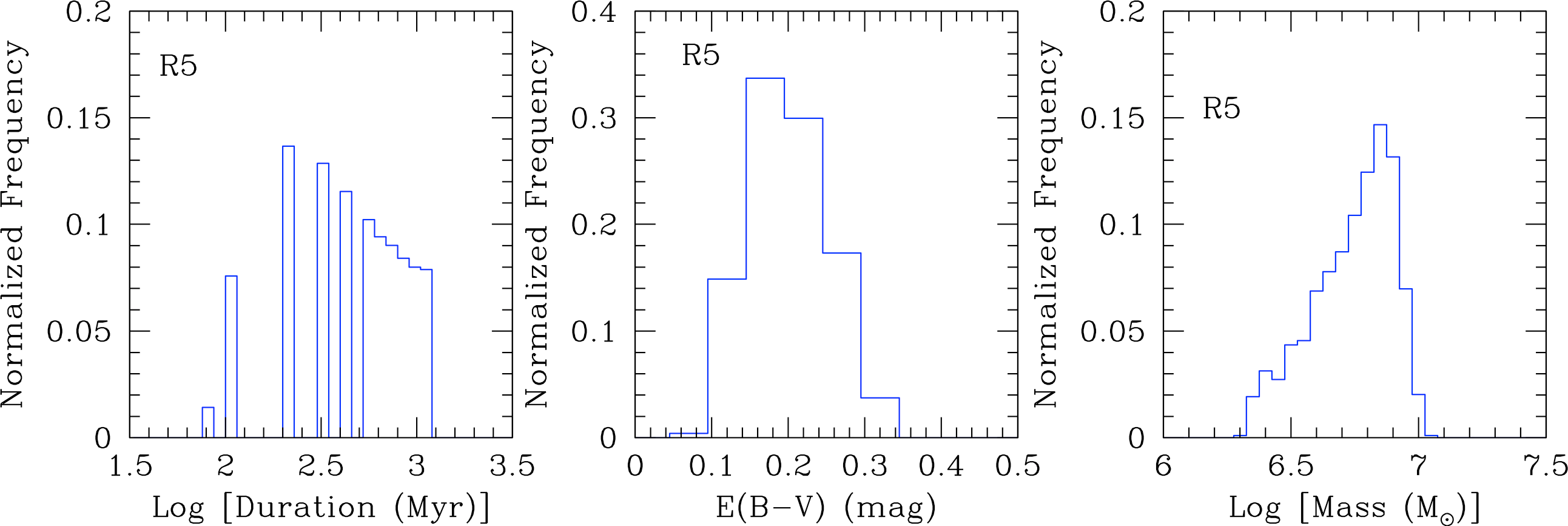}
\caption{The same as Figure~\ref{fig:distrib_1}, but for R5. This is an example of constant star formation with long duration.\label{fig:distrib_4}}
\end{figure}

\bibliography{ms_v6_accepted.bib}{}

\begin{thebibliography}{}
\expandafter\ifx\csname natexlab\endcsname\relax\def\natexlab#1{#1}\fi

\bibitem[{{Adamo} {et~al.}(2017){Adamo}, {Ryon}, {Messa}, {Kim}, {Grasha},
  {Cook}, {Calzetti}, {Lee}, {Whitmore}, {Elmegreen}, {Ubeda}, {Smith},
  {Bright}, {Runnholm}, {Andrews}, {Fumagalli}, {Gouliermis}, {Kahre}, {Nair},
  {Thilker}, {Walterbos}, {Wofford}, {Aloisi}, {Ashworth}, {Brown}, {Chandar},
  {Christian}, {Cignoni}, {Clayton}, {Dale}, {de Mink}, {Dobbs}, {Elmegreen},
  {Evans}, {Gallagher}, {Grebel}, {Herrero}, {Hunter}, {Johnson}, {Kennicutt},
  {Krumholz}, {Lennon}, {Levay}, {Martin}, {Nota}, {{\"O}stlin}, {Pellerin},
  {Prieto}, {Regan}, {Sabbi}, {Sacchi}, {Schaerer}, {Schiminovich}, {Shabani},
  {Tosi}, {Van Dyk}, \& {Zackrisson}}]{Adamo+2017}
{Adamo}, A., {Ryon}, J.~E., {Messa}, M., {et~al.} 2017, \apj, 841, 131

\bibitem[{{Aniano} {et~al.}(2011){Aniano}, {Draine}, {Gordon}, \&
  {Sandstrom}}]{Aniano+2011}
{Aniano}, G., {Draine}, B.~T., {Gordon}, K.~D., \& {Sandstrom}, K. 2011, \pasp,
  123, 1218

\bibitem[{{Aniano} {et~al.}(2020){Aniano}, {Draine}, {Hunt}, {Sandstrom},
  {Calzetti}, {Kennicutt}, {Dale}, {Galametz}, {Gordon}, {Leroy}, {Smith},
  {Roussel}, {Sauvage}, {Walter}, {Armus}, {Bolatto}, {Boquien}, {Crocker}, {De
  Looze}, {Donovan Meyer}, {Helou}, {Hinz}, {Johnson}, {Koda}, {Miller},
  {Montiel}, {Murphy}, {Rela{\~n}o}, {Rix}, {Schinnerer}, {Skibba}, {Wolfire},
  \& {Engelbracht}}]{Aniano+2020}
{Aniano}, G., {Draine}, B.~T., {Hunt}, L.~K., {et~al.} 2020, \apj, 889, 150

\bibitem[{{Battisti} {et~al.}(2016){Battisti}, {Calzetti}, \&
  {Chary}}]{Battisti+2016}
{Battisti}, A.~J., {Calzetti}, D., \& {Chary}, R.~R. 2016, \apj, 818, 13

\bibitem[{{Battisti} {et~al.}(2017{\natexlab{a}}){Battisti}, {Calzetti}, \&
  {Chary}}]{Battisti+2017b}
---. 2017{\natexlab{a}}, \apj, 851, 90

\bibitem[{{Battisti} {et~al.}(2017{\natexlab{b}}){Battisti}, {Calzetti}, \&
  {Chary}}]{Battisti+2017a}
---. 2017{\natexlab{b}}, \apj, 840, 109

\bibitem[{{Battisti} {et~al.}(2020){Battisti}, {Cunha}, {Shivaei}, \&
  {Calzetti}}]{Battisti+2020}
{Battisti}, A.~J., {Cunha}, E.~d., {Shivaei}, I., \& {Calzetti}, D. 2020, \apj,
  888, 108

\bibitem[{{Bohlin} {et~al.}(1978){Bohlin}, {Savage}, \& {Drake}}]{Bohlin+1978}
{Bohlin}, R.~C., {Savage}, B.~D., \& {Drake}, J.~F. 1978, \apj, 224, 132

\bibitem[{{B{\"o}ker} {et~al.}(2008){B{\"o}ker}, {Falc{\'o}n-Barroso},
  {Schinnerer}, {Knapen}, \& {Ryder}}]{Boeker+2008}
{B{\"o}ker}, T., {Falc{\'o}n-Barroso}, J., {Schinnerer}, E., {Knapen}, J.~H.,
  \& {Ryder}, S. 2008, \aj, 135, 479

\bibitem[{{Boquien} {et~al.}(2019){Boquien}, {Burgarella}, {Roehlly}, {Buat},
  {Ciesla}, {Corre}, {Inoue}, \& {Salas}}]{Boquien+2019}
{Boquien}, M., {Burgarella}, D., {Roehlly}, Y., {et~al.} 2019, \aap, 622, A103

\bibitem[{{Boquien} {et~al.}(2012){Boquien}, {Buat}, {Boselli}, {Baes},
  {Bendo}, {Ciesla}, {Cooray}, {Cortese}, {Eales}, {Gavazzi}, {Gomez},
  {Lebouteiller}, {Pappalardo}, {Pohlen}, {Smith}, \&
  {Spinoglio}}]{Boquien+2012}
{Boquien}, M., {Buat}, V., {Boselli}, A., {et~al.} 2012, \aap, 539, A145

\bibitem[{{Bournaud} {et~al.}(2005){Bournaud}, {Combes}, \&
  {Semelin}}]{Bournaud+2005}
{Bournaud}, F., {Combes}, F., \& {Semelin}, B. 2005, \mnras, 364, L18

\bibitem[{{Bouwens} {et~al.}(2020){Bouwens}, {Gonzalez-Lopez}, {Aravena},
  {Decarli}, {Novak}, {Stefanon}, {Walter}, {Boogaard}, {Carilli},
  {Dudzeviciute}, {Smail}, {Daddi}, {da Cunha}, {Ivison}, {Nanayakkara},
  {Cortes}, {Cox}, {Inami}, {Oesch}, {Popping}, {Riechers}, {van der Werf},
  {Weiss}, {Fudamoto}, \& {Wagg}}]{Bouwens+2020}
{Bouwens}, R., {Gonzalez-Lopez}, J., {Aravena}, M., {et~al.} 2020, arXiv
  e-prints, arXiv:2009.10727

\bibitem[{{Bouwens} {et~al.}(2016){Bouwens}, {Aravena}, {Decarli}, {Walter},
  {da Cunha}, {Labb{\'e}}, {Bauer}, {Bertoldi}, {Carilli}, {Chapman}, {Daddi},
  {Hodge}, {Ivison}, {Karim}, {Le Fevre}, {Magnelli}, {Ota}, {Riechers},
  {Smail}, {van der Werf}, {Weiss}, {Cox}, {Elbaz}, {Gonzalez-Lopez},
  {Infante}, {Oesch}, {Wagg}, \& {Wilkins}}]{Bouwens+2016}
{Bouwens}, R.~J., {Aravena}, M., {Decarli}, R., {et~al.} 2016, \apj, 833, 72

\bibitem[{{Bressan} {et~al.}(2012){Bressan}, {Marigo}, {Girardi}, {Salasnich},
  {Dal Cero}, {Rubele}, \& {Nanni}}]{Bressan+2012}
{Bressan}, A., {Marigo}, P., {Girardi}, L., {et~al.} 2012, \mnras, 427, 127

\bibitem[{{Buat} {et~al.}(2002){Buat}, {Boselli}, {Gavazzi}, \&
  {Bonfanti}}]{Buat+2002}
{Buat}, V., {Boselli}, A., {Gavazzi}, G., \& {Bonfanti}, C. 2002, \aap, 383,
  801

\bibitem[{{Buat} {et~al.}(2005){Buat}, {Iglesias-P{\'a}ramo}, {Seibert},
  {Burgarella}, {Charlot}, {Martin}, {Xu}, {Heckman}, {Boissier}, {Boselli},
  {Barlow}, {Bianchi}, {Byun}, {Donas}, {Forster}, {Friedman}, {Jelinski},
  {Lee}, {Madore}, {Malina}, {Milliard}, {Morissey}, {Neff}, {Rich},
  {Schiminovitch}, {Siegmund}, {Small}, {Szalay}, {Welsh}, \&
  {Wyder}}]{Buat+2005}
{Buat}, V., {Iglesias-P{\'a}ramo}, J., {Seibert}, M., {et~al.} 2005, \apjl,
  619, L51

\bibitem[{{Calapa} {et~al.}(2014){Calapa}, {Calzetti}, {Draine}, {Boquien},
  {Kramer}, {Xilouris}, {Verley}, {Braine}, {Rela{\~n}o}, {van der Werf},
  {Israel}, {Hermelo}, \& {Albrecht}}]{Calapa+2014}
{Calapa}, M.~D., {Calzetti}, D., {Draine}, B.~T., {et~al.} 2014, \apj, 784, 130

\bibitem[{{Calzetti}(2001)}]{Calzetti2001}
{Calzetti}, D. 2001, \pasp, 113, 1449

\bibitem[{{Calzetti}(2013)}]{Calzetti2013}
---. 2013, {Star Formation Rate Indicators}, ed. J.~{Falc{\'o}n-Barroso} \&
  J.~H. {Knapen}, 419

\bibitem[{{Calzetti} {et~al.}(2000){Calzetti}, {Armus}, {Bohlin}, {Kinney},
  {Koornneef}, \& {Storchi-Bergmann}}]{Calzetti+2000}
{Calzetti}, D., {Armus}, L., {Bohlin}, R.~C., {et~al.} 2000, \apj, 533, 682

\bibitem[{{Calzetti} {et~al.}(1994){Calzetti}, {Kinney}, \&
  {Storchi-Bergmann}}]{Calzetti+1994}
{Calzetti}, D., {Kinney}, A.~L., \& {Storchi-Bergmann}, T. 1994, \apj, 429, 582

\bibitem[{{Calzetti} {et~al.}(1996){Calzetti}, {Kinney}, \&
  {Storchi-Bergmann}}]{Calzetti+1996}
---. 1996, \apj, 458, 132

\bibitem[{{Calzetti} {et~al.}(2005){Calzetti}, {Kennicutt}, {Bianchi},
  {Thilker}, {Dale}, {Engelbracht}, {Leitherer}, {Meyer}, {Sosey}, {Mutchler},
  {Regan}, {Thornley}, {Armus}, {Bendo}, {Boissier}, {Boselli}, {Draine},
  {Gordon}, {Helou}, {Hollenbach}, {Kewley}, {Madore}, {Martin}, {Murphy},
  {Rieke}, {Rieke}, {Roussel}, {Sheth}, {Smith}, {Walter}, {White}, {Yi},
  {Scoville}, {Polletta}, \& {Lindler}}]{Calzetti+2005}
{Calzetti}, D., {Kennicutt}, R.~C., J., {Bianchi}, L., {et~al.} 2005, \apj,
  633, 871

\bibitem[{{Calzetti} {et~al.}(2007){Calzetti}, {Kennicutt}, {Engelbracht},
  {Leitherer}, {Draine}, {Kewley}, {Moustakas}, {Sosey}, {Dale}, {Gordon},
  {Helou}, {Hollenbach}, {Armus}, {Bendo}, {Bot}, {Buckalew}, {Jarrett}, {Li},
  {Meyer}, {Murphy}, {Prescott}, {Regan}, {Rieke}, {Roussel}, {Sheth}, {Smith},
  {Thornley}, \& {Walter}}]{Calzetti+2007}
{Calzetti}, D., {Kennicutt}, R.~C., {Engelbracht}, C.~W., {et~al.} 2007, \apj,
  666, 870

\bibitem[{{Calzetti} {et~al.}(2015){Calzetti}, {Lee}, {Sabbi}, {Adamo},
  {Smith}, {Andrews}, {Ubeda}, {Bright}, {Thilker}, {Aloisi}, {Brown},
  {Chandar}, {Christian}, {Cignoni}, {Clayton}, {da Silva}, {de Mink}, {Dobbs},
  {Elmegreen}, {Elmegreen}, {Evans}, {Fumagalli}, {Gallagher}, {Gouliermis},
  {Grebel}, {Herrero}, {Hunter}, {Johnson}, {Kennicutt}, {Kim}, {Krumholz},
  {Lennon}, {Levay}, {Martin}, {Nair}, {Nota}, {{\"O}stlin}, {Pellerin},
  {Prieto}, {Regan}, {Ryon}, {Schaerer}, {Schiminovich}, {Tosi}, {Van Dyk},
  {Walterbos}, {Whitmore}, \& {Wofford}}]{Calzetti+2015}
{Calzetti}, D., {Lee}, J.~C., {Sabbi}, E., {et~al.} 2015, \aj, 149, 51

\bibitem[{{Calzetti} {et~al.}(2018){Calzetti}, {Wilson}, {Draine}, {Roussel},
  {Johnson}, {Heyer}, {Wall}, {Grasha}, {Battisti}, {Andrews}, {Kirkpatrick},
  {Rosa Gonz{\'a}lez}, {Vega}, {Puschnig}, {Yun}, {{\"O}stlin}, {Evans},
  {Tang}, {Lowenthal}, \& {S{\'a}nchez-Arguelles}}]{Calzetti+2018}
{Calzetti}, D., {Wilson}, G.~W., {Draine}, B.~T., {et~al.} 2018, \apj, 852, 106

\bibitem[{{Capak} {et~al.}(2015){Capak}, {Carilli}, {Jones}, {Casey},
  {Riechers}, {Sheth}, {Carollo}, {Ilbert}, {Karim}, {Lefevre}, {Lilly},
  {Scoville}, {Smolcic}, \& {Yan}}]{Capak+2015}
{Capak}, P.~L., {Carilli}, C., {Jones}, G., {et~al.} 2015, \nat, 522, 455

\bibitem[{{Cardelli} {et~al.}(1989){Cardelli}, {Clayton}, \&
  {Mathis}}]{Cardelli+1989}
{Cardelli}, J.~A., {Clayton}, G.~C., \& {Mathis}, J.~S. 1989, \apj, 345, 245

\bibitem[{{Casey} {et~al.}(2014){Casey}, {Narayanan}, \& {Cooray}}]{Casey+2014}
{Casey}, C.~M., {Narayanan}, D., \& {Cooray}, A. 2014, \physrep, 541, 45

\bibitem[{{Casey} {et~al.}(2018){Casey}, {Zavala}, {Spilker}, {da Cunha},
  {Hodge}, {Hung}, {Staguhn}, {Finkelstein}, \& {Drew}}]{Casey+2018}
{Casey}, C.~M., {Zavala}, J.~A., {Spilker}, J., {et~al.} 2018, \apj, 862, 77

\bibitem[{{Cervi{\~n}o} {et~al.}(2002){Cervi{\~n}o}, {Valls-Gabaud},
  {Luridiana}, \& {Mas-Hesse}}]{Cervino+2002}
{Cervi{\~n}o}, M., {Valls-Gabaud}, D., {Luridiana}, V., \& {Mas-Hesse}, J.~M.
  2002, \aap, 381, 51

\bibitem[{{Charlot} \& {Fall}(2000)}]{CharlotFall2000}
{Charlot}, S., \& {Fall}, S.~M. 2000, \apj, 539, 718

\bibitem[{{Chevallard} {et~al.}(2013){Chevallard}, {Charlot}, {Wandelt}, \&
  {Wild}}]{Chevallard+2013}
{Chevallard}, J., {Charlot}, S., {Wandelt}, B., \& {Wild}, V. 2013, \mnras,
  432, 2061

\bibitem[{{Cignoni} {et~al.}(2019){Cignoni}, {Sacchi}, {Tosi}, {Aloisi},
  {Cook}, {Calzetti}, {Lee}, {Sabbi}, {Thilker}, {Adamo}, {Dale}, {Elmegreen},
  {Gallagher}, {Grebel}, {Johnson}, {Messa}, {Smith}, \&
  {Ubeda}}]{Cignoni+2019}
{Cignoni}, M., {Sacchi}, E., {Tosi}, M., {et~al.} 2019, \apj, 887, 112

\bibitem[{{Combes} \& {Gerin}(1985)}]{Combes+1985}
{Combes}, F., \& {Gerin}, M. 1985, \aap, 150, 327

\bibitem[{{Conroy}(2013)}]{Conroy2013}
{Conroy}, C. 2013, \araa, 51, 393

\bibitem[{{Conroy} {et~al.}(2010){Conroy}, {Schiminovich}, \&
  {Blanton}}]{Conroy+2010}
{Conroy}, C., {Schiminovich}, D., \& {Blanton}, M.~R. 2010, \apj, 718, 184

\bibitem[{{Cook} {et~al.}(2014){Cook}, {Dale}, {Johnson}, {Van Zee}, {Lee},
  {Kennicutt}, {Calzetti}, {Staudaher}, \& {Engelbracht}}]{Cook+2014}
{Cook}, D.~O., {Dale}, D.~A., {Johnson}, B.~D., {et~al.} 2014, \mnras, 445, 899

\bibitem[{{Cortese} {et~al.}(2006){Cortese}, {Boselli}, {Buat}, {Gavazzi},
  {Boissier}, {Gil de Paz}, {Seibert}, {Madore}, \& {Martin}}]{Cortese+2006}
{Cortese}, L., {Boselli}, A., {Buat}, V., {et~al.} 2006, \apj, 637, 242

\bibitem[{{Croxall} {et~al.}(2016){Croxall}, {Pogge}, {Berg}, {Skillman}, \&
  {Moustakas}}]{Croxall+2016}
{Croxall}, K.~V., {Pogge}, R.~W., {Berg}, D.~A., {Skillman}, E.~D., \&
  {Moustakas}, J. 2016, \apj, 830, 4

\bibitem[{{da Cunha} {et~al.}(2008){da Cunha}, {Charlot}, \&
  {Elbaz}}]{daCunha+2008}
{da Cunha}, E., {Charlot}, S., \& {Elbaz}, D. 2008, \mnras, 388, 1595

\bibitem[{{Dale} \& {Helou}(2002)}]{DaleHelou2002}
{Dale}, D.~A., \& {Helou}, G. 2002, \apj, 576, 159

\bibitem[{{Dale} {et~al.}(2009){Dale}, {Cohen}, {Johnson}, {Schuster},
  {Calzetti}, {Engelbracht}, {Gil de Paz}, {Kennicutt}, {Lee}, {Begum},
  {Block}, {Dalcanton}, {Funes}, {Gordon}, {Johnson}, {Marble}, {Sakai},
  {Skillman}, {van Zee}, {Walter}, {Weisz}, {Williams}, {Wu}, \&
  {Wu}}]{Dale+2009}
{Dale}, D.~A., {Cohen}, S.~A., {Johnson}, L.~C., {et~al.} 2009, \apj, 703, 517

\bibitem[{{De Barros} {et~al.}(2016){De Barros}, {Reddy}, \&
  {Shivaei}}]{DeBarros+2016}
{De Barros}, S., {Reddy}, N., \& {Shivaei}, I. 2016, \apj, 820, 96

\bibitem[{{D{\'\i}az} {et~al.}(2007){D{\'\i}az}, {Terlevich}, {Castellanos}, \&
  {H{\"a}gele}}]{Diaz+2007}
{D{\'\i}az}, {\'A}.~I., {Terlevich}, E., {Castellanos}, M., \& {H{\"a}gele},
  G.~F. 2007, \mnras, 382, 251

\bibitem[{{Draine}(2003)}]{Draine2003}
{Draine}, B.~T. 2003, \araa, 41, 241

\bibitem[{{Draine} \& {Li}(2007)}]{DraineLi2007}
{Draine}, B.~T., \& {Li}, A. 2007, \apj, 657, 810

\bibitem[{{Draine} {et~al.}(2007){Draine}, {Dale}, {Bendo}, {Gordon}, {Smith},
  {Armus}, {Engelbracht}, {Helou}, {Kennicutt}, {Li}, {Roussel}, {Walter},
  {Calzetti}, {Moustakas}, {Murphy}, {Rieke}, {Bot}, {Hollenbach}, {Sheth}, \&
  {Teplitz}}]{Draine+2007}
{Draine}, B.~T., {Dale}, D.~A., {Bendo}, G., {et~al.} 2007, \apj, 663, 866

\bibitem[{{Draine} {et~al.}(2014){Draine}, {Aniano}, {Krause}, {Groves},
  {Sandstrom}, {Braun}, {Leroy}, {Klaas}, {Linz}, {Rix}, {Schinnerer},
  {Schmiedeke}, \& {Walter}}]{Draine+2014}
{Draine}, B.~T., {Aniano}, G., {Krause}, O., {et~al.} 2014, \apj, 780, 172

\bibitem[{{Ferguson} {et~al.}(1996){Ferguson}, {Wyse}, {Gallagher}, \&
  {Hunter}}]{Ferguson+1996}
{Ferguson}, A. M.~N., {Wyse}, R. F.~G., {Gallagher}, J.~S., I., \& {Hunter},
  D.~A. 1996, \aj, 111, 2265

\bibitem[{{Ferland} {et~al.}(2013){Ferland}, {Porter}, {van Hoof}, {Williams},
  {Abel}, {Lykins}, {Shaw}, {Henney}, \& {Stancil}}]{Ferland+2013}
{Ferland}, G.~J., {Porter}, R.~L., {van Hoof}, P.~A.~M., {et~al.} 2013, \rmxaa,
  49, 137

\bibitem[{{Fischera} \& {Dopita}(2011)}]{Fischera+2011}
{Fischera}, J., \& {Dopita}, M. 2011, \aap, 533, A117

\bibitem[{{Fitzpatrick}(1999)}]{Fitzpatrick1999}
{Fitzpatrick}, E.~L. 1999, \pasp, 111, 63

\bibitem[{{Fitzpatrick} \& {Massa}(2007)}]{Fitzpatrick+2007}
{Fitzpatrick}, E.~L., \& {Massa}, D. 2007, \apj, 663, 320

\bibitem[{{Fitzpatrick} {et~al.}(2019){Fitzpatrick}, {Massa}, {Gordon},
  {Bohlin}, \& {Clayton}}]{Fitzpatrick+2019}
{Fitzpatrick}, E.~L., {Massa}, D., {Gordon}, K.~D., {Bohlin}, R., \& {Clayton},
  G.~C. 2019, \apj, 886, 108

\bibitem[{{Forrest} {et~al.}(2016){Forrest}, {Tran}, {Tomczak}, {Broussard},
  {Labb{\'e}}, {Papovich}, {Kriek}, {Allen}, {Cowley}, {Dickinson},
  {Glazebrook}, {van Houdt}, {Inami}, {Kacprzak}, {Kawinwanichakij}, {Kelson},
  {McCarthy}, {Monson}, {Morrison}, {Nanayakkara}, {Persson}, {Quadri},
  {Spitler}, {Straatman}, \& {Tilvi}}]{Forrest+2016}
{Forrest}, B., {Tran}, K.-V.~H., {Tomczak}, A.~R., {et~al.} 2016, \apjl, 818,
  L26

\bibitem[{{Freedman} {et~al.}(2001){Freedman}, {Madore}, {Gibson}, {Ferrarese},
  {Kelson}, {Sakai}, {Mould}, {Kennicutt}, {Ford}, {Graham}, {Huchra},
  {Hughes}, {Illingworth}, {Macri}, \& {Stetson}}]{Freedman+2001}
{Freedman}, W.~L., {Madore}, B.~F., {Gibson}, B.~K., {et~al.} 2001, \apj, 553,
  47

\bibitem[{{Fudamoto} {et~al.}(2020){Fudamoto}, {Oesch}, {Magnelli},
  {Schinnerer}, {Liu}, {Lang}, {Jim{\'e}nez-Andrade}, {Groves}, {Leslie}, \&
  {Sargent}}]{Fudamoto+2020}
{Fudamoto}, Y., {Oesch}, P.~A., {Magnelli}, B., {et~al.} 2020, \mnras, 491,
  4724

\bibitem[{{George} {et~al.}(2019){George}, {Joseph}, {Mondal}, {Subramanian},
  {Subramaniam}, \& {Paul}}]{George+2019}
{George}, K., {Joseph}, P., {Mondal}, C., {et~al.} 2019, \aap, 621, L4

\bibitem[{{Gil de Paz} {et~al.}(2007){Gil de Paz}, {Boissier}, {Madore},
  {Seibert}, {Joe}, {Boselli}, {Wyder}, {Thilker}, {Bianchi}, {Rey}, {Rich},
  {Barlow}, {Conrow}, {Forster}, {Friedman}, {Martin}, {Morrissey}, {Neff},
  {Schiminovich}, {Small}, {Donas}, {Heckman}, {Lee}, {Milliard}, {Szalay}, \&
  {Yi}}]{GildePaz+2007}
{Gil de Paz}, A., {Boissier}, S., {Madore}, B.~F., {et~al.} 2007, \apjs, 173,
  185

\bibitem[{{Girardi} {et~al.}(2000){Girardi}, {Bressan}, {Bertelli}, \&
  {Chiosi}}]{Girardi+2000}
{Girardi}, L., {Bressan}, A., {Bertelli}, G., \& {Chiosi}, C. 2000, \aaps, 141,
  371

\bibitem[{{Goldader} {et~al.}(2002){Goldader}, {Meurer}, {Heckman}, {Seibert},
  {Sanders}, {Calzetti}, \& {Steidel}}]{Goldader+2002}
{Goldader}, J.~D., {Meurer}, G., {Heckman}, T.~M., {et~al.} 2002, \apj, 568,
  651

\bibitem[{{Gordon} {et~al.}(1997){Gordon}, {Calzetti}, \& {Witt}}]{Gordon+1997}
{Gordon}, K.~D., {Calzetti}, D., \& {Witt}, A.~N. 1997, \apj, 487, 625

\bibitem[{{Gordon} {et~al.}(2003){Gordon}, {Clayton}, {Misselt}, {Land olt}, \&
  {Wolff}}]{Gordon+2003}
{Gordon}, K.~D., {Clayton}, G.~C., {Misselt}, K.~A., {Land olt}, A.~U., \&
  {Wolff}, M.~J. 2003, \apj, 594, 279

\bibitem[{{Goulding} \& {Alexander}(2009)}]{Goulding+2009}
{Goulding}, A.~D., \& {Alexander}, D.~M. 2009, \mnras, 398, 1165

\bibitem[{{Grasha} {et~al.}(2013){Grasha}, {Calzetti}, {Andrews}, {Lee}, \&
  {Dale}}]{Grasha+2013}
{Grasha}, K., {Calzetti}, D., {Andrews}, J.~E., {Lee}, J.~C., \& {Dale}, D.~A.
  2013, \apj, 773, 174

\bibitem[{{Grasha} {et~al.}(2018){Grasha}, {Calzetti}, {Bittle}, {Johnson},
  {Donovan Meyer}, {Kennicutt}, {Elmegreen}, {Adamo}, {Krumholz}, {Fumagalli},
  {Grebel}, {Gouliermis}, {Cook}, {Gallagher}, {Aloisi}, {Dale}, {Linden},
  {Sacchi}, {Thilker}, {Walterbos}, {Messa}, {Wofford}, \&
  {Smith}}]{Grasha+2018}
{Grasha}, K., {Calzetti}, D., {Bittle}, L., {et~al.} 2018, \mnras, 481, 1016

\bibitem[{{Grasha} {et~al.}(2019){Grasha}, {Calzetti}, {Adamo}, {Kennicutt},
  {Elmegreen}, {Messa}, {Dale}, {Fedorenko}, {Mahadevan}, {Grebel},
  {Fumagalli}, {Kim}, {Dobbs}, {Gouliermis}, {Ashworth}, {Gallagher}, {Smith},
  {Tosi}, {Whitmore}, {Schinnerer}, {Colombo}, {Hughes}, {Leroy}, \&
  {Meidt}}]{Grasha+2019}
{Grasha}, K., {Calzetti}, D., {Adamo}, A., {et~al.} 2019, \mnras, 483, 4707

\bibitem[{{Grier} {et~al.}(2011){Grier}, {Mathur}, {Ghosh}, \&
  {Ferrarese}}]{Grier+2011}
{Grier}, C.~J., {Mathur}, S., {Ghosh}, H., \& {Ferrarese}, L. 2011, \apj, 731,
  60

\bibitem[{{Hannon} {et~al.}(2019){Hannon}, {Lee}, {Whitmore}, {Chandar},
  {Adamo}, {Mobasher}, {Aloisi}, {Calzetti}, {Cignoni}, {Cook}, {Dale},
  {Deger}, {Della Bruna}, {Elmegreen}, {Gouliermis}, {Grasha}, {Grebel},
  {Herrero}, {Hunter}, {Johnson}, {Kennicutt}, {Kim}, {Sacchi}, {Smith},
  {Thilker}, {Turner}, {Walterbos}, \& {Wofford}}]{Hannon+2019}
{Hannon}, S., {Lee}, J.~C., {Whitmore}, B.~C., {et~al.} 2019, \mnras, 490, 4648

\bibitem[{{Heckman}(2000)}]{Heckman2000}
{Heckman}, T. 2000, {Starburst Galaxies}, ed. P.~{Murdin}, 1583

\bibitem[{{Helou} {et~al.}(2004){Helou}, {Roussel}, {Appleton}, {Frayer},
  {Stolovy}, {Storrie-Lombardi}, {Hurt}, {Lowrance}, {Makovoz}, {Masci},
  {Surace}, {Gordon}, {Alonso-Herrero}, {Engelbracht}, {Misselt}, {Rieke},
  {Rieke}, {Willner}, {Pahre}, {Ashby}, {Fazio}, \& {Smith}}]{Helou+2004}
{Helou}, G., {Roussel}, H., {Appleton}, P., {et~al.} 2004, \apjs, 154, 253

\bibitem[{{Hunt} \& {Hirashita}(2009)}]{Hunt+2009}
{Hunt}, L.~K., \& {Hirashita}, H. 2009, \aap, 507, 1327

\bibitem[{{Hunt} {et~al.}(2020){Hunt}, {Tortora}, {Ginolfi}, \&
  {Schneider}}]{Hunt2020}
{Hunt}, L.~K., {Tortora}, C., {Ginolfi}, M., \& {Schneider}, R. 2020, arXiv
  e-prints, arXiv:2010.02919

\bibitem[{{Hunt} {et~al.}(2019){Hunt}, {De Looze}, {Boquien}, {Nikutta},
  {Rossi}, {Bianchi}, {Dale}, {Granato}, {Kennicutt}, {Silva}, {Ciesla},
  {Rela{\~n}o}, {Viaene}, {Brandl}, {Calzetti}, {Croxall}, {Draine},
  {Galametz}, {Gordon}, {Groves}, {Helou}, {Herrera-Camus}, {Hinz}, {Koda},
  {Salim}, {Sandstrom}, {Smith}, {Wilson}, \& {Zibetti}}]{Hunt+2019}
{Hunt}, L.~K., {De Looze}, I., {Boquien}, M., {et~al.} 2019, \aap, 621, A51

\bibitem[{{Iyer} {et~al.}(2019){Iyer}, {Gawiser}, {Faber}, {Ferguson},
  {Kartaltepe}, {Koekemoer}, {Pacifici}, \& {Somerville}}]{Iyer+2019}
{Iyer}, K.~G., {Gawiser}, E., {Faber}, S.~M., {et~al.} 2019, \apj, 879, 116

\bibitem[{{Kenney} {et~al.}(1992){Kenney}, {Wilson}, {Scoville}, {Devereux}, \&
  {Young}}]{Kenney+1992}
{Kenney}, J. D.~P., {Wilson}, C.~D., {Scoville}, N.~Z., {Devereux}, N.~A., \&
  {Young}, J.~S. 1992, \apjl, 395, L79

\bibitem[{{Kennicutt}(1984)}]{Kennicutt1984}
{Kennicutt}, R.~C., J. 1984, \apj, 287, 116

\bibitem[{{Kennicutt} {et~al.}(1989){Kennicutt}, {Keel}, \&
  {Blaha}}]{Kennicutt+1989}
{Kennicutt}, Robert~C., J., {Keel}, W.~C., \& {Blaha}, C.~A. 1989, \aj, 97,
  1022

\bibitem[{{Kennicutt} {et~al.}(2003){Kennicutt}, {Armus}, {Bendo}, {Calzetti},
  {Dale}, {Draine}, {Engelbracht}, {Gordon}, {Grauer}, {Helou}, {Hollenbach},
  {Jarrett}, {Kewley}, {Leitherer}, {Li}, {Malhotra}, {Regan}, {Rieke},
  {Rieke}, {Roussel}, {Smith}, {Thornley}, \& {Walter}}]{Kennicutt+2003}
{Kennicutt}, Robert~C., J., {Armus}, L., {Bendo}, G., {et~al.} 2003, \pasp,
  115, 928

\bibitem[{{Kennicutt} \& {Evans}(2012)}]{KennicuttEvans2012}
{Kennicutt}, R.~C., \& {Evans}, N.~J. 2012, \araa, 50, 531

\bibitem[{{Kennicutt} {et~al.}(2011){Kennicutt}, {Calzetti}, {Aniano},
  {Appleton}, {Armus}, {Beir{\~a}o}, {Bolatto}, {Brandl}, {Crocker}, {Croxall},
  {Dale}, {Donovan Meyer}, {Draine}, {Engelbracht}, {Galametz}, {Gordon},
  {Groves}, {Hao}, {Helou}, {Hinz}, {Hunt}, {Johnson}, {Koda}, {Krause},
  {Leroy}, {Li}, {Meidt}, {Montiel}, {Murphy}, {Rahman}, {Rix}, {Roussel},
  {Sandstrom}, {Sauvage}, {Schinnerer}, {Skibba}, {Smith}, {Srinivasan},
  {Vigroux}, {Walter}, {Wilson}, {Wolfire}, \& {Zibetti}}]{Kennicutt+2011}
{Kennicutt}, R.~C., {Calzetti}, D., {Aniano}, G., {et~al.} 2011, \pasp, 123,
  1347

\bibitem[{{Kinney} {et~al.}(1993){Kinney}, {Bohlin}, {Calzetti}, {Panagia}, \&
  {Wyse}}]{Kinney+1993}
{Kinney}, A.~L., {Bohlin}, R.~C., {Calzetti}, D., {Panagia}, N., \& {Wyse}, R.
  F.~G. 1993, \apjs, 86, 5

\bibitem[{{Knapen}(2005)}]{Knapen2005}
{Knapen}, J.~H. 2005, \aap, 429, 141

\bibitem[{{Kong} {et~al.}(2004){Kong}, {Charlot}, {Brinchmann}, \&
  {Fall}}]{Kong+2004}
{Kong}, X., {Charlot}, S., {Brinchmann}, J., \& {Fall}, S.~M. 2004, \mnras,
  349, 769

\bibitem[{{Kormendy} \& {Kennicutt}(2004)}]{Kormendy+2004}
{Kormendy}, J., \& {Kennicutt}, Robert~C., J. 2004, \araa, 42, 603

\bibitem[{{Kreckel} {et~al.}(2013){Kreckel}, {Groves}, {Schinnerer}, {Johnson},
  {Aniano}, {Calzetti}, {Croxall}, {Draine}, {Gordon}, {Crocker}, {Dale},
  {Hunt}, {Kennicutt}, {Meidt}, {Smith}, \& {Tabatabaei}}]{Kreckel+2013}
{Kreckel}, K., {Groves}, B., {Schinnerer}, E., {et~al.} 2013, \apj, 771, 62

\bibitem[{{Kroupa}(2001)}]{Kroupa2001}
{Kroupa}, P. 2001, \mnras, 322, 231

\bibitem[{{Krumholz} {et~al.}(2019){Krumholz}, {McKee}, \&
  {Bland-Hawthorn}}]{Krumholz+2019}
{Krumholz}, M.~R., {McKee}, C.~F., \& {Bland-Hawthorn}, J. 2019, \araa, 57, 227

\bibitem[{{Leaman} {et~al.}(2019){Leaman}, {Fragkoudi}, {Querejeta}, {Leung},
  {Gadotti}, {Husemann}, {Falc{\'o}n-Barroso}, {S{\'a}nchez-Bl{\'a}zquez}, {van
  de Ven}, {Kim}, {Coelho}, {Lyubenova}, {de Lorenzo-C{\'a}ceres}, {Martig},
  {Martinez-Valpuesta}, {Neumann}, {P{\'e}rez}, \& {Seidel}}]{Leaman+2019}
{Leaman}, R., {Fragkoudi}, F., {Querejeta}, M., {et~al.} 2019, \mnras, 488,
  3904

\bibitem[{{Leitherer} {et~al.}(1999){Leitherer}, {Schaerer}, {Goldader},
  {Delgado}, {Robert}, {Kune}, {de Mello}, {Devost}, \&
  {Heckman}}]{Leitherer+1999}
{Leitherer}, C., {Schaerer}, D., {Goldader}, J.~D., {et~al.} 1999, \apjs, 123,
  3

\bibitem[{{Leja} {et~al.}(2019){Leja}, {Carnall}, {Johnson}, {Conroy}, \&
  {Speagle}}]{Leja+2019}
{Leja}, J., {Carnall}, A.~C., {Johnson}, B.~D., {Conroy}, C., \& {Speagle},
  J.~S. 2019, \apj, 876, 3

\bibitem[{{Leja} {et~al.}(2017){Leja}, {Johnson}, {Conroy}, {van Dokkum}, \&
  {Byler}}]{Leja+2017}
{Leja}, J., {Johnson}, B.~D., {Conroy}, C., {van Dokkum}, P.~G., \& {Byler}, N.
  2017, \apj, 837, 170

\bibitem[{{Leroy} {et~al.}(2009){Leroy}, {Walter}, {Bigiel}, {Usero}, {Weiss},
  {Brinks}, {de Blok}, {Kennicutt}, {Schuster}, {Kramer}, {Wiesemeyer}, \&
  {Roussel}}]{Leroy+2009}
{Leroy}, A.~K., {Walter}, F., {Bigiel}, F., {et~al.} 2009, \aj, 137, 4670

\bibitem[{{Liang} {et~al.}(2020){Liang}, {Feldmann}, {Hayward}, {Narayanan},
  {{\c{C}}atmabacak}, {Kere{\v{s}}}, {Faucher-Gigu{\`e}re}, \&
  {Hopkins}}]{Liang+2020}
{Liang}, L., {Feldmann}, R., {Hayward}, C.~C., {et~al.} 2020, arXiv e-prints,
  arXiv:2009.13522

\bibitem[{{Lin} {et~al.}(2018){Lin}, {Davies}, {Hicks}, {Burtscher},
  {Contursi}, {Genzel}, {Koss}, {Lutz}, {Maciejewski},
  {M{\"u}ller-S{\'a}nchez}, {Orban de Xivry}, {Ricci}, {Riffel}, {Riffel},
  {Rosario}, {Schartmann}, {Schnorr-M{\"u}ller}, {Shimizu}, {Sternberg},
  {Sturm}, {Storchi-Bergmann}, {Tacconi}, \& {Veilleux}}]{Lin+2018}
{Lin}, M.-Y., {Davies}, R.~I., {Hicks}, E.~K.~S., {et~al.} 2018, \mnras, 473,
  4582

\bibitem[{{Linden} {et~al.}(2020){Linden}, {Murphy}, {Dong}, {Momjian},
  {Kennicutt}, {Meier}, {Schinnerer}, \& {Turner}}]{Linden+2020}
{Linden}, S.~T., {Murphy}, E.~J., {Dong}, D., {et~al.} 2020, \apjs, 248, 25

\bibitem[{{Liu} {et~al.}(2011){Liu}, {Koda}, {Calzetti}, {Fukuhara}, \&
  {Momose}}]{Liu+2011}
{Liu}, G., {Koda}, J., {Calzetti}, D., {Fukuhara}, M., \& {Momose}, R. 2011,
  \apj, 735, 63

\bibitem[{{Lo Faro} {et~al.}(2017){Lo Faro}, {Buat}, {Roehlly},
  {Alvarez-Marquez}, {Burgarella}, {Silva}, \& {Efstathiou}}]{LoFaro+2017}
{Lo Faro}, B., {Buat}, V., {Roehlly}, Y., {et~al.} 2017, \mnras, 472, 1372

\bibitem[{{Lower} {et~al.}(2020){Lower}, {Narayanan}, {Leja}, {Johnson},
  {Conroy}, \& {Dav{\'e}}}]{Lower+2020}
{Lower}, S., {Narayanan}, D., {Leja}, J., {et~al.} 2020, \apj, 904, 33

\bibitem[{{Lutz}(2014)}]{Lutz2014}
{Lutz}, D. 2014, \araa, 52, 373

\bibitem[{{Madau} \& {Dickinson}(2014)}]{MadauDickinson2014}
{Madau}, P., \& {Dickinson}, M. 2014, \araa, 52, 415

\bibitem[{{Marigo} {et~al.}(2017){Marigo}, {Girardi}, {Bressan}, {Rosenfield},
  {Aringer}, {Chen}, {Dussin}, {Nanni}, {Pastorelli}, {Rodrigues}, {Trabucchi},
  {Bladh}, {Dalcanton}, {Groenewegen}, {Montalb{\'a}n}, \&
  {Wood}}]{Marigo+2017}
{Marigo}, P., {Girardi}, L., {Bressan}, A., {et~al.} 2017, \apj, 835, 77

\bibitem[{{Martin} {et~al.}(2005){Martin}, {Fanson}, {Schiminovich},
  {Morrissey}, {Friedman}, {Barlow}, {Conrow}, {Grange}, {Jelinsky},
  {Milliard}, {Siegmund}, {Bianchi}, {Byun}, {Donas}, {Forster}, {Heckman},
  {Lee}, {Madore}, {Malina}, {Neff}, {Rich}, {Small}, {Surber}, {Szalay},
  {Welsh}, \& {Wyder}}]{Martin+2005}
{Martin}, D.~C., {Fanson}, J., {Schiminovich}, D., {et~al.} 2005, \apjl, 619,
  L1

\bibitem[{{Masters} {et~al.}(2010){Masters}, {Mosleh}, {Romer}, {Nichol},
  {Bamford}, {Schawinski}, {Lintott}, {Andreescu}, {Campbell}, {Crowcroft},
  {Doyle}, {Edmondson}, {Murray}, {Raddick}, {Slosar}, {Szalay}, \&
  {Vandenberg}}]{Masters+2010}
{Masters}, K.~L., {Mosleh}, M., {Romer}, A.~K., {et~al.} 2010, \mnras, 405, 783

\bibitem[{{Masters} {et~al.}(2011){Masters}, {Nichol}, {Hoyle}, {Lintott},
  {Bamford}, {Edmondson}, {Fortson}, {Keel}, {Schawinski}, {Smith}, \&
  {Thomas}}]{Masters+2011}
{Masters}, K.~L., {Nichol}, R.~C., {Hoyle}, B., {et~al.} 2011, \mnras, 411,
  2026

\bibitem[{{Mathis}(1990)}]{Mathis1990}
{Mathis}, J.~S. 1990, \araa, 28, 37

\bibitem[{{McLure} {et~al.}(2018){McLure}, {Dunlop}, {Cullen}, {Bourne},
  {Best}, {Khochfar}, {Bowler}, {Biggs}, {Geach}, {Scott}, {Micha{\l}owski},
  {Rujopakarn}, {van Kampen}, {Kirkpatrick}, \& {Pope}}]{McLure+2018}
{McLure}, R.~J., {Dunlop}, J.~S., {Cullen}, F., {et~al.} 2018, \mnras, 476,
  3991

\bibitem[{{Meidt} {et~al.}(2012){Meidt}, {Schinnerer}, {Knapen}, {Bosma},
  {Athanassoula}, {Sheth}, {Buta}, {Zaritsky}, {Laurikainen}, {Elmegreen},
  {Elmegreen}, {Gadotti}, {Salo}, {Regan}, {Ho}, {Madore}, {Hinz}, {Skibba},
  {Gil de Paz}, {Mu{\~n}oz-Mateos}, {Men{\'e}ndez-Delmestre}, {Seibert}, {Kim},
  {Mizusawa}, {Laine}, \& {Comer{\'o}n}}]{Meidt+2012}
{Meidt}, S.~E., {Schinnerer}, E., {Knapen}, J.~H., {et~al.} 2012, \apj, 744, 17

\bibitem[{{Meurer} {et~al.}(1999){Meurer}, {Heckman}, \&
  {Calzetti}}]{Meurer+1999}
{Meurer}, G.~R., {Heckman}, T.~M., \& {Calzetti}, D. 1999, \apj, 521, 64

\bibitem[{{Moustakas} {et~al.}(2010){Moustakas}, {Kennicutt}, {Tremonti},
  {Dale}, {Smith}, \& {Calzetti}}]{Moustakas+2010}
{Moustakas}, J., {Kennicutt}, Robert~C., J., {Tremonti}, C.~A., {et~al.} 2010,
  \apjs, 190, 233

\bibitem[{{Narayanan} {et~al.}(2018){Narayanan}, {Conroy}, {Dav{\'e}},
  {Johnson}, \& {Popping}}]{Narayanan+2018}
{Narayanan}, D., {Conroy}, C., {Dav{\'e}}, R., {Johnson}, B.~D., \& {Popping},
  G. 2018, \apj, 869, 70

\bibitem[{{Nersesian} {et~al.}(2020){Nersesian}, {Verstocken}, {Viaene},
  {Baes}, {Xilouris}, {Bianchi}, {Casasola}, {Clark}, {Davies}, {De Looze}, {De
  Vis}, {Dobbels}, {Fritz}, {Galametz}, {Galliano}, {Jones}, {Madden},
  {Mosenkov}, {Tr{\v{c}}ka}, \& {Ysard}}]{Nersesian+2020}
{Nersesian}, A., {Verstocken}, S., {Viaene}, S., {et~al.} 2020, \aap, 637, A25

\bibitem[{{Noll} {et~al.}(2009){Noll}, {Burgarella}, {Giovannoli}, {Buat},
  {Marcillac}, \& {Mu{\~n}oz-Mateos}}]{Noll+2009}
{Noll}, S., {Burgarella}, D., {Giovannoli}, E., {et~al.} 2009, \aap, 507, 1793

\bibitem[{{Oey} {et~al.}(2007){Oey}, {Meurer}, {Yelda}, {Furst},
  {Caballero-Nieves}, {Hanish}, {Levesque}, {Thilker}, {Walth},
  {Bland-Hawthorn}, {Dopita}, {Ferguson}, {Heckman}, {Doyle}, {Drinkwater},
  {Freeman}, {Kennicutt}, {Kilborn}, {Knezek}, {Koribalski}, {Meyer}, {Putman},
  {Ryan-Weber}, {Smith}, {Staveley-Smith}, {Webster}, {Werk}, \&
  {Zwaan}}]{Oey+2007}
{Oey}, M.~S., {Meurer}, G.~R., {Yelda}, S., {et~al.} 2007, \apj, 661, 801

\bibitem[{{Osterbrock} \& {Ferland}(2006)}]{Osterbrock+2006}
{Osterbrock}, D.~E., \& {Ferland}, G.~J. 2006, {Astrophysics of gaseous nebulae
  and active galactic nuclei}

\bibitem[{{Overzier} {et~al.}(2011){Overzier}, {Heckman}, {Wang}, {Armus},
  {Buat}, {Howell}, {Meurer}, {Seibert}, {Siana}, {Basu-Zych}, {Charlot},
  {Gon{\c{c}}alves}, {Martin}, {Neill}, {Rich}, {Salim}, \&
  {Schiminovich}}]{Overzier+2011}
{Overzier}, R.~A., {Heckman}, T.~M., {Wang}, J., {et~al.} 2011, \apjl, 726, L7

\bibitem[{{Pei}(1992)}]{Pei1992}
{Pei}, Y.~C. 1992, \apj, 395, 130

\bibitem[{{Peng} {et~al.}(2010){Peng}, {Lilly}, {Kova{\v{c}}}, {Bolzonella},
  {Pozzetti}, {Renzini}, {Zamorani}, {Ilbert}, {Knobel}, {Iovino}, {Maier},
  {Cucciati}, {Tasca}, {Carollo}, {Silverman}, {Kampczyk}, {de Ravel},
  {Sanders}, {Scoville}, {Contini}, {Mainieri}, {Scodeggio}, {Kneib}, {Le
  F{\`e}vre}, {Bardelli}, {Bongiorno}, {Caputi}, {Coppa}, {de la Torre},
  {Franzetti}, {Garilli}, {Lamareille}, {Le Borgne}, {Le Brun}, {Mignoli},
  {Perez Montero}, {Pello}, {Ricciardelli}, {Tanaka}, {Tresse}, {Vergani},
  {Welikala}, {Zucca}, {Oesch}, {Abbas}, {Barnes}, {Bordoloi}, {Bottini},
  {Cappi}, {Cassata}, {Cimatti}, {Fumana}, {Hasinger}, {Koekemoer},
  {Leauthaud}, {Maccagni}, {Marinoni}, {McCracken}, {Memeo}, {Meneux}, {Nair},
  {Porciani}, {Presotto}, \& {Scaramella}}]{Peng2010}
{Peng}, Y.-j., {Lilly}, S.~J., {Kova{\v{c}}}, K., {et~al.} 2010, \apj, 721, 193

\bibitem[{{Pope} {et~al.}(2017){Pope}, {Monta{\~n}a}, {Battisti}, {Limousin},
  {Marchesini}, {Wilson}, {Alberts}, {Aretxaga}, {Avila-Reese}, {Ram{\'o}n
  Bermejo-Climent}, {Brammer}, {Bravo-Alfaro}, {Calzetti}, {Chary}, {Cybulski},
  {Giavalisco}, {Hughes}, {Kado-Fong}, {Keller}, {Kirkpatrick}, {Labbe},
  {Lange-Vagle}, {Lowenthal}, {Murphy}, {Oesch}, {Rosa Gonzalez},
  {S{\'a}nchez-Arg{\"u}elles}, {Shipley}, {Stefanon}, {Vega}, {Whitaker},
  {Williams}, {Yun}, {Zavala}, \& {Zeballos}}]{Pope+2017}
{Pope}, A., {Monta{\~n}a}, A., {Battisti}, A., {et~al.} 2017, \apj, 838, 137

\bibitem[{{Popping} {et~al.}(2017){Popping}, {Puglisi}, \&
  {Norman}}]{Popping+2017}
{Popping}, G., {Puglisi}, A., \& {Norman}, C.~A. 2017, \mnras, 472, 2315

\bibitem[{{Reddy} {et~al.}(2012{\natexlab{a}}){Reddy}, {Dickinson}, {Elbaz},
  {Morrison}, {Giavalisco}, {Ivison}, {Papovich}, {Scott}, {Buat},
  {Burgarella}, {Charmandaris}, {Daddi}, {Magdis}, {Murphy}, {Altieri},
  {Aussel}, {Dannerbauer}, {Dasyra}, {Hwang}, {Kartaltepe}, {Leiton},
  {Magnelli}, \& {Popesso}}]{Reddy+2012}
{Reddy}, N., {Dickinson}, M., {Elbaz}, D., {et~al.} 2012{\natexlab{a}}, \apj,
  744, 154

\bibitem[{{Reddy} {et~al.}(2010){Reddy}, {Erb}, {Pettini}, {Steidel}, \&
  {Shapley}}]{Reddy+2010}
{Reddy}, N.~A., {Erb}, D.~K., {Pettini}, M., {Steidel}, C.~C., \& {Shapley},
  A.~E. 2010, \apj, 712, 1070

\bibitem[{{Reddy} {et~al.}(2012{\natexlab{b}}){Reddy}, {Pettini}, {Steidel},
  {Shapley}, {Erb}, \& {Law}}]{Reddy+2012b}
{Reddy}, N.~A., {Pettini}, M., {Steidel}, C.~C., {et~al.} 2012{\natexlab{b}},
  \apj, 754, 25

\bibitem[{{Reddy} {et~al.}(2006){Reddy}, {Steidel}, {Fadda}, {Yan}, {Pettini},
  {Shapley}, {Erb}, \& {Adelberger}}]{Reddy+2006}
{Reddy}, N.~A., {Steidel}, C.~C., {Fadda}, D., {et~al.} 2006, \apj, 644, 792

\bibitem[{{Reddy} {et~al.}(2015){Reddy}, {Kriek}, {Shapley}, {Freeman},
  {Siana}, {Coil}, {Mobasher}, {Price}, {Sanders}, \& {Shivaei}}]{Reddy+2015}
{Reddy}, N.~A., {Kriek}, M., {Shapley}, A.~E., {et~al.} 2015, \apj, 806, 259

\bibitem[{{Reddy} {et~al.}(2018){Reddy}, {Oesch}, {Bouwens}, {Montes},
  {Illingworth}, {Steidel}, {van Dokkum}, {Atek}, {Carollo}, {Cibinel},
  {Holden}, {Labb{\'e}}, {Magee}, {Morselli}, {Nelson}, \&
  {Wilkins}}]{Reddy+2018}
{Reddy}, N.~A., {Oesch}, P.~A., {Bouwens}, R.~J., {et~al.} 2018, \apj, 853, 56

\bibitem[{{Regan} {et~al.}(2001){Regan}, {Thornley}, {Helfer}, {Sheth}, {Wong},
  {Vogel}, {Blitz}, \& {Bock}}]{Regan+2001}
{Regan}, M.~W., {Thornley}, M.~D., {Helfer}, T.~T., {et~al.} 2001, \apj, 561,
  218

\bibitem[{{Regan} {et~al.}(2006){Regan}, {Thornley}, {Vogel}, {Sheth},
  {Draine}, {Hollenbach}, {Meyer}, {Dale}, {Engelbracht}, {Kennicutt}, {Armus},
  {Buckalew}, {Calzetti}, {Gordon}, {Helou}, {Leitherer}, {Malhotra}, {Murphy},
  {Rieke}, {Rieke}, \& {Smith}}]{Regan+2006}
{Regan}, M.~W., {Thornley}, M.~D., {Vogel}, S.~N., {et~al.} 2006, \apj, 652,
  1112

\bibitem[{{Sabbi} {et~al.}(2018){Sabbi}, {Calzetti}, {Ubeda}, {Adamo},
  {Cignoni}, {Thilker}, {Aloisi}, {Elmegreen}, {Elmegreen}, {Gouliermis},
  {Grebel}, {Messa}, {Smith}, {Tosi}, {Dolphin}, {Andrews}, {Ashworth},
  {Bright}, {Brown}, {Chandar}, {Christian}, {Clayton}, {Cook}, {Dale}, {de
  Mink}, {Dobbs}, {Evans}, {Fumagalli}, {Gallagher}, {Grasha}, {Herrero},
  {Hunter}, {Johnson}, {Kahre}, {Kennicutt}, {Kim}, {Krumholz}, {Lee},
  {Lennon}, {Martin}, {Nair}, {Nota}, {{\"O}stlin}, {Pellerin}, {Prieto},
  {Regan}, {Ryon}, {Sacchi}, {Schaerer}, {Schiminovich}, {Shabani}, {Van Dyk},
  {Walterbos}, {Whitmore}, \& {Wofford}}]{Sabbi+2018}
{Sabbi}, E., {Calzetti}, D., {Ubeda}, L., {et~al.} 2018, \apjs, 235, 23

\bibitem[{{Salim} \& {Boquien}(2019)}]{Salim+2019}
{Salim}, S., \& {Boquien}, M. 2019, \apj, 872, 23

\bibitem[{{Salim} {et~al.}(2018){Salim}, {Boquien}, \& {Lee}}]{Salim+2018}
{Salim}, S., {Boquien}, M., \& {Lee}, J.~C. 2018, \apj, 859, 11

\bibitem[{{Salim} \& {Narayanan}(2020)}]{Salim+2020}
{Salim}, S., \& {Narayanan}, D. 2020, arXiv e-prints, arXiv:2001.03181

\bibitem[{{Salmon} {et~al.}(2016){Salmon}, {Papovich}, {Long}, {Willner},
  {Finkelstein}, {Ferguson}, {Dickinson}, {Duncan}, {Faber}, {Hathi},
  {Koekemoer}, {Kurczynski}, {Newman}, {Pacifici}, {P{\'e}rez-Gonz{\'a}lez}, \&
  {Pforr}}]{Salmon+2016}
{Salmon}, B., {Papovich}, C., {Long}, J., {et~al.} 2016, \apj, 827, 20

\bibitem[{{Schruba} {et~al.}(2011){Schruba}, {Leroy}, {Walter}, {Bigiel},
  {Brinks}, {de Blok}, {Dumas}, {Kramer}, {Rosolowsky}, {Sand strom},
  {Schuster}, {Usero}, {Weiss}, \& {Wiesemeyer}}]{Schruba+2011}
{Schruba}, A., {Leroy}, A.~K., {Walter}, F., {et~al.} 2011, \aj, 142, 37

\bibitem[{{Schulz} {et~al.}(2020){Schulz}, {Popping}, {Pillepich}, {Nelson},
  {Vogelsberger}, {Marinacci}, \& {Hernquist}}]{Schulz+2020}
{Schulz}, S., {Popping}, G., {Pillepich}, A., {et~al.} 2020, \mnras, 497, 4773

\bibitem[{{Scoville} {et~al.}(2015){Scoville}, {Faisst}, {Capak}, {Kakazu},
  {Li}, \& {Steinhardt}}]{Scoville+2015}
{Scoville}, N., {Faisst}, A., {Capak}, P., {et~al.} 2015, \apj, 800, 108

\bibitem[{{Seon} \& {Draine}(2016)}]{Seon+2016}
{Seon}, K.-I., \& {Draine}, B.~T. 2016, \apj, 833, 201

\bibitem[{{S{\'e}rsic} \& {Pastoriza}(1965)}]{Sersic+1965}
{S{\'e}rsic}, J.~L., \& {Pastoriza}, M. 1965, \pasp, 77, 287

\bibitem[{{Shivaei} {et~al.}(2020{\natexlab{a}}){Shivaei}, {Darvish},
  {Sattari}, {Chartab}, {Mobasher}, {Scoville}, \& {Rieke}}]{Shivaei+2020b}
{Shivaei}, I., {Darvish}, B., {Sattari}, Z., {et~al.} 2020{\natexlab{a}},
  \apjl, 903, L28

\bibitem[{{Shivaei} {et~al.}(2015){Shivaei}, {Reddy}, {Steidel}, \&
  {Shapley}}]{Shivaei+2015}
{Shivaei}, I., {Reddy}, N.~A., {Steidel}, C.~C., \& {Shapley}, A.~E. 2015,
  \apj, 804, 149

\bibitem[{{Shivaei} {et~al.}(2016){Shivaei}, {Kriek}, {Reddy}, {Shapley},
  {Barro}, {Conroy}, {Coil}, {Freeman}, {Mobasher}, {Siana}, {Sanders},
  {Price}, {Azadi}, {Pasha}, \& {Inami}}]{Shivaei+2016}
{Shivaei}, I., {Kriek}, M., {Reddy}, N.~A., {et~al.} 2016, \apjl, 820, L23

\bibitem[{{Shivaei} {et~al.}(2020{\natexlab{b}}){Shivaei}, {Reddy}, {Rieke},
  {Shapley}, {Kriek}, {Battisti}, {Mobasher}, {Sanders}, {Fetherolf}, {Azadi},
  {Coil}, {Freeman}, {de Groot}, {Leung}, {Price}, {Siana}, \&
  {Zick}}]{Shivaei+2020a}
{Shivaei}, I., {Reddy}, N., {Rieke}, G., {et~al.} 2020{\natexlab{b}}, \apj,
  899, 117

\bibitem[{{Soifer} {et~al.}(1984){Soifer}, {Helou}, {Lonsdale}, {Neugebauer},
  {Hacking}, {Houck}, {Low}, {Rice}, \& {Rowan-Robinson}}]{Soifer+1984}
{Soifer}, B.~T., {Helou}, G., {Lonsdale}, C.~J., {et~al.} 1984, \apjl, 283, L1

\bibitem[{{Spinoso} {et~al.}(2017){Spinoso}, {Bonoli}, {Dotti}, {Mayer},
  {Madau}, \& {Bellovary}}]{Spinoso+2017}
{Spinoso}, D., {Bonoli}, S., {Dotti}, M., {et~al.} 2017, \mnras, 465, 3729

\bibitem[{{Stanway} {et~al.}(2020){Stanway}, {Chrimes}, {Eldridge}, \&
  {Stevance}}]{Stanway+2020}
{Stanway}, E.~R., {Chrimes}, A.~A., {Eldridge}, J.~J., \& {Stevance}, H.~F.
  2020, \mnras, 495, 4605

\bibitem[{{Storchi-Bergmann} {et~al.}(1995){Storchi-Bergmann}, {Kinney}, \&
  {Challis}}]{StorchiBergmann+1995}
{Storchi-Bergmann}, T., {Kinney}, A.~L., \& {Challis}, P. 1995, \apjs, 98, 103

\bibitem[{{Teklu} {et~al.}(2020){Teklu}, {Lin}, {Kong}, {Wang}, {Gao}, {Liu},
  {Hu}, \& {Liu}}]{Teklu+2020}
{Teklu}, B.~B., {Lin}, Z., {Kong}, X., {et~al.} 2020, \apj, 893, 94

\bibitem[{{Theios} {et~al.}(2019){Theios}, {Steidel}, {Strom}, {Rudie},
  {Trainor}, \& {Reddy}}]{Theios+2019}
{Theios}, R.~L., {Steidel}, C.~C., {Strom}, A.~L., {et~al.} 2019, \apj, 871,
  128

\bibitem[{{Tr{\v{c}}ka} {et~al.}(2020){Tr{\v{c}}ka}, {Baes}, {Camps}, {Meidt},
  {Trayford}, {Bianchi}, {Casasola}, {Cassar{\`a}}, {De Looze}, {De Vis},
  {Dobbels}, {Fritz}, {Galametz}, {Galliano}, {Katsianis}, {Madden},
  {Mosenkov}, {Nersesian}, {Viaene}, \& {Xilouris}}]{Trcka+2020}
{Tr{\v{c}}ka}, A., {Baes}, M., {Camps}, P., {et~al.} 2020, \mnras, 494, 2823

\bibitem[{{Turner} {et~al.}(2021){Turner}, {Dale}, {Lee}, {Boquien}, {Chandar},
  {Deger}, {Larson}, {Mok}, {Thilker}, {Ubeda}, {Whitmore}, {Belfiore},
  {Bigiel}, {Blanc}, {Emsellem}, {Grasha}, {Groves}, {Klessen}, {Kreckel},
  {Kruijssen}, {Leroy}, {Rosolowsky}, {Sanchez-Blazquez}, {Schinnerer},
  {Schruba}, {Van Dyk}, \& {Williams}}]{Turner+2021}
{Turner}, J.~A., {Dale}, D.~A., {Lee}, J.~C., {et~al.} 2021, arXiv e-prints,
  arXiv:2101.02134

\bibitem[{{Utomo} {et~al.}(2014){Utomo}, {Kriek}, {Labb{\'e}}, {Conroy}, \&
  {Fumagalli}}]{Utomo+2014}
{Utomo}, D., {Kriek}, M., {Labb{\'e}}, I., {Conroy}, C., \& {Fumagalli}, M.
  2014, \apjl, 783, L30

\bibitem[{{van der Laan} {et~al.}(2013){van der Laan}, {Schinnerer},
  {Emsellem}, {Hunt}, {McDermid}, \& {Liu}}]{vanderLaan+2013}
{van der Laan}, T.~P.~R., {Schinnerer}, E., {Emsellem}, E., {et~al.} 2013,
  \aap, 551, A81

\bibitem[{{V{\'a}zquez} \& {Leitherer}(2005)}]{Vazquez+2005}
{V{\'a}zquez}, G.~A., \& {Leitherer}, C. 2005, \apj, 621, 695

\bibitem[{{Weingartner} \& {Draine}(2001)}]{Weingartner+2001}
{Weingartner}, J.~C., \& {Draine}, B.~T. 2001, \apj, 548, 296

\bibitem[{{Wild} {et~al.}(2011){Wild}, {Charlot}, {Brinchmann}, {Heckman},
  {Vince}, {Pacifici}, \& {Chevallard}}]{Wild+2011}
{Wild}, V., {Charlot}, S., {Brinchmann}, J., {et~al.} 2011, \mnras, 417, 1760

\bibitem[{{Witt} {et~al.}(1992){Witt}, {Thronson}, \& {Capuano}}]{Witt+1992}
{Witt}, A.~N., {Thronson}, Harley~A., J., \& {Capuano}, John~M., J. 1992, \apj,
  393, 611

\bibitem[{{Wofford} {et~al.}(2016){Wofford}, {Charlot}, {Bruzual}, {Eldridge},
  {Calzetti}, {Adamo}, {Cignoni}, {de Mink}, {Gouliermis}, {Grasha}, {Grebel},
  {Lee}, {{\"O}stlin}, {Smith}, {Ubeda}, \& {Zackrisson}}]{Wofford+2016}
{Wofford}, A., {Charlot}, S., {Bruzual}, G., {et~al.} 2016, \mnras, 457, 4296

\bibitem[{{Zackrisson} {et~al.}(2011){Zackrisson}, {Rydberg}, {Schaerer},
  {{\"O}stlin}, \& {Tuli}}]{Zackrisson+2011}
{Zackrisson}, E., {Rydberg}, C.-E., {Schaerer}, D., {{\"O}stlin}, G., \&
  {Tuli}, M. 2011, \apj, 740, 13

\bibitem[{{Zaritsky} {et~al.}(2004){Zaritsky}, {Harris}, {Thompson}, \&
  {Grebel}}]{Zaritsky+2004}
{Zaritsky}, D., {Harris}, J., {Thompson}, I.~B., \& {Grebel}, E.~K. 2004, \aj,
  128, 1606

\end{thebibliography}
\bibliographystyle{aasjournal}

\end{document}